\documentclass[final,3p,times]{elsarticle}

\usepackage{algorithmic}
\usepackage{graphicx}
\usepackage{textcomp}
\usepackage[table,xcdraw]{xcolor}
\usepackage{soul}
\usepackage{multirow}
\usepackage{url}
\usepackage{subcaption}
\usepackage{graphicx}
\usepackage{pgfplots}
\usepackage{tikz}
\usepackage{footmisc}
\usepackage{amssymb}
\usepackage{booktabs}
\usepackage{adjustbox}
\usepackage[hidelinks]{hyperref}
\usepackage{listings}
\usepackage{mdframed}
\usepackage{amsmath}
\usepackage{forest}
\usepackage{extracommands}
\usepackage{longtable}
\usepackage{tabularray}

\usepackage{bm}

\usepackage{amsthm}
\newtheorem{definition}{Definition}

\newcommand{\Rlogo}{\protect\includegraphics[height=1.8ex,keepaspectratio]{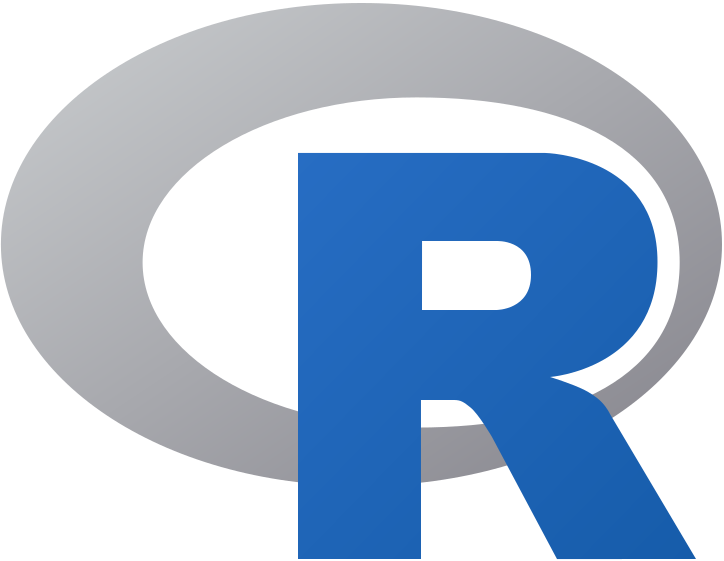}}

\journal{Information Software Technology}

\begin{document}

\begin{frontmatter}

\title{Causal Reasoning in Software Quality Assurance: A Systematic Review}

\author[]{Luca Giamattei\corref{cor1}}
\ead{luca.giamattei@unina.it - Phone: +39 0817683820 - Fax: +39 0817683816}
\cortext[cor1]{Corresponding author}

\author[]{Antonio Guerriero, Roberto Pietrantuono, Stefano Russo\vspace{-3pt}}
\address{DIETI, Università degli Studi di Napoli Federico II, Via Claudio 21, 80125, Napoli, Italy}

\begin{abstract}

\noindent \textbf{Context:} Software Quality Assurance (SQA) is a fundamental part of software engineering to ensure stakeholders that software products work as expected after release in operation. 
Machine Learning (ML) has proven to be able to boost SQA activities and contribute to the development of quality software systems.
In this context, \textit{Causal Reasoning} is gaining increasing interest as a methodology to go beyond a purely data-driven approach by exploiting the use of causality for more effective SQA strategies.

\noindent \textbf{Objective:} Provide a broad and detailed overview of the use of causal reasoning for SQA activities, in order to support researchers to access this research field, identifying room for application, main challenges and research opportunities.

\noindent \textbf{Methods:} A systematic review of the scientific literature on causal reasoning for SQA. The study has found, classified, and analyzed 86 articles, according to established guidelines for software engineering secondary studies.

\noindent \textbf{Results:} 
Results highlight the primary areas within SQA where causal reasoning has been applied, the predominant methodologies used, and the level of maturity of the proposed solutions. Fault localization is the activity where causal reasoning is more exploited, especially in the web services/microservices domain, but other tasks like testing are rapidly gaining popularity. Both causal inference and causal discovery are exploited, with the Pearl's graphical formulation of causality being preferred, likely due to its intuitiveness. Tools to favour their application are appearing at a fast pace -- most of them after 2021.

\noindent \textbf{Conclusions:} The findings show that causal reasoning is a valuable means for SQA tasks with respect to multiple quality attributes, especially during V\&V, evolution and maintenance to ensure reliability, while it is not yet fully exploited for phases like requirements engineering and design. We give a picture of the current landscape, pointing out exciting possibilities for future research.  %

\end{abstract}


\end{frontmatter}

\section{Introduction}
\label{sec:introduction}
Software engineering is an intellectually demanding and creative activity involving complex interdependent tasks aimed at building software products and ensuring their quality. 
The advancement of Machine Learning (ML) fostered a human-machine co-design view to develop dependable systems \cite{Avizienis04}: 
ML algorithms are able to search for significant patterns in large historical datasets gathered throughout the system life cycle, 
thus supporting several software engineering tasks, aimed for instance at fault avoidance (e.g., testing), fault removal (e.g., debugging) and prediction.

While recognizing patterns in data is a fundamental tool for decision-making, well supported by ML, engineers do much more when building and validating a system. They tend to infer cause-effect relationships among the involved variables, and, based on that, infer hypotheses, simulate possible actions, and derive explanations to then support decisions – in other words, they \textit{reason} on what have learned. 
Researchers have been trying to explain causality using statistical and ML methods for years. However, these methods are able to identify connections between variables like correlation and regression but fall short in detecting causality. Techniques that merely recognize patterns in data only get us to what Pearl and Mackenzie \cite{Pearl18} call the first rung of the causation ladder (i.e., “association”). At this level, we are limited to reason about what has been observed.

\begin{figure}[t]
	\centering 	\includegraphics[width=.94\textwidth]{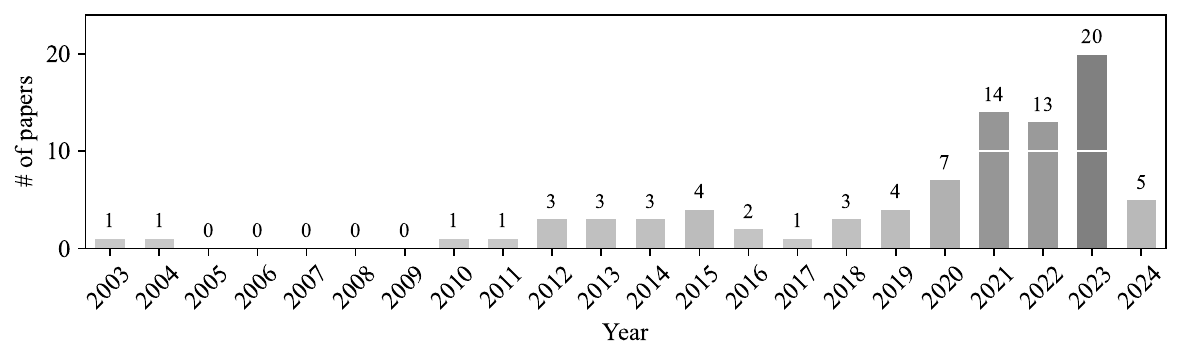}
	\vspace{-9pt}
	\caption{Number of papers on Causal Reasoning in SQA per year.}
	\label{fig:years}
	\vspace{-6pt}
\end{figure}

Currently, software engineering researchers are exploring the usage of \textit{Causal Reasoning} (CR) methodologies to go beyond a purely data-driven approach and to exploit the use of causality for finer-tuned strategies. Through the years, researchers have designed a number of techniques able to capture the essence of CR and provide mathematical frameworks for it. This enables the replication of human reasoning on modern machines, greatly enhancing it with computational power. 
While CR finds consolidated ground in many domains (e.g., epidemiology, economics, social sciences, etc.), it has recently raised interest in software engineering, particularly in the context of software quality assurance (SQA), mainly due to the seminal work of the Turing award-winner Pearl \cite{pearl2009}. Figure \ref{fig:years} shows the number of papers on the topic of CR in SQA between 2003 and (June) 2024 (we explain the search and collection process of these papers in Section \ref{sec:methodology}). Notably, almost 70\% of the papers have been published in the last three years (i.e., starting from 2021), highlighting the emergence of CR as a methodology for SQA practices.

Given its inherent multi-disciplinary history, the study of causality is very fragmented and this has been reversed also in the software engineering literature. For this reason, systematic reviews can greatly contribute to a better understanding of the current landscape and guide future research directions.
A recent study by Siebert \cite{Siebert23} reviews the use of CR in software engineering, covering 25 papers published between 2010 and 2022. They focus on the use of graphical causal models (introduced by Wright \cite{Wright1921} and more recently by Pearl \cite{pearl2009}) and notice an increasing trend, with works mostly pertaining to root cause analysis and fault localization. 

We present a systematic review aiming to comprehensively investigate diverse causal models and their applications across various dimensions of SQA within the software engineering process.
Starting from an initial set of 770 papers, we ended up with a final list of 86 papers that depict the current landscape of ideas, trends, and challenges in using CR for a variety of quality-related tasks (e.g., fault prevention, prediction, tolerance, removal) and attributes (e.g.,  reliability, security, performance), along the whole lifecycle. 

The study investigates three research questions concerning: the activities CR is used for along the software life cycle (RQ1), how CR is used (RQ2), and if the CR-based solutions are experimentally validated, and in which domain (RQ3).
The work provides a comprehensive overview of the state of the art in the use of CR in quality-related activities in software engineering. 

Besides corroborating and updating the trends identified by Siebert \cite{Siebert23}, results show the main areas of SQA in which CR has been employed, identifying the main methodologies for causal discovery and inference, and the maturity of the proposals. Additionally, we provide an extensive discussion highlighting the main findings, open challenges and research opportunities.
The findings provide evidence that the software engineering community today considers CR as a valuable means for several SQA tasks, including testing, fault localization, fault and quality attributes prediction.

The paper is structured as follows. Section \ref{sec:background} provides background on CR and SQA. Section \ref{sec:methodology} presents the three research questions and the methodology followed from the search to the analysis of the literature. Sections \ref{sec:rq1}-\ref{sec:rq3} show the results for the research questions. Section \ref{sec:discussion} discusses the main findings of the analysis, as well as open challenges and research opportunities. Section \ref{sec:threats} analyzes threats to validity. Section \ref{sec:related} reviews the related secondary studies. Section \ref{sec:conclusions} provides the conclusions.
We release data and scripts to replicate this study in the online repository.\footnote{{Replication package available at: \url{https://zenodo.org/doi/10.5281/zenodo.12515969.}}\label{note1}}

\section{Background}%
\label{sec:background}
 This Section sets out the background concepts and the terminology used throughout the paper, with respect to the two drivers of this work: \textbf{causal reasoning} and \textbf{software quality assurance}.

\subsection{Causal Reasoning}

CR has its roots in the $20^{th}$ century. It had its first applications in a variety of domains, like epidemiology \cite{Hernan00}, economics \cite{Imbens04}, social and medical sciences \cite{Morgan2014,Mani01}, education \cite{Rajeev99, LaLonde86}, with significant impact on better understanding and explaining complex phenomena, making predictions, and managing decision processes. 
CR mimic something natural in human logic, namely the identification of causes and effects and the answering of “\textit{what if}” (causal) questions. Pearl and Mackenzie \cite{Pearl18} have conceptualized a framework describing three levels of understanding and reasoning about causation, called the “ladder of causation”. 
The first level is association (correlation), based only on what we observe. It is a basic understanding that two variables tend to co-occur, making no claims about causation. %
CR allows to stair up to the second and third rung of the ladder, by performing, respectively, “interventions” (evaluating the impact of actively manipulating one variable to observe the effect on another) and “counterfactuals” (exploring hypothetical scenarios, that did not occur but help understand what might have happened under different conditions).

Through the years a number of frameworks have been proposed to allow machines to reason with causality, by modeling cause-effect relations and quantifying the effects of causes through what is called “causal inference”.

\subsubsection{Causal frameworks}
\label{sec:background_causal_frameworks}
Causal frameworks are mathematical and/or graphical representations of causal relationships within an individual system or population \cite{wire,stanford-causal-models}. They embed some form of external knowledge, like prior knowledge of the data generation mechanism and assumptions about plausible causal mechanisms, that distinguishes them from associational methods \cite{sharmaBook}.
They are used to learn and estimate causal effects between involved variables of a certain system, facilitating causal inference. One of the first conceptual frameworks is the “potential outcome"  by Neyman \cite{Neyman90,Neyman1923} and Rubin \cite{Rubin74} (also called Neyman-Rubin Causal Model). 
It allows estimating counterfactuals, i.e., estimating the causal effect of a treatment on an individual by comparing what happened (\textit{observed outcome}) to what would have happened in the absence of that treatment (\textit{potential outcome}).
The formal definitions follow \cite{pearl2009, wire,Guo2020}.

\begin{definition}
	\textbf{Potential Outcome (PO)} \\A potential outcome $Y_{T=t}(u)$ of a variable Y is the value that Y would have taken for individual $u$, had T (treatment variable) assumed the value t.
	\label{def:po}
\end{definition}
A potential outcome is fundamentally distinct from an actual outcome. The first is a hypothetical outcome that, differently from the second, has not been observed.
For instance, in case of a binary treatment $T=\{0,1\}$, if we observe an outcome $Y_{T=1}(u)$, we cannot observe $Y_{T=0}(u)$ at the same time -- e.g., if an individual took a medicine ($T=1$), we cannot observe the case in which (s)he did not take it ($T=0$)). This is a potential outcome (also called \textit{counterfactual outcome}), and one retrospectively reasons about \textit{“what would have happened if"}, as opposed to the observed outcome (or \textit{factual outcome}). 

The PO framework, defined at the level of an individual, aims to estimate a potential outcome and then compute the treatment effect. An example metric is the \textit{Individual Treatment Effect} (ITE):
\begin{equation}
	ITE(u) = Y_{T=1}(u) - Y_{T=0}(u)
\end{equation}

The impossibility of simultaneously observing both potential outcomes, and consequently observing the effect of a treatment \textit{t} on an individual \textit{u}, is the \textit{fundamental problem of causal inference} \cite{Holland86}. However, statistical solutions allow to replace this impossible problem of observing the causal effect of \textit{t} on a specific individual with the feasible estimation of the average causal effect of \textit{t} over a population of individuals.
At the population level, we define the \textit{treated} group, including all the treated individuals ($T=1$), and the \textit{control} group ($T=0$). For simplicity, we use a binary treatment; otherwise multiple treated groups are formed.
The ITE computed at population level is called \textit{Average Treatment Effect} (ATE):

\begin{equation}
	ATE = E[Y_{T=1} - Y_{T=0}]
\end{equation}

ATE is also often called Average Causal Effect (ACE) in the counterfactual literature \cite{Morgan2014}. In general, ACE refers to the estimation of the effect of the change of value of a certain cause to a variable of interest \cite{pearl2009}. More broadly average causal effects include a variety of metrics,
such as Average Treatment effect on the Treated group (ATT), and Conditional Average Treatment Effect (CATE) \cite{wire}.

To compute (when feasible) an unbiased estimation of these metrics, randomized controlled trials (i.e., random assignment of \textit{control} and \textit{treatment} groups) are performed. 
Otherwise, in presence of observational data, a number of causal inference methods can be employed (introduced in the next section).
To ease the estimation of treatment effects, some assumptions are usually made, such as the Stable Unit Treatment Value Assumption (SUTVA) \cite{Cox1958,Rubin80}. SUTVA specifies that the value of Y for unit \textit{u} when exposed to treatment \textit{t} will be the same no matter what mechanism is used to assign treatment \textit{t} to unit \textit{u} and no matter what treatments the other units receive.

Another powerful way of dealing with causality is to use graphical models - usually in the form of Direct Acyclic Graphs (DAG). In a CausalDAG, nodes are the variables of the system and edges are interpreted as causal relationships between them \cite{Guo2020}.

\begin{definition}
	\textbf{Causal Direct Acyclic Graph (CausalDAG)} \\A CausalDAG $\mathcal{G}=(\bm{X}, \mathcal{E})$ is a directed acyclic graph that describes the causal effects between variables, where $\bm{X}$ is the node set and $\mathcal{E}$ the edge set. In a CausalDAG, each node represents a random variable including the treatment, the outcome, and other observed and unobserved variables. A directed edge $x\rightarrow y$ denotes a causal effect of x on y.
	\label{def:causaldag}
\end{definition}

A causal framework building upon CausalDAGs are \textit{Causal Bayesian Networks (CBN)}. A CBN can be viewed as a CausalDAG where nodes are variables and edges represent cause-effect relationships through conditional probabilities. It  provides a causal interpretation of a conventional Bayesian Network (BN). Specifically, a BN is defined as follows \cite{wire}:%

\begin{definition}
	\textbf{Bayesian Network (BN)} \\A BN is a pair $ \mathcal{B} = (\mathcal{G}, P)$ where $P$ factorizes over the graph $\mathcal{G}$ and is expressed as a product of conditional probability distributions associated with $\mathcal{G}$'s nodes $P(X_1,...,X_n)=\prod_{i=1}^{n}P(X_i|Pa_{X_i}^\mathcal{G})$ where \( X_i\) are the nodes of $\mathcal{G}$ and \( Pa_{X_i}^\mathcal{G}\) are their parents, namely the nodes with directed edges pointing to \( X_i \).
	
	\label{def:bn}
\end{definition}

While BN embeds probabilistic knowledge, this is not necessarily \textit{causal} knowledge.  Indeed, \textit{BN} are primarily for estimating the likelihood of one event occurring based on the observation of another (i.e., first level of the Pearl's ladder of causation).
Instead, in causal inference we are interested in the distribution of an outcome variable $X_i$ after \textit{actively setting} another variable $X_k$ to a certain value $x$ (i.e., \textit{doing} an intervention, applying a treatment). 
Pearl introduced the  \textit{do-operator}, a mathematical representation of physical intervention, written as $P(X_i|do(X_k=x))$ \cite{pearl2009,Pearl1999}. The $do$-calculus, along with estimation methods \cite{wire}, supports the inference of type $P(X_i|do(X_k = x))$.  
An intervention $do(X_k=x)$  changes the DAG (hence the distribution), by removing the causal relations with its predecessors (i.e.,  deleting the $Pa(X_k)\rightarrow X_k$ edges), meaning that $X_k$ is no longer affected by any other variable.
A \textit{do}-intervention changes the data generative process, thus: $P(X_i|do(X_k=x)) \neq P(X_i|(X_k=x))$.
To account for \textit{do}-interventions, and to go over the first level of the Pearl's ladder of causation, CBN provide a causal interpretation of BN where the factorization formula in Definition \ref{def:bn} is extended to accommodate interventions through a \textit{truncated factorization}: $P(X_i|do(X_k=x))=\prod_{i \neq k}P(X_i|Pa_{X_i}^\mathcal{G})$.
In any BN, the edges into $X_i$ mean that the probability of $X_i$ is governed by the conditional probability tables for $X_i$, given observations of its parent variables. %
In CBN, the conditional probability tables specify the probability of $X_i$ given interventions on the parent variables \cite{pearl2009,Pearl18}.

Another important framework based on CausalDAG is the Functional Causal Model (FCM). In FCMs the value of each variable $X_i$ is assumed to be a deterministic function of its parents $Pa(X_i)$ and of the unmeasured disturbance $U_i$ ($X_i = f(Pa(X_i), U_i)$) - the same idea of representing stochasticity as in Variational Auto-Encoders with the \textit{reparametrization trick} \cite{Kingma22}.  These are a non-linear non-parametric generalization of linear Structural Equation Models (SEMs). 
A CausalDAG using an FCM for conditional distributions is a Structural Causal Model (SCM) \cite{pearl2009}.  

\begin{definition}
	\textbf{Structural Causal Model (SCM)} \\A Structural Causal Model is a causalDAG $\mathcal{G}=(\bm{X}, \mathcal{E})$ where causal relationships $\mathcal{E}$ are described as a collection of structural assignments $X_i:=f_i(Pa(X_i), U_i)$ that define the (endogenous) random variables $X_i$ as a function of their parents $Pa(X_i)$ and of (exogenous) independent random noise variables $U_i$. 
	\label{def:scm}
\end{definition}

An SCM enables the estimation of the effects of both interventions and counterfactuals. Although modeling frameworks based on PO and SCM are logically equivalent when addressing counterfactual questions \cite{pearl2009}, they differ in their underlying assumptions and treatment of counterfactuals. In the PO framework, counterfactuals are treated as undefined primitives. In contrast, SCMs derive counterfactuals from more fundamental concepts, such as causal mechanisms and their structure \cite{Pearl18}.

In an SCM, there are no conditional probability tables as in \textit{BN}, the edges simply mean that $X_i$ is a function of its parents, as well as the exogenous variable $U_{X_i}$ \cite{Pearl18}. Interventions are modeled as follows \cite{Peters17}. 
\begin{definition} \textbf{Intervention distribution.} The probability $P(X_i|do(X_k=x))$ over an SCM is the distribution entailed by the SCM obtained by replacing the definition $X_k:=f_k(Pa(X_k), U_k)$ with $X_k:=x$. 
	\label{def:2}
\end{definition}

A number of other models and frameworks to deal with causality have been conceptualized through the years. 
Difference-in-Differences (DiD) \cite{Lechner11}, a common framework in econometrics and health care for estimating the ATE, compares the changes in outcomes over time between a treatment group and a control group \cite{wire}. DiD is based on typical assumptions made also for the PO framework, like SUTVA, but requires additional key assumptions. For instance, it assumes that in the absence of treatment, the average treatment and control groups would have followed parallel trends, and that in the pre-treatment period the treatment had no effect on the pre-treatment population \cite{Lechner11}. Fuzzy Cognitive Maps \cite{Kosko1986} are graph structures for representing causal relationships, with their strength encoded as fuzzy values. It is especially applicable to soft knowledge domains, in presence of uncertain system concepts and relationships. It implements a fuzzy causal algebra aiming at estimating, through backward and forward chaining, \textit{direct} and \textit{indirect} effects between variables \cite{Kosko1986}. However, differently from previous frameworks, its purpose is not to provide rigorous and precise causal methods, but ‘‘semi-quantitative" (i.e. using and producing indicative rather than predictive numerical values) \cite{Barbrook-Johnson22}. Finally, Causal trees \cite{Athey16} are types of decision tree structure that incorporates causal modeling principles. They are based on the PO framework and estimate treatment effects (usually Heterogeneous Treatment Effect \cite{Wager17}) with the use of tree-based methods, like classification and regression trees \cite{Athey16,Tran19}.

\subsubsection{Causal Inference (CI)}

CI aims to estimate the causal effect of a specific variable (referred to as the treatment) on an outcome of interest \cite{wire}. There are two primary strategies for quantifying causal effects: performing randomized experiments, where treatment is randomly assigned, or using CI methods to estimate the treatment effect from observational data. Randomized experiments are the gold standard, they enable causal identification by design but can be too costly, unethical or otherwise not feasible for certain situations \cite{Yao21}. In contrast, methods that rely on observational data have gained popularity due to the abundance of available data. This study focuses on CI methods that leverage observational data.

Given a causal framework, the fundamental objective of CI is to estimate how the outcome variable \textit{Y} changes when we intervene to assign a specific value \textit{t} to the treatment variable \textit{T} (i.e., $P(Y|do(T=t))$). 

\begin{figure}[t]
	\centering 	\includegraphics[width=.91\textwidth]{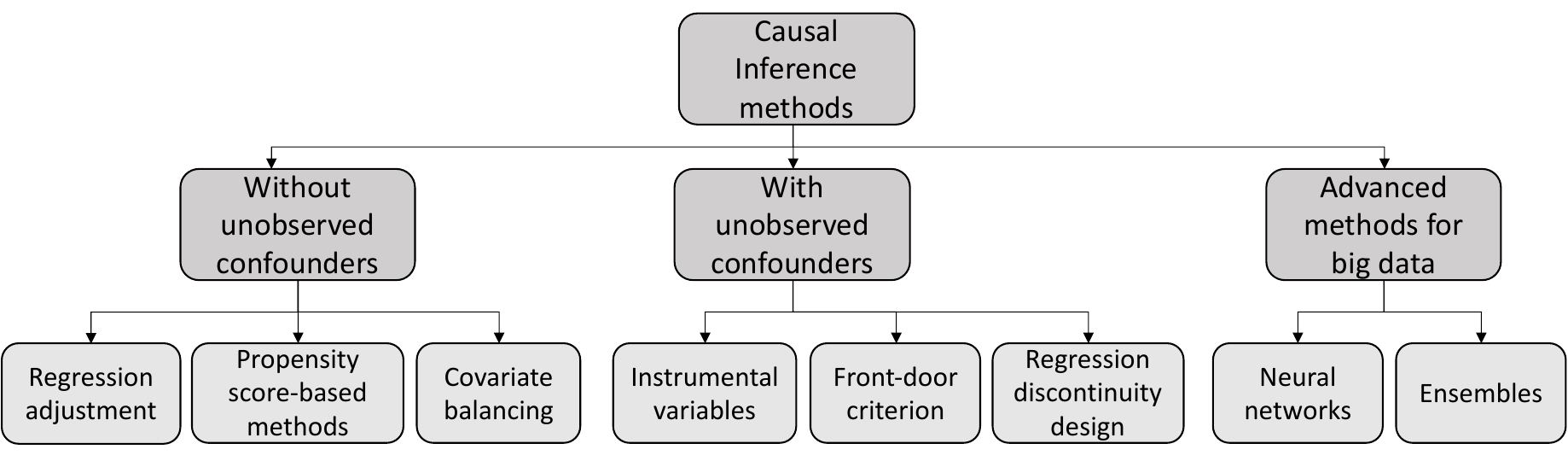}
	\vspace{-3pt}
	\caption{Taxonomy of Causal Inference methods \cite{Guo2020}.}
	\label{fig:ci_methods}
	\vspace{-6pt}
\end{figure}

Reaching this objective involves several steps. The first step is the \textit{identification}, which consists in distinguishing spurious correlations from true causal effects and determining whether and how a causal effect can be estimated from available data. 
It consists in finding the mathematical formula that generates the answer to a causal query or, in other words,
a ‘‘statistical quantity to be estimated from the data that, once estimated, can legitimately represent the answer to the query" \cite{Pearl18}.
However, it may not always be possible to identify this quantity: 
for example, imagine an unmeasured variable \textit{Z} that affects both \textit{X} and \textit{Y}, with \textit{X} also having a direct effect on \textit{Y}. In this case, identifying the causal effect $P(Y|do(X))$ is challenging due to the confounding introduced by \textit{Z}. To address this, model refinement or additional assumptions (e.g., assuming \textit{Z}'s effect is negligible) may be necessary. In graphical causal models, the \textit{do}-calculus provides a set of rules that leverage the structure of the causal graph to transform an interventional/counterfactual query into an observational query. In general, a number of methods can be exploited to control for \textit{confounding bias} (e.g., the \textit{back-door}, \textit{front-door}, \textit{mediators, instrumental variables} \cite{Pearl18,pearl2009,imbens20}).
The second step is the \textit{estimation} of a causal effect from observational data. Once identification is established, a variety of statistical methods can be used to 
this purpose \cite{wire}, including propensity-based stratification \cite{Rosenbaum84}, propensity score matching \cite{Caliendo08}, inverse propensity weighting \cite{vanderwal11}, regression discontinuity \cite{Thistlethwaite1960}, two-stage least square \cite{Theil61}, and generalized linear models \cite{wire}. 
For our investigation, we consider the classification given by Guo \textit{et al.} \cite{Guo2020} on methods to estimate causal effects starting from observational data\footnote{In absence or impossibility to gain experimental data (e.g., randomized controlled trials - Section \ref{sec:background_causal_frameworks})}, shown in Figure \ref{fig:ci_methods}. 
Depending on the assumptions of data, CI methods are separated into two categories: observational data with unconfoundness and with confounding (i.e, without and with unobserved confounders). 

Methods without unobserved confounders include three categories of \textit{adjustment}: \textit{regression adjustment}, \textit{propensity score-based methods}, and \textit{covariate balancing}. Adjustments eliminate the confounding bias based on a set of observed features. In essence, these methods isolate the true causal effect of the treatment by statistically controlling the influence of confounding variables, assuming that all confounder variables are among the observed ones. 

Methods with unobserved confounders relax the assumption of unconfoundness, making them more related to real-world scenarios where unobserved confounders may play a role. In terms of SCM, the assumption on unconfoundness implies that conditioning on a subset of observed variables blocks all back-door paths. These are indirect paths between the treatment and outcome through confounders, which, if not blocked, introduce bias into the causal effect estimation. Without the unconfoundness assumption, conditioning is insufficient, and alternative information must be utilized. These methods include \textit{instrumental variables}, \textit{front-door criterion}, and \textit{regression discontinuity design}. 

With the increasing availability of Big Data, machine learning-based methods have been designed for learning causal effects, mainly through the use of neural networks and ensembles. To go beyond the unconfoundeness assumption, most of these methods learn representations of confounders, under the assumption that they can be learnt from observational data \cite{Guo2020}. For instance, Louizos \textit{et al.} \cite{louizos17} propose a Causal Effect Variational Autoencoder (CEVAE), aiming at estimating a latent-variable model where they simultaneously discover the hidden confounders and infer how they affect treatment and outcome.

Finally, following the Pearlian inference methodology, recently Blöbaum \textit{et al.} \cite{dowhyGCM} have proposed
a \textit{simulation-based} method called \textit{Do-Sampler}, based on sampling from post-intervention distribution of variables. Given a set of interventional variables, \textit{Do-Sampler}: \textit{i)} cuts all their incoming edges (they are intervened, hence no longer causally affected by any other), \textit{ii)} sets the values of these variables to their interventional quantities, and \textit{iii)} propagates those values through the causal graph to compute interventional outcomes with a sampling procedure.
Comprehensive discussions on CI methods have been published by Guo \textit{et al.} \cite{Guo2020}, Yao \textit{et al.} \cite{Yao21}, and Nogueira \textit{et al.} \cite{wire}. 

\subsubsection{Causal Discovery (CD)}

CD aims at learning causal relationships between variables \cite{wire} to build a causal framework explicitly embedding cause-effect relationships, i.e., graphical models.\footnote{Non-graphical causal frameworks, such as PO, do not require CD.} These
can be built in three main ways: by randomized controlled experiments, manually with domain knowledge, or by employing CD algorithms on observational data \cite{Glymour19}. As for CI methods, the focus of this section is on the latter, that aims at extracting a graphical causal structure from observational data, avoiding controlled experiments, which may be too expensive or even technically infeasible \cite{Sjoberg03}. To achieve this, CD algorithms assume that causality can be derived from statistical dependencies. 
These algorithms rely (to a different extent) on various subsets of assumptions, the main ones being the \textit{Causal Markov}, \textit{Faithfulness}, and \textit{Sufficiency} assumptions \cite{Eberhardt2017,Glymour19,wire,wang24}, to derive correspondences between the (conditional) independence in the probability distribution and the causal connectivity relationships in the generated DAG \cite{Eberhardt2017}. The first two conditions are the most important ones and respectively state that: \textit{i)} every vertex X in the graph G is probabilistically independent of its non-descendants given its parents; \textit{ii)} if a variable X is independent of Y given a conditioning set Z
in the probability distribution,\footnote{Roughly, X and Y are independent conditional on a set of variables Z if knowledge about X gives no extra information about Y once you have knowledge of Z.} then X is \textit{d}-separated\footnote{Let $X$,$Y$, and $Z$ be disjoint subsets of all the vertex in the DAG. $Z$ \textit{d-separates} $X$ and $Y$ just in case every path from a variable in $X$ to a variable in $Y$ contains at least one vertex $X_i$ such that either: \textit{i}) $X_i$ is a \textit{collider} (i.e. the arrows converge on $X_i$ in the path), and no descendant of $X_i$ (including $X_i$) is in $Z$; or \textit{ii}) $X_i$ is not a collider, and $X_i$ is in $Z$ \cite{stanford-causal-models}.} from Y given Z in the DAG (in other words, the statistical dependence between variables estimated from the data does not violate the independence defined by any
causal graph that generates the data \cite{Guo2020}). Markov and faithfulness conditions are sufficient to define an equivalence structure over directed acyclic graphs, where graphs that are in the same Markov equivalence class have the same (conditional) independence structure. Sufficiency requires that, for a pair of observed variables, all their common causes must also be observed in the data.

\begin{figure}[t]
	\centering 	\includegraphics[width=.66\textwidth]{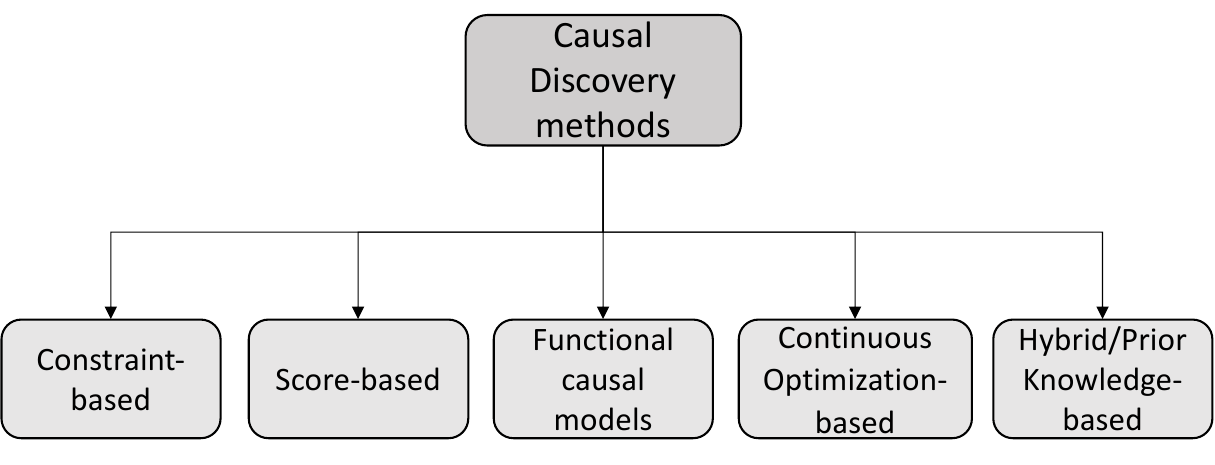}
	\vspace{-3pt}
	\caption{Taxonomy of Causal Discovery methods \cite{wang24}.}
	\label{fig:cd_methods}
	\vspace{-6pt}
\end{figure}

Wang \textit{et al.} \cite{wang24} provide a classification of CD methods, shown in Figure \ref{fig:cd_methods}. In particular, they are divided in \textit{contstraint-based}, \textit{score-based}, \textit{FCM-based} (Functional Causal Models), Continuous Optimization-based, and Hybrid or Prior Knowledge-based methods.

\textit{Constraint-based} algorithms use conditional independence tests on observed data to identify a set of edge constraints \cite{Spirtes21}.
These algorithms have the advantage of being generally applicable, despite being based on the strong assumption of causal faithfulness, therefore often requiring large sample sizes to perform well \cite{Glymour19}.  
Typical (conditional independence) constraint-based algorithms are PC, FCI \cite{Spirtes21} and its improvement RFCI \cite{Colombo12}. 

\textit{Score-based} algorithms optimize a score assigned to candidate graphs, exploiting adjustment techniques such as the Bayesian Information Criterion, that approximates the posterior probability of the model given the data (assuming a uniform prior probability distribution over the DAG space) \cite{Malinsky18}. They relax the faithfulness assumption by replacing conditional independence tests with the goodness-of-fit tests. They are computationally expensive, as they enumerate (and score) every possible graph for the given variables. 
Well known score-based algorithms are GES \cite{Chickering02} and its successor FGES \cite{Ramsey17}, that use parallelization to optimize performance, and GFCI, that combines FCI and FGES \cite{Ogarrio16}. 

\textit{FCM-based} algorithms determine the causal direction of edges, identifying the true causal structure out of all the graphs within a Markov equivalence class. The most noticeable example, working for continuous variables, is LiNGAM, proposed by Shimizu \textit{et al.} \cite{Shimizu06}, where the model is assumed to be linear and non-Gaussian. Under the  causal Markov assumption, acyclicity and a linear non-Gaussian parameterization (i.e., each variable is determined by a linear function of the values of its parents and an additive non-Gaussian noise term), it has been proved that the causal structure can be uniquely determined \cite{Spirtes2016}. FCM-based algorithms have the advantage of not relying on the faithfulness assumption and of learning substantially more about the causal structure, sometimes even determining a unique model. However, 
to relax faithfulness, they introduce other assumptions that can be controversial or difficult to test \cite{Malinsky18} and generally require a larger sample size to be accurate. 

Continuous optimization-based algorithms combine the benefits of score-based methods with gradient optimization by converting the discrete DAG search space into a continuous and optimizable constraint space \cite{wang24}. Usually these methods are machine learning-based and the most representative ones are NOTEARS \cite{Zheng18} and DAG-GNN \cite{Yu19}.

Many CD algorithms allow to specify some constraints (e.g., required or forbidden causal relationships), based on prior human knowledge. The category of Hybrid or Prior Knowledge-based methods like Joint Causal Inference (JCI) \cite{Mooij20} and Max-Min Hill-Climbing (MMHC) \cite{Tsamardinos06} aim to build on top of these algorithms and incorporate specific domain knowledge in addition to observational data in an effective way.

\noindent \textbf{CD for time series.} The conditional-based algorithms like PC, FCI, LiNGAM, although applicable to extract causal relationships between time series, are not specifically designed to this aim. 
Many extensions and specialized algorithms have been developed to address this specific task \cite{Moraffah21}. Among these, a strategy relying on Granger causality \cite{granger69} and its generalization based on (Multivariate) Transfer Entropy \cite{Barnett09} are common choices. Given two time series X and Y, the idea behind Granger causality is to investigate whether the prediction of the current value of time series X improves by incorporating Y's past into its own past \cite{Moraffah21}. In this case, we say that Y \textit{Granger causes} X.

Extensive discussions on CD algorithms have been published by Glymour \textit{et al.} \cite{Glymour19}, Nogueira \textit{et al.} \cite{wire}, Guo \textit{et al.} \cite{Guo2020}, Wang \textit{et al.} \cite{wang24}, and Moraffah \textit{et al.} \cite{Moraffah21}.

\subsection{Software Quality Assurance}
Software quality  is the ``capability of a software product to conform to requirements" \cite{ISO9000,ISO24765}. 
Software Quality Assurance (SQA) 
activities aim to support a justified confidence that the software product conforms to its established requirements \cite{IEEE730}. %
The \textit{SQA process} refers to  
``a set of activities that assess adherence to, and the adequacy of the software processes used to develop and
modify software products. SQA also determines the degree to which the desired results from software quality control are being obtained" \cite{ISO24765}.

A failure of a system is a deviation from its intended service. It is due to the presence of \textit{faults} or \textit{defects} which, once activated, corrupt the system state and may provoke a failure if they reach the interface. Since in the wide literature about software quality some terms have different meanings in different contexts, we hereafter report the meaning of the basic terms as used throughout this paper. We refer to the terminology adopted by the IFIP WG 10.4 on Dependable Computing and Fault Tolerance \footnote{IFIP Working Group 10.4 on Dependable Computing and Fault Tolerance. \url{https://www.dependability.org/?page_id=265}.}
-- due to the foundational work on dependability by Avizienis, Laprie, \textit{et al.} \cite{Avizienis04} - as it better reflects the one adopted in the surveyed papers. 
\begin{description}
	\item A \textbf{Failure} is a deviation of the delivered service  from correct service. 
	\item An \textbf{Error} is that part of the system state that may cause a subsequent failure: a failure occurs when an error reaches the service interface, making the service deviate from correct service.\footnote{The term error in the literature  refers also to the human mistake that introduced a fault in the system \cite{ISO24765,IEEE1044}.}
	\item A \textbf{Fault} is the adjudged or hypothesized cause of an error. An \textit{active} fault produces an error; a fault not activated is said to be \textit{dormant}.  Throughout this paper, with \textit{fault} we implicitly refer to \textit{software faults} if not differently specified. In addition, a fault  in the code, i.e., a fault introduced by a human mistake at some point in the life cycle and that slipped through the code, is also called a \textit{software defect} \cite{IEEE610} or a \textit{bug}.\footnote{The IEEE definition for \textit{fault} is different, as it refers to a defect being activated. %
	} 
\end{description}

\begin{figure}[t]
	\centering 	\includegraphics[width=.63\textwidth]{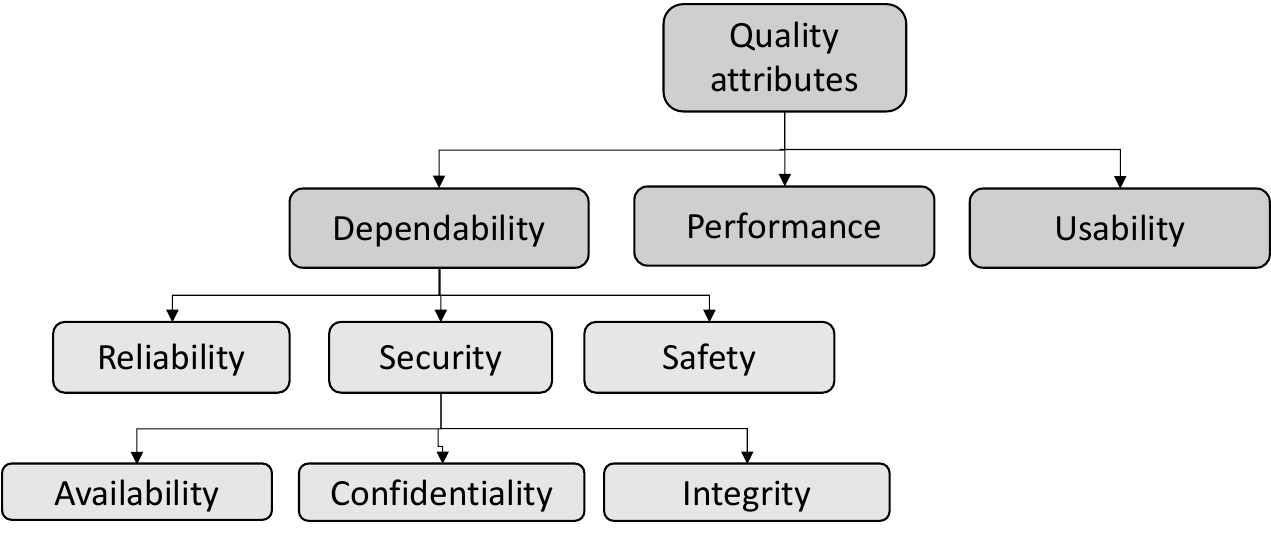}
	\caption{Taxonomy of quality attributes \cite{Avizienis04}.}
	\label{fig:taxonomy}
	\vspace{-6pt}
\end{figure}

The SQA process entails activities spanning the whole life cycle, concerning both \textit{functional} and \textit{non-functional} requirements. %
We categorize activities and quality attributes by borrowing again the taxonomy of dependability and security by Avizienis, Laprie, \textit{et al.} \cite{Avizienis04}, extending it with tasks and attributes not related to dependability and security as reported in Figure \ref{fig:taxonomy}.

As for the means for assuring quality related to \textit{software} faults, we refer to the following categories:  
\begin{description}
	\item \textbf{Fault avoidance/prevention}, to limit the introduction of faults. This is typically achieved by good engineering practices, such as requirements engineering, design modularity, encapsulation and information hiding, reuse. %
	\item \textbf{Fault removal}, to reduce the number and
	severity of faults. This refers to activities aiming at detecting the presence of faults (e.g., testing), localizing them (fault localization) and subsequently correcting the code (debugging). 
	\item \textbf{Fault tolerance}, 
	to tolerate failures, through techniques such as error detection and recovery. 
	
	\item \textbf{(Fault) forecasting}, to \textit{i)} estimate the present number, the future incidence, and the likely consequences of faults, as well as \textit{ii)} to predict  key performance indicators (KPIs) not necessarily related to faults, but whose monitoring and prediction supports future quality improvement actions (hence the parentheses around ``Fault"). Strictly, the category departs from the task of forecasting a fault \cite{Avizienis04}; it is included in our analysis since we target quality attributes beyond dependability and security (e.g., performance, usability). 
\end{description}

As for the quality attributes addressed, we include: 
\begin{itemize}
	\item \textbf{Dependability attributes}.
	Dependability is the \textit{trustworthiness of a computing system which allows reliance to be justifiably placed on the service it delivers}.$^4$ It is an umbrella term encompassing concepts such as reliability, safety, availability and security. We borrow from Avizienis, Laprie, \textit{et al.} \cite{Avizienis04} the following definitions for dependability quality attributes: 
	\begin{description}
		\item \textbf{Availability} refers to the readiness for correct service, defined as 
		``the ability of a system to be in a state to perform a required function at a given instant of time" (recovery after failure is allowed). It is expressed by the probability of failure-free operation at a given point in time. 
		\item \textbf{Reliability} refers to the continuity of a service, that is, for how long it is provided without failures; precisely, it is ``the ability of a system to perform a required function under given conditions for a given time interval". Once the system fails, no recovery takes place (while there can be recovery upon component failures). It is expressed by the probability of failure-free operation in a time interval. 
		Note that in the ISO/IEC 25010 \cite{iso25010} standard reliability has a broader meaning, encompassing concepts like availability and maturity. 
		\item \textbf{Safety} refers to the absence of catastrophic consequences on the user(s) and the environment; it is the probability of catastrophic-failure-free operation during mission time interval. 
		\item \textbf{Maintainability} is the ability to undergo modifications
		and repairs. 
		\item \textbf{Integrity} is the absence of improper system alterations. 
		\item \textbf{Confidentiality} is the absence of unauthorized disclosure of information. 
	\end{description}
	\noindent \textbf{Security} is a macroattribute encompassing Availability, Integrity and Confidentiality. 
\end{itemize}

This taxonomy 
provides quantitative (probability-based) definitions of dependability  attributes and suits the analysis of the papers regarding dependability and security. The same concepts are also present in the ISO-25010 standard \cite{iso25010}, %
although with slightly different definitions (e.g., for reliability).  
For other attributes, we adopt the ISO-25010 definitions: 
\begin{itemize}
	\item \textbf{Performance} is the degree to which \textit{i)} the response and processing times and throughput rates, \textit{ii)} the amounts and types of resources used, and \textit{iii)} the maximum limits of a product or system parameter, when performing its functions, meet requirements. 
	
	\item \textbf{Usability} is the degree to which a product or system can be used by specified users to achieve specified goals with effectiveness, efficiency and satisfaction in a specified context of use \cite{iso25010}. %
	
\end{itemize}

\section{Methodology}
\label{sec:methodology}
\subsection{Research questions}
This study aims to answer the following research questions (\textbf{RQ}) about CR \textbf{in the context of SQA.}
\begin{description}
	\item \textbf{RQ1:} \textbf{Which SQA activities is CR used for and for what quality objective?} \\
	This RQ characterizes \textit{what} CR is used for, namely what are the quality-related tasks and quality objectives addressed by researchers by means of CR (CI and/or CD) methods.
	
	\item \textbf{RQ2:} \textbf{How is CR used?} \\
	This RQ aims to characterize \textit{how} researchers adopted CR for quality improvement and assessment. We investigate what causal task is performed (i.e., CD and/or CI), the causal framework, the CD and CI methodologies, and the tools used. %
	
	\item \textbf{RQ3:} \textbf{Are the solutions using CR experimentally validated?}\\ 
	We finally aim to check the maturity of the proposal; being it a relatively new field of application, we aim to check how many solutions are experimentally validated and, if any, how many solutions are being experimented in industrial environments, and in what domain.
\end{description}

\subsection{Search and selection}

\begin{figure}[t]
	\centering 	
	\includegraphics[width=.77\textwidth]{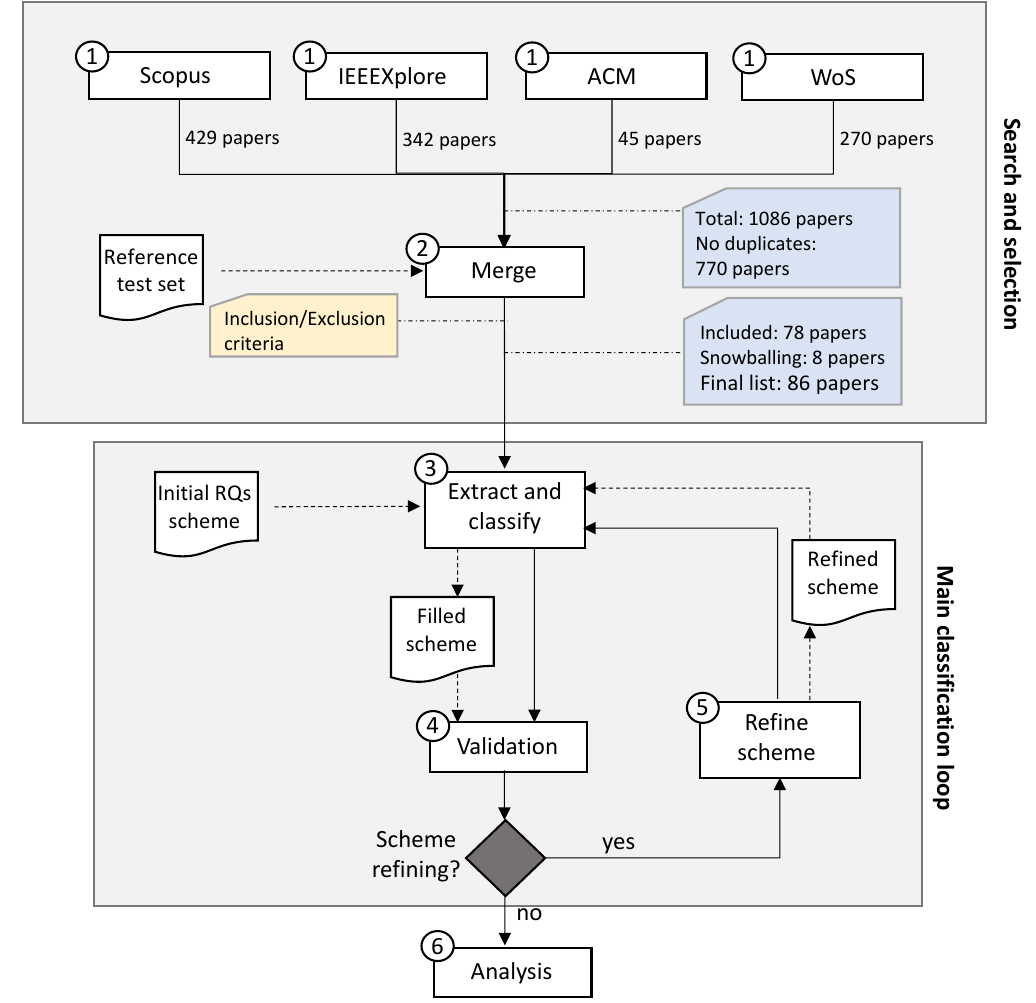}
	\vspace{-3pt}
	\caption{Literature search protocol.}
	\label{fig:protocol}
	\vspace{-3pt}
\end{figure}

To address the above questions, we conducted the research according to the protocol shown in Figure \ref{fig:protocol}. 

The search and selection process %
started from the formulation of the query shown in Listing \ref{lst:searchstring}. We executed the query in June 2024 on four of the most popular digital libraries - \texttt{Scopus}, \texttt{IEEEXplore}, \texttt{ACM Digital Library}, \texttt{ISI Web of Science} %
(step {\textcircled{\raisebox{-.9pt} {1}}}). 
Then, we merged the results and removed duplicates (step {\textcircled{\raisebox{-.9pt} {2}}}), obtaining an initial list of 770 papers.

\lstset{frame=single, backgroundcolor=\color{white}}
\lstset{basicstyle=\fontsize{8}{8}\selectfont\ttfamily}
\lstset{literate={*}{{*}}1}
{\centering
	
	\begin{lstlisting}[extendedchars=false, breaklines, breakatwhitespace, showstringspaces=false, stringstyle=\ttfamily, xleftmargin=.03\textwidth, xrightmargin=.03\textwidth, caption={Literature search string over the main digital libraries}, label={lst:searchstring}]
		("quality" OR "testing" OR "dependability" OR "reliability" OR "availability" OR "performance" OR "security" OR "safety" OR "integrity" OR "confidentiality" OR "usability" OR "maintainability" OR "fault tolerance" OR "fault removal" OR "fault prediction" OR "fault prevention" OR "fault avoidance" OR "fault localization" OR "root cause analysis" OR "debugging") AND ("software") AND ("causality" OR "causal reasoning" OR "causal inference" OR "causal discovery" OR ("pearl" AND ("counterfact*" OR "intervention")))
	\end{lstlisting}
}

\vspace{6pt}
To the list of unique 770 papers we applied the following inclusion and exclusion criteria.
\begin{itemize}
	\item Inclusion criteria
	\begin{itemize}
		
		\item IC1: the paper’s primary goal is software quality assurance/assessment (namely, activities to assure/assess quality related to software faults according to the categories defined in Section 2.2) with the use of CR.

		\item IC2: the target of the software quality assurance/assessment activity is a software application, or a software-enabled component (e.g., a machine learning model like a Deep Neural Network) or system  where software quality has a non-negligible impact on system quality.

\end{itemize}
\item Exclusion criteria
\begin{itemize}
	\item EC1: the paper is a thesis, book, or review.
	\item EC2:        the paper is not subject to peer review, e.g., it is a technical report, a white paper, a project report, government document (i.e., grey literature) or it is a preprint.
	
	\item EC3: the paper presents a CI or CD method, algorithm, library or tool not used for an SQA task. 
	
	\item EC4: the paper is a secondary or tertiary study. %
	\item EC5: the paper is not written in English.
	
\end{itemize}
\end{itemize}
Two researchers were involved in applying the criteria to the entire set of papers, with the support of a senior author as arbiter. Since we only have two raters evaluating identical items, a good fit is to use the Cohen’s kappa to assess the level of agreement. According to the common interpretation in literature \cite{Fleiss03}, the results (k = 0.778) indicate substantial agreement.
This inclusion/exclusion process allowed us to exclude 693 papers while including 76, listed in Table \ref{tab:papers}. Most of the papers were excluded due to violation of IC1 (a few IC2), as they focus on CR techniques applied in healthcare and economics, not SQA. A subset of papers were excluded due to EC3 and EC4. Additionally, we compiled a list of papers we knew and believed should be part of the list, creating a reference \textit{test set}. We then compared the selected papers with this reference set, following established guidelines for secondary studies, as recommended by Petersen \textit{et al.} \cite{Petersen15}, Kitchenham and Brereton \cite{Kitchenham13}. The reference test set was composed by 21 papers. 17/21 papers are selected from a set of 25 papers reviewed in a similar previous survey by Siebert \cite{Siebert23}, specifically, by taking those that match also our inclusion/exclusion criteria (P1, P8, P11, P15-17, P20-21, P24, P26-28, P30, P32, P44, P48, P86 - Table \ref{tab:papers}). To this, we added 4 further papers that we expected to be in the search (P31, P33, P34, and P42 - Table \ref{tab:papers}). 19/21 papers were included in the list of papers selected by our query, ensuring representativeness of the selection --  the remaining two papers (i.e., P17 and P86) did not appear in the query results since they do not contain the words included in the last two “AND" of the query (i.e.,“software" and “caus*") in their title, abstract, or keywords. We considered these terms pivotal for the scope of the search, and decided to not further refine the query to avoid side effects, e.g., explosion of false positives. 
Thus, considering the list of 76 papers resulting from the query and the list of 21 papers from the test set (19 of which are in both lists), we got to 
78 included papers at this stage. Starting from the list of 78 included papers we included additional 8 papers (i.e., P35, P38-41, P49, P77-78) by snowballing, for a \textbf{final list of 86 papers}.

{\footnotesize
\renewcommand{\arraystretch}{1.05}
\begin{longtable}{l|p{14cm}|l}
	\caption{Full list of selected papers.}
	\label{tab:papers} \\
	\toprule
	\textbf{ID} & \textbf{Title} & \textbf{Ref.} \\
	\midrule
	\endfirsthead
	
	\multicolumn{3}{c}%
	{{\tablename\ \thetable{}: Selected papers -- continued from previous page}} \\
	\toprule
	\textbf{ID} & \textbf{Title} & \textbf{Ref.} \\
	\midrule
	\endhead
	
	\midrule \multicolumn{3}{r}{{Continued on next page}} \\ \midrule
	\endfoot
	
	\bottomrule
	\endlastfoot
	P1 & \begin{tabular}[c]{@{}l@{}}Effectively sampling higher order mutants using causal effect\end{tabular} & \cite{1} \\ \hline
	P2 & \begin{tabular}[c]{@{}l@{}}Requirements-driven adaptive security: Protecting variable assets at runtime\end{tabular} & \cite{2} \\ \hline
	P3 & \begin{tabular}[c]{@{}l@{}}A practical approach to explaining defect proneness of code commits by causal discovery\end{tabular} & \cite{3} \\ \hline
	P4 & \begin{tabular}[c]{@{}l@{}}DataPrism: Exposing Disconnect between Data and Systems\end{tabular} & \cite{4} \\ \hline
	P5 & \begin{tabular}[c]{@{}l@{}}Online reasoning about the root causes of software rollout failures in the smart grid\end{tabular} & \cite{5} \\ \hline
	P6 & \begin{tabular}[c]{@{}l@{}}FALCON
		fault localization for SDN control plane\end{tabular} & \cite{6} \\ \hline
	P7 & \begin{tabular}[c]{@{}l@{}}Comparative causality: Explaining the differences between executions\end{tabular} & \cite{7} \\ \hline
	P8 & \begin{tabular}[c]{@{}l@{}}Reducing confounding bias in predicate-level statistical debugging metrics\end{tabular} & \cite{8} \\ \hline
	P9 & \begin{tabular}[c]{@{}l@{}}Engineering for a science-centric experimentation platform\end{tabular} & \cite{9} \\ \hline
	P10 & \begin{tabular}[c]{@{}l@{}}An Empirical Study on the Impact of Refactoring on Quality Metrics in Android Applications\end{tabular} & \cite{10} \\ \hline
	P11 & \begin{tabular}[c]{@{}l@{}}MFL: Method-level fault localization with causal inference\end{tabular} & \cite{11} \\ \hline
	P12 & \begin{tabular}[c]{@{}l@{}}Thinking inside the Box: Differential fault localization for SDN control plane\end{tabular} & \cite{12} \\ \hline
	P13 & \begin{tabular}[c]{@{}l@{}}Bayesian propensity score matching in automotive embedded software engineering\end{tabular} & \cite{13} \\ \hline
	P14 & \begin{tabular}[c]{@{}l@{}}Heterogeneous Effects of Software Patches in a Multiplayer Online Battle Arena Game\end{tabular} & \cite{14} \\ \hline
	P15 & \begin{tabular}[c]{@{}l@{}}NUMFL: Localizing faults in numerical software using a value-based causal model\end{tabular} & \cite{15} \\ \hline
	P16 & \begin{tabular}[c]{@{}l@{}}Causal inference based fault localization for numerical software with NUMFL\end{tabular} & \cite{16} \\ \hline
	P17 & \begin{tabular}[c]{@{}l@{}}Inforence: effective fault localization based on information-theoretic analysis and statistical causal inference\end{tabular} & \cite{17} \\ \hline
	P18 & \begin{tabular}[c]{@{}l@{}}CaRE: Finding Root Causes of Configuration Issues in Highly-Configurable Robots\end{tabular} & \cite{18} \\ \hline
	P19 & \begin{tabular}[c]{@{}l@{}}Dataflow graphs as complete causal graphs\end{tabular} & \cite{19} \\ \hline
	P20 & \begin{tabular}[c]{@{}l@{}}Unicorn: Reasoning about Confgurable System Performance through the Lens of Causality\end{tabular} & \cite{20} \\ \hline
	P21 & \begin{tabular}[c]{@{}l@{}}CounterFault: Value-Based Fault Localization by Modeling and Predicting Counterfactual Outcomes\end{tabular} & \cite{21} \\ \hline
	P22 & \begin{tabular}[c]{@{}l@{}}What will affect software reuse: A causal model analysis\end{tabular} & \cite{22} \\ \hline
	P23 & \begin{tabular}[c]{@{}l@{}}Exploring the Effect of NULL Usage in Source Code\end{tabular} & \cite{23} \\ \hline
	P24 & \begin{tabular}[c]{@{}l@{}}Mitigating the confounding effects of program dependences for effective fault localization\end{tabular} & \cite{24} \\ \hline
	P25 & \begin{tabular}[c]{@{}l@{}}LADDERS: Log Based Anomaly Detection and Diagnosis for Enterprise Systems\end{tabular} & \cite{25} \\ \hline
	P26 & \begin{tabular}[c]{@{}l@{}}Improving fault localization by integrating value and predicate based causal inference techniques\end{tabular} & \cite{26} \\ \hline
	P27 & \begin{tabular}[c]{@{}l@{}}Properties of Effective Metrics for Coverage-Based Statistical Fault Localization\end{tabular} & \cite{27} \\ \hline
	P28 & \begin{tabular}[c]{@{}l@{}}Causal inference for statistical fault localization\end{tabular} & \cite{28} \\ \hline
	P29 & \begin{tabular}[c]{@{}l@{}}An Empirical Study of Software Testing Quality based on Natural Experiments\end{tabular} & \cite{29} \\ \hline
	P30 & \begin{tabular}[c]{@{}l@{}}The Importance of Being Positive in Causal Statistical Fault Localization: Important Properties of Baah et al.'s CSFL \\Regression Model\end{tabular} & \cite{30} \\ \hline
	P31 & \begin{tabular}[c]{@{}l@{}}Root Cause Analysis of Failures in Microservices through Causal Discovery\end{tabular} & \cite{31} \\ \hline
	P32 & \begin{tabular}[c]{@{}l@{}}Causal Inference Based Service Dependency Graph for Statistical Service Fault Localization\end{tabular} & \cite{32} \\ \hline
	P33 & \begin{tabular}[c]{@{}l@{}}Metamorphic Testing with Causal Graphs\end{tabular} & \cite{33} \\ \hline
	P34 & \begin{tabular}[c]{@{}l@{}}Reasoning-Based Software Testing\end{tabular} & \cite{34} \\ \hline
	P35 & \begin{tabular}[c]{@{}l@{}}Causal Inference Techniques for Microservice Performance Diagnosis: Evaluation and Guiding Recommendations\end{tabular} & \cite{35} \\ \hline
	P36 & \begin{tabular}[c]{@{}l@{}}A Quantitative Causal Analysis for Network Log Data\end{tabular} & \cite{36} \\ \hline
	P37 & \begin{tabular}[c]{@{}l@{}}Causality in Configurable Software Systems\end{tabular} & \cite{37} \\ \hline
	P38 & \begin{tabular}[c]{@{}l@{}}MicroDiag: Fine-grained Performance Diagnosis for Microservice Systems\end{tabular} & \cite{38} \\ \hline
	P39 & \begin{tabular}[c]{@{}l@{}}Multi-indicators prediction in microservice using Granger causality test and Attention LSTM\end{tabular} & \cite{39} \\ \hline
	P40 & \begin{tabular}[c]{@{}l@{}}Leveraging Causal Inference for Explainable Automatic Program Repair\end{tabular} & \cite{40} \\ \hline
	P41 & \begin{tabular}[c]{@{}l@{}}Helpfulness Prediction for VR Application Reviews: Exploring Topic Signals for Causal Inference\end{tabular} & \cite{41} \\ \hline
	P42 & \begin{tabular}[c]{@{}l@{}}CC: Causality-Aware Coverage Criterion for Deep Neural Networks\end{tabular} & \cite{42} \\ \hline
	P43 & \begin{tabular}[c]{@{}l@{}}Causal Fault Localisation in Dataflow Systems\end{tabular} & \cite{43} \\ \hline
	P44 & \begin{tabular}[c]{@{}l@{}}Testing Causality in Scientific Modelling Software\end{tabular} & \cite{44} \\ \hline
	P45 & \begin{tabular}[c]{@{}l@{}}Perfce: Performance Debugging on Databases with Chaos Engineering-Enhanced Causality Analysis\end{tabular} & \cite{45} \\ \hline
	P46 & \begin{tabular}[c]{@{}l@{}}FCA: A Causal Inference Based Method for Analyzing the Failure Causes of Object Detection Algorithms\end{tabular} & \cite{46} \\ \hline
	P47 & \begin{tabular}[c]{@{}l@{}}Causal Testing: Understanding Defects' Root Causes\end{tabular} & \cite{47} \\ \hline
	P48 & \begin{tabular}[c]{@{}l@{}}Mitigating the Dependence Confounding Effect for Effective Predicate-Based Statistical Fault Localization\end{tabular} & \cite{48} \\ \hline
	P49 & \begin{tabular}[c]{@{}l@{}}CausalRCA: Causal inference based precise fine-grained root cause localization for microservice applications\end{tabular} & \cite{49} \\ \hline
	P50 & \begin{tabular}[c]{@{}l@{}}Fault Localization for Microservice Applications with System Logs and Monitoring Metrics\end{tabular} & \cite{50} \\ \hline
	P51 & \begin{tabular}[c]{@{}l@{}}Causality-Based Neural Network Repair\end{tabular} & \cite{51} \\ \hline
	P52 & \begin{tabular}[c]{@{}l@{}}Causal Models to Support Scenario-Based Testing of ADAS\end{tabular} & \cite{52} \\ \hline
	P53 & \begin{tabular}[c]{@{}l@{}}Latent Hazard Notification for Highly Automated Driving: Expected Safety Benefits and Driver Behavioral Adaptation\end{tabular} & \cite{53} \\ \hline
	P54 & \begin{tabular}[c]{@{}l@{}}Causality-Aided Trade-Off Analysis for Machine Learning Fairness\end{tabular} & \cite{54} \\ \hline
	P55 & \begin{tabular}[c]{@{}l@{}}CauseInfer: Automated End-to-End Performance Diagnosis with Hierarchical Causality Graph in Cloud Environment\end{tabular} & \cite{55} \\ \hline
	P56 & \begin{tabular}[c]{@{}l@{}}Mutation-Based Graph Inference for Fault Localization\end{tabular} & \cite{56} \\ \hline
	P57 & \begin{tabular}[c]{@{}l@{}}CloudRanger: Root Cause Identification for Cloud Native Systems\end{tabular} & \cite{57} \\ \hline
	P58 & \begin{tabular}[c]{@{}l@{}}FUNNEL: Assessing Software Changes in Web-Based Services\end{tabular} & \cite{58} \\ \hline
	P59 & \begin{tabular}[c]{@{}l@{}}PUS: A Fast and Highly Efficient Solver for Inclusion-based Pointer Analysis\end{tabular} & \cite{59} \\ \hline
	P60 & \begin{tabular}[c]{@{}l@{}}Two Sides of the Same Coin: Exploiting the Impact of Identifiers in Neural Code Comprehension\end{tabular} & \cite{60} \\ \hline
	P61 & \begin{tabular}[c]{@{}l@{}}Building causal models for finding actual causes of unmanned aerial vehicle failures\end{tabular} & \cite{61} \\ \hline
	P62 & \begin{tabular}[c]{@{}l@{}}Detecting multi-sensor fusion errors in advanced driver-assistance systems\end{tabular} & \cite{62} \\ \hline
	P63 & \begin{tabular}[c]{@{}l@{}}Causality-Guided Adaptive Interventional Debugging\end{tabular} & \cite{63} \\ \hline
	P64 & \begin{tabular}[c]{@{}l@{}}Causality-driven Testing of Autonomous Driving Systems\end{tabular} & \cite{64} \\ \hline
	P65 & \begin{tabular}[c]{@{}l@{}}Towards Causal Deep Learning for Vulnerability Detection\end{tabular} & \cite{65} \\ \hline
	P66 & \begin{tabular}[c]{@{}l@{}}Uncovering Causal Relationships between Software Metrics and Bugs\end{tabular} & \cite{66} \\ \hline
	P67 & \begin{tabular}[c]{@{}l@{}}Granger Causality-Aware Prediction and Diagnosis of Software Degradation\end{tabular} & \cite{67} \\ \hline
	P68 & \begin{tabular}[c]{@{}l@{}}Performance degradation analysis of a supercomputer\end{tabular} & \cite{68} \\ \hline
	P69 & \begin{tabular}[c]{@{}l@{}}Prevent: An Unsupervised Approach to Predict Software Failures in Production\end{tabular} & \cite{69} \\ \hline
	P70 & \begin{tabular}[c]{@{}l@{}}DyCause: Crowdsourcing to Diagnose Microservice Kernel Failure\end{tabular} & \cite{70} \\ \hline
	P71 & \begin{tabular}[c]{@{}l@{}}GCFormer: Granger Causality based Attention Mechanism for Multivariate Time Series Anomaly Detection\end{tabular} & \cite{71} \\ \hline
	P72 & \begin{tabular}[c]{@{}l@{}}Adaptive Incremental Learning for Software Reliability Growth Models\end{tabular} & \cite{72} \\ \hline
	P73 & \begin{tabular}[c]{@{}l@{}}Faster, deeper, easier: Crowdsourcing diagnosis of microservice kernel failure from user space\end{tabular} & \cite{73} \\ \hline
	P74 & \begin{tabular}[c]{@{}l@{}}An empirical investigation on the relationship between design and architecture smells\end{tabular} & \cite{74} \\ \hline
	P75 & \begin{tabular}[c]{@{}l@{}}Predicting software defects with causality tests\end{tabular} & \cite{75} \\ \hline
	P76 & \begin{tabular}[c]{@{}l@{}}Statistical causality analysis of INFOSEC alert data\end{tabular} & \cite{76} \\ \hline
	P77 & \begin{tabular}[c]{@{}l@{}}A Comparative Analysis of Software Aging in Image Classifiers on Cloud and Edge\end{tabular} & \cite{77} \\ \hline
	P78 & \begin{tabular}[c]{@{}l@{}}Testing the resilience of MEC-based IoT applications against resource exhaustion attacks\end{tabular} & \cite{78} \\ \hline
	P79 & \begin{tabular}[c]{@{}l@{}}Memory Degradation Analysis in Private and Public Cloud Environments\end{tabular} & \cite{79} \\ \hline
	P80 & \begin{tabular}[c]{@{}l@{}}ViSRE: A Unified Visual Analysis Dashboard for Proactive Cloud Outage Management\end{tabular} & \cite{80} \\ \hline
	P81 & \begin{tabular}[c]{@{}l@{}}Explaining Regressions via Alignment Slicing and Mending\end{tabular} & \cite{81} \\ \hline
	P82 & \begin{tabular}[c]{@{}l@{}}Fine-Grained Causality Extraction from Natural Language Requirements Using Recursive Neural Tensor Networks\end{tabular} & \cite{82} \\ \hline
	P83 & \begin{tabular}[c]{@{}l@{}}Enabling Runtime Verification of Causal Discovery Algorithms with Automated Conditional Independence Reasoning\end{tabular} & \cite{83} \\ \hline
	P84 & \begin{tabular}[c]{@{}l@{}}An information flow-based feature selection method for cross-project defect prediction\end{tabular} & \cite{84} \\ \hline
	P85 & \begin{tabular}[c]{@{}l@{}}Rapid and robust impact assessment of software changes in large internet-based services\end{tabular} & \cite{85} \\ \hline
	P86 & \begin{tabular}[c]{@{}l@{}}An empirical study of Linespots: A novel past-fault algorithm\end{tabular} & \cite{86} \\ \bottomrule
\end{longtable}
}

\begin{table}[hb]
\renewcommand{\arraystretch}{1.1}
\centering
\caption{The data extraction form of this study.}
\label{tab:dimensions}
\resizebox{\textwidth}{!}{%
	\begin{tabular}{l|l|l|l} 
		\toprule
		\textbf{Category} & \textbf{Dimensions} & \textbf{Definition} & \textbf{Type/Domain} \\ 
		\midrule
		\multirow{6}{*}{\begin{tabular}[c]{@{}c@{}}Metadata\end{tabular}}  & Id & The internally used ID of the tool & “P”+[numeric] \\ \cline{2-4}
		& Article title & The name of the article & Free text \\ \cline{2-4}
		& Source title & Name of publication venue & Free text \\ \cline{2-4}
		& Year & Year of publication & Free text \\ \cline{2-4}
		& Document type & Where the paper has been published & \{Conference, Journal, Workshop\} \\ \cline{2-4}
		& Availability & Whether the paper publicly provides the implemented code & {[}Yes, No] \\ 
		\hline
		\multirow{2}{*}{\begin{tabular}[c]{@{}l@{}}What is CR\\used for?\end{tabular}}   & Task & What task of quality assurance/assessment is covered & Free text \\ \cline{2-4}
		& Quality attributes~ & What quality attributes are targeted & Figure \\ 
		\hline
		\multirow{5}{*}{\begin{tabular}[c]{@{}l@{}}How is CR\\used?\end{tabular}}  & Task & If the paper apply CD, CI, or both & \{Discovery, Inference, Both\} \\ \cline{2-4}
		& Model & What kind of causal model the paper builds & Free text \\ \cline{2-4}
		& Discovery methodology & Which CD algorithm the paper use & Free Text \\ \cline{2-4}
		& Inference methodology & What CI methods the paper use & Free Text \\ \cline{2-4}
		& Tools/libraries & What tools and libraries the paper use for CD and/or CI & Free Text \\
		\hline
		\multirow{3}{*}{\begin{tabular}[c]{@{}l@{}}How is CR\\validated?\end{tabular}} & Experiment & An experimental validation is conducted & {[}Yes, No] \\ \cline{2-4}
		& Industrial application & Whether or not the paper presents an industrial application & {[}Yes, No] \\ \cline{2-4}
		& Domain & In what domain & Free text \\ 
		\bottomrule
		
	\end{tabular}
}
\end{table}

\subsection{Data extraction, classification, and analysis}
The data extraction step {\textcircled{\raisebox{-.9pt} {3}}} provides key information about each paper, useful for classification.  
Starting from the RQs, we defined an \textit{initial RQs scheme} (Figure \ref{fig:protocol}).
Then, to refine and agree on the scheme, we proceeded iteratively, beginning with a \textit{horizontal} classification. We divided the list of papers into equal subsets and assigned them to three authors, who independently classified them according to all dimensions of the scheme. With the classified papers (i.e., the \textit{filled scheme}), we validated the classification in plenary meetings. The validation phase was useful to decide whether refining the scheme was necessary (e.g., adding, removing, or modifying dimensions). In such cases, a refined scheme was agreed upon in phase {\textcircled{\raisebox{-.9pt} {5}}}. This iterative process concluded with the agreement on the final scheme with 17 dimensions, as reported in Table \ref{tab:dimensions}. Finally, to ensure homogeneous classification, the papers were re-classified \textit{vertically}: each author was assigned a subset of sub-dimensions and classified all 86 papers for only the assigned dimensions. This favored the detection of inconsistent classifications performed during the horizontal classification phase. Dedicated meetings solved such disagreements.

On the classified papers, we present in Sections \ref{sec:rq1}-\ref{sec:rq3} the analysis (step {\textcircled{\raisebox{-.9pt} {6}}}) we performed to answer the RQs.
We highlight the main findings, implications, and open challenges derived by the analysis in Section \ref{sec:discussion}.

\subsection{Results overview}

\begin{figure}[t]
\centering 	
\includegraphics[width=0.44\textwidth]{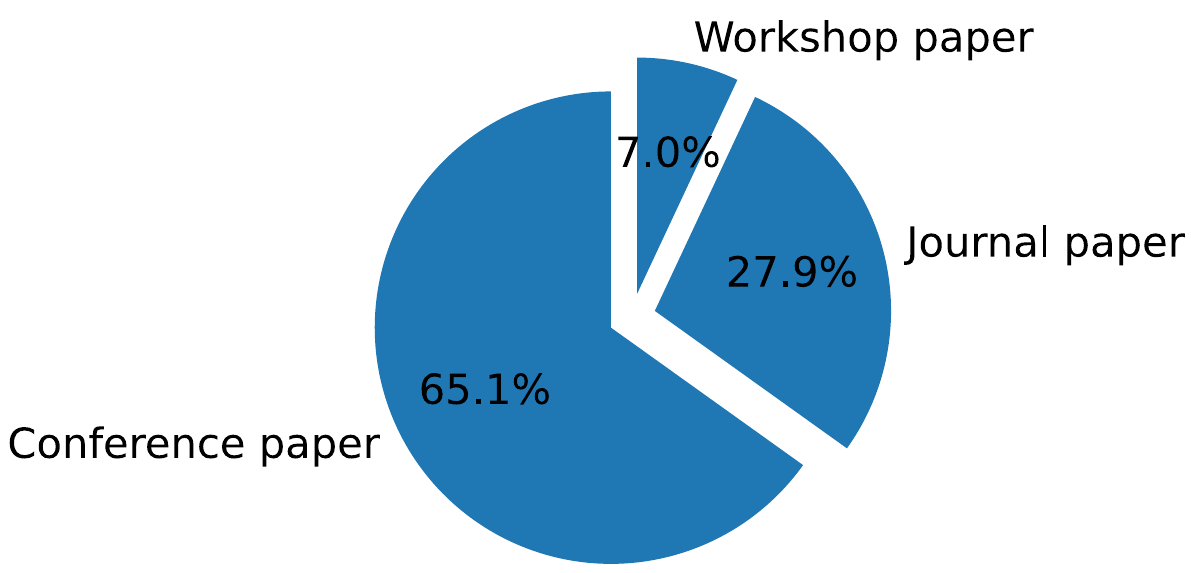}
\vspace{-3pt}
\caption{Number of papers by type of publication venue.}
\vspace{-6pt}	
\label{fig:venues}
\end{figure}

The search in the digital libraries produced a total of 1,086 papers: 429 papers retrieved from \texttt{Scopus}, 342 from \texttt{IEEExplore}, 45 from \texttt{ACM DL}, and 270 from \texttt{Web of Science}. After removing duplicates, we obtained a list of 770 papers, on which we applied the inclusion/exclusion criteria. Following this step, 78 papers were included, and an additional 8 were added through the snowballing phase, resulting in a total of 86 papers for classification. Figure \ref{fig:venues} shows the distribution of the 86 included papers by publication type. The majority of papers (56) are conference papers, followed by journal papers (24), and workshop papers (6).

\begin{figure}[hb]
\centering
\begin{subfigure}[b]{0.46\textwidth}
	\centering
	\includegraphics[width=\textwidth]{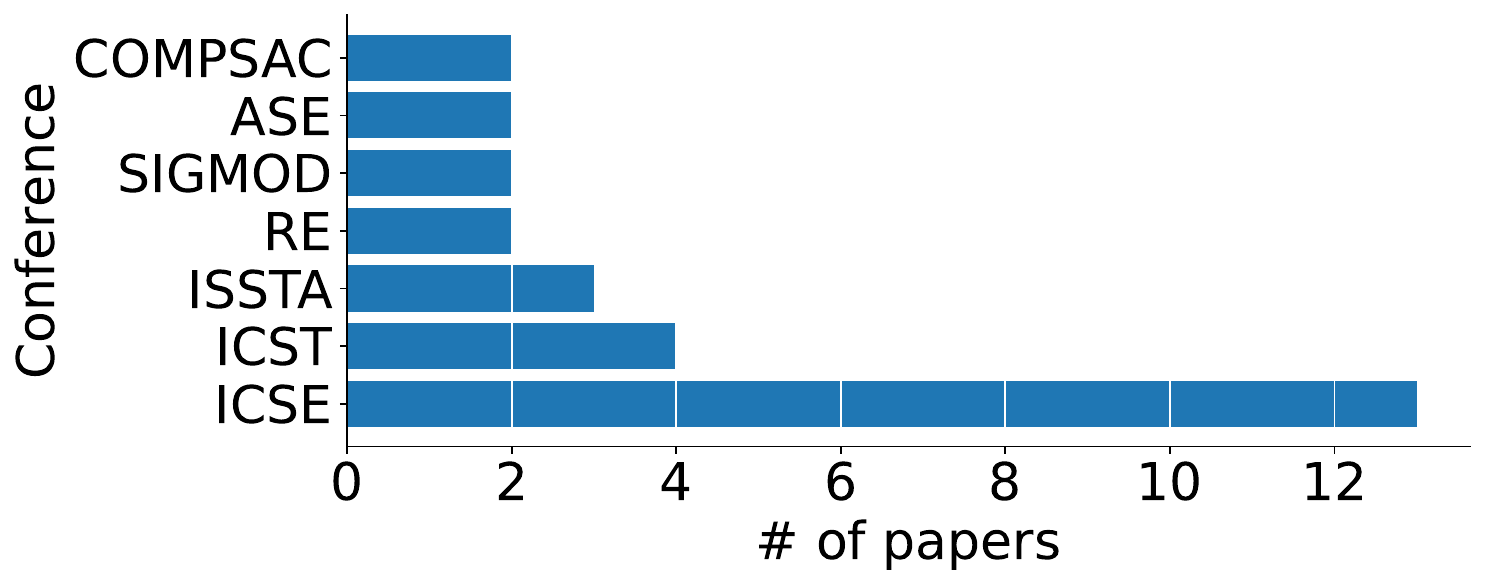}
	\caption{}
	\label{fig:conferences}
\end{subfigure}
\hfill
\begin{subfigure}[b]{0.46\textwidth}
	\centering
	\includegraphics[width=\textwidth]{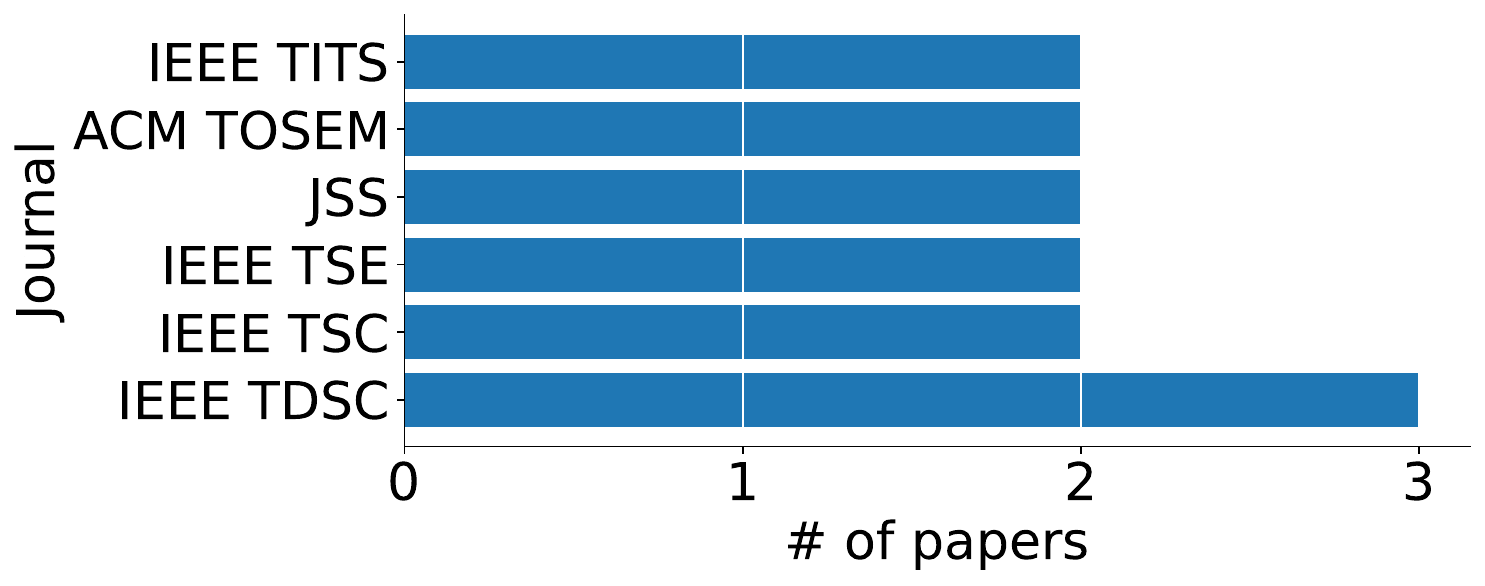}
	\caption{}
	\label{fig:journals}
\end{subfigure}
\vspace{-6pt}
\caption{Number of papers per publication venue (with at least 2 papers): (a) conference, (b) journal).}
\label{fig:conference_journal_comparison}
\end{figure}

We further analyzed the metadata of the included papers by examining the major venues. 
Figure \ref{fig:conferences} shows the number of papers for all conferences with at least 2 papers. The results indicate that the most common conferences are in the Software Engineering and Testing fields. Notably, the International Conference on Software Engineering (ICSE) has the largest number of papers (13), followed by the International Conference on Software Testing (ICST) with 4 papers, and the International Symposium on Software Testing and Analysis (ISSTA) with 3 papers. We found four other conferences with 2 papers each, namely the International Requirements Engineering Conference (RE), the International Conference on Automated Software Engineering (ASE), the International Conference on Management of Data (SIGMOD), and the International Computer Software and Applications Conference (COMPSAC).

Figure \ref{fig:journals} shows the number of papers for all journals with at least 2 papers. The journal with the greatest number of papers (3) is the IEEE Transactions on Dependable and Secure Computing (IEEE TDSC). Then, other 5 journal finds 2 papers each: IEEE Transactions on Services Computing (IEEE TSC), IEEE Transactions on Software Engineering (IEEE TSE), Journal of Systems and Software (JSS), ACM Transactions on Software Engineering and Methodology (ACM TOSEM), and IEEE Transactions on Intelligent Transportation Systems (IEEE TITS).

\section{RQ1: Quality assurance tasks and quality attributes}
\label{sec:rq1}

\subsection{Dimensions}\label{sec:dims}
This question investigates two dimensions of the scheme defined in Table \ref{tab:dimensions}: \textbf{Task},  %
and \textbf{Quality attribute}.  
Considering the above-defined quality assurance means (Section \ref{sec:background}), we hereafter list the specific \textbf{tasks} we have found after classifying the surveyed papers, categorized by quality assurance means: 
\begin{description}
	\item \textit{Fault avoidance/prevention}: %
	\begin{itemize}
		\item \textbf{Threats modeling.} An approach used to identify, quantify, and address potential security threats and vulnerabilities in a system.
	\end{itemize}
	
	\item \textit{Fault removal}: 
	\begin{itemize}
		\item \textbf{Testing and analysis}. Complementary activities aimed to expose failures through code execution (that is \textit{testing}) and to verify properties (e.g., correctness, security) on (a model of) the source code (that is \textit{static analysis}) or on execution traces (\textit{dynamic analysis}) \cite{pezze}.%
	\item \textbf{Fault localization} (\textit{Debugging}). This is the task of locating %
	the cause that originated the failure. 
	It is referred to as \textit{diagnosis} in Avizienis, Laprie, \textit{et al.} \cite{Avizienis04}. Its objective can be twofold: \textit{i)} the identification of the line(s) of code containing the fault and of the correction action; in this case, we tag the task as \textit{debugging}, clearly a fault removal means; \textit{ii)} the identification of a faulty unit in a running system (e.g., the failure culprit microservice in a service-based system), to design \textit{fault tolerance} means (referred to as \textit{fault handling} in Ref. \cite{Avizienis04})
	or to support \textit{post mortem} analysis and repair; in this case, we tag the task as \textit{root cause analysis}, the common term used in these works to spot the failure-causing system components.
	Therefore, this \textit{fault localization} category is repeated under the next category of \textit{fault tolerance} means. 
	
\end{itemize}

\item \textit{Fault tolerance}: 
\begin{itemize}
	\item \textbf{Fault localization}, for Root Cause Analysis (RCA). The process of identifying the specific location or component within a system where a defect or malfunction has occurred.
	\item \textbf{Anomaly detection}. The dependability classification in Ref. \cite{Avizienis04} considers \textit{error detection} in the fault tolerance means; anomaly detection is here referred to the attempt of performing error detection. 
\end{itemize}

\item \textit{(Fault) forecasting}: 
\begin{itemize}
	\item \textbf{Fault prediction}, also referred to as \textbf{defect/bug prediction}. The process using statistical and machine learning techniques to predict the location and occurrence of defects in software.
	
	\item \textbf{KPI prediction}. As discussed in Section \ref{sec:background}, this category refers to the works exploiting CR for predicting quality-related indicators other than faults in the strict sense, but whose monitoring and prediction support quality improvement actions, like performance, energy or maintainability indicators. 
\end{itemize}

\end{description}

As for the \textbf{quality attributes}, we consider the ones defined in Section \ref{sec:background}, namely the listed dependability and security attributes, plus performance and usability. %

Based on the previous definitions, we also distinguish between papers using CR for \textit{retrospective} analysis and for \textit{forecasting}. The former refers to the solutions that act \textit{after} an event of interest occurs (e.g., a failure), for instance, to localize the corresponding fault, to tolerate and/or to remove it. Hence, these papers use solutions falling into the \textit{fault removal} and \textit{fault tolerance} categories. The latter refers to those solutions using CR for predicting future events of interest, and then acting consequently for instance in the development or maintenance of the system. Hence, these papers use solutions falling into the \textit{fault avoidance/prevention} and \textit{(fault) forecasting} categories in different phases of the software system life cycle.

\subsection{Results}

\begin{figure}[t]
\centering 	\includegraphics[width=.74\textwidth]{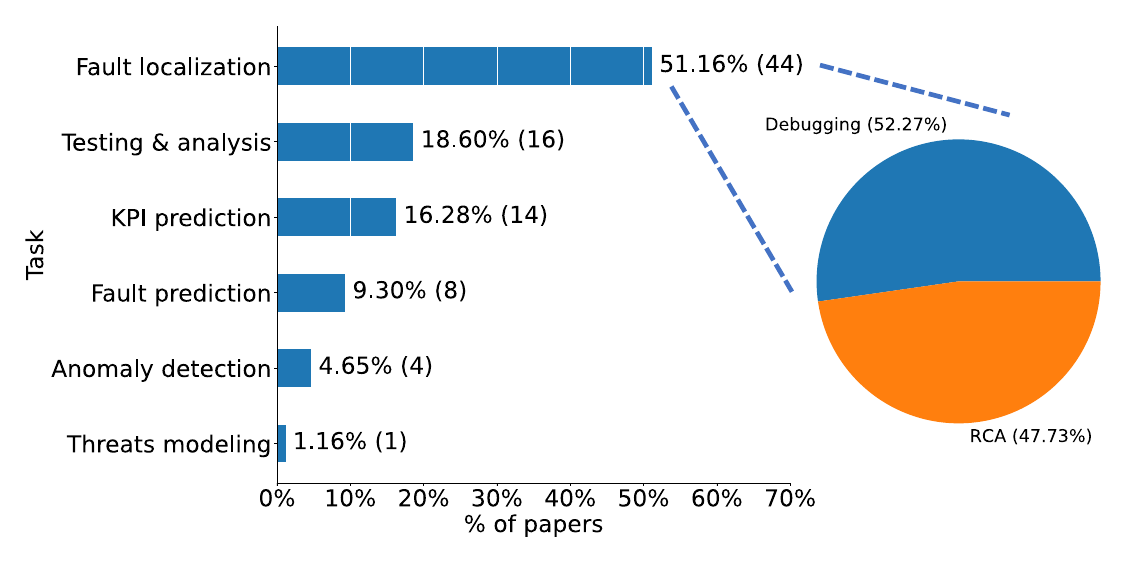}
\vspace{-6pt}
\caption{RQ1: Percentage of papers per SQA task.}
\label{fig:task}
\end{figure}

Figure \ref{fig:task} shows the percentage of papers using CR per task, namely QA activity (a paper can be related to more tasks\footnote{The total percentage is greater than $100\%$ because paper P62 is on both Testing and Fault localization.}). The detail about which paper falls into which category is reported in the online material.\footref{note1} %
The Figure shows that most of the papers use CR for fault localization, be it for \textit{debugging} or for \textit{root cause analysis}. This is probably due to how CR is perceived, namely as a method to \textit{explain} the reasons for an event, and, as such, is particularly prone to represent the causal chain that leads from a fault to a failure. The employment of CR for other tasks is a more recent application. More details are in the following subsections.

The most considered phases for CR application are V\&V (39 papers) and Evolution \& Maintenance (39). 
Most of the papers apply CR for tasks concerning the V\&V (39 papers) and Evolution \& Maintenance (39) phases. This is in line with the results on the activity, as \textit{i)} fault localization happens in the V\&V phase and during maintenance; \textit{ii)} several other activities are carried out in the post-coding phases, such as testing, anomaly detection, KPI prediction based on field data, which can rely on the information available at those stages, such as the structure of the software system established during the design, the events log collected during testing and during the operational phase.

\begin{figure}[!ht]
\centering 	\includegraphics[width=0.66\textwidth]{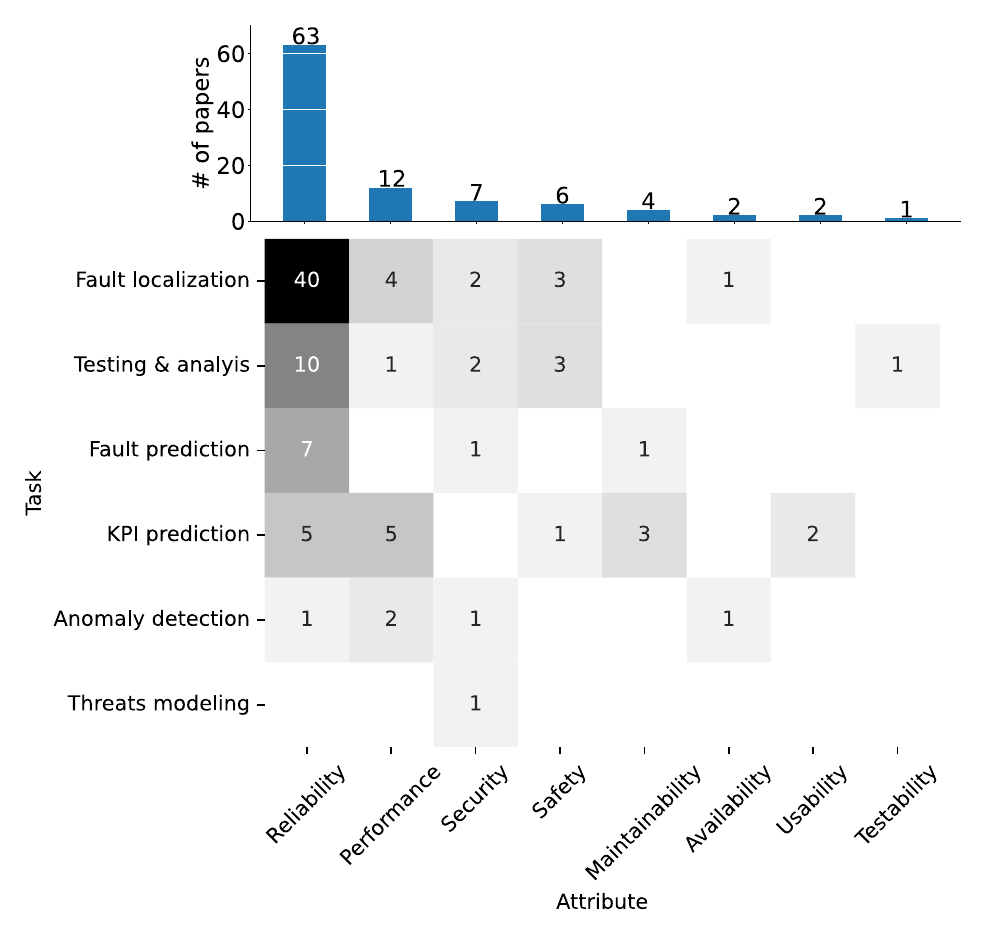}
\vspace{-9pt}
\caption{RQ1: Number of papers per quality attribute. \textit{(Note: some papers cover multiple tasks and/or multiple quality attributes per task.)}}
\label{fig:attr}
\vspace{-6pt}
\end{figure}

Figure \ref{fig:attr} reports the number of papers by quality attribute -- a single paper can address more than one task/attribute (nine papers address multiple quality attributes).\footnote{Safety is addressed in 6 papers. P62 caused it to be counted twice in the heatmap both for testing and fault localization.} %

Reliability and Performance are the most present attributes with 63 and 12 occurrences, respectively. 
Papers targeting reliability include those addressing failures in operation and those on debugging. This quality attribute is mainly targeted in papers on fault localization and testing, aiming to improve the reliability of software systems by detecting and resolving faults. Performance is targeted in KPI prediction, fault localization, and anomaly detection tasks also in conjunction with reliability (P4, P9, P85), availability (P25), and safety (P13). Safety and Security are covered by seven papers each. Threat modeling paper targets security, while safety is not related to a specific task. Safety is more related to the application domain (safety-critical): e.g., P18 focuses on robots, while P13, P52, and P53 on automotive. Maintainability is addressed in four papers, Availability and Usability are treated by two papers each, while only one paper is on testability. 
In the following, we discuss papers based on the task they cover 
and for which quality attribute. According to the distinction reported in Section \ref{sec:dims}, we split the papers in two groups concerning CR for \textit{retrospective} analysis and for \textit{forecasting}, respectively.

\subsubsection{Causal Reasoning for retrospective analysis}
The intrinsic nature of CR makes it suitable to perform tasks aiming to trace causal relationships among software system components to find the “reasons” behind certain undesired behaviors. In the following, we discuss the main tasks performed when something wrong happens in the software systems. 

\paragraph{Fault localization \& Anomaly detection} 
For fault localization, we mean all the techniques, methods, and tools to localize issues (namely, software faults or failures' root causes) causing undesired behaviors of a target software system. As mentioned, fault localization is at the base of debugging  and of root cause analysis processes. 
Anomaly detection aims to highlight working conditions potentially yielding to failures or to degradation of the observed software system.

As shown in Figure \ref{fig:task}, these two categories together cover more than $50\%$ of the papers. %

The possibility to model the software system structure and/or development process by casual models, such as to model program dependencies (P24) or the connection between dataflow architectures and causal graphs in Flow-based programming (P19), has enabled a  plethora of CR solutions for the localization of causes of unexpected behavior. 

High effectiveness
has been achieved in studies where the causal model is extracted from existing (white-box) knowledge, like for delta debugging of Software Defined Networks (P6, P12) or for faulty services localization (P33).

Although fault localization is mainly related to finding reliability issues, CR is also used to detect causes of performance (P4), availability (P5), security (P5, P51), and safety (P18, P51) issues.

We classify fault localization in two categories: %
\textit{debugging} and RCA. 
The count by these two categories is highlighted in Figure \ref{fig:task}.
Many fault localization techniques (in 24 papers) are aimed at code debugging actions (P24, P47). The main applications are \textit{Automatic Program Repair} (P40), also concerning Neural Networks repair (P51), \textit{delta debugging} (P6, P7, P12), evolution by using casual difference graph (P81), and \textit{Statistical Fault Localization} (SFL) (P8, P11, P15, P16,  P17, P21, P26, P27, P28, P30, P48, P63).
Just one of the paper focuses on localizing code smells and anti-patterns (P23).

Other specific applications regard the application of causal graphs obtained through mutation testing for debugging (P56), the debugging of conditional independence tests (P83).

Most of the techniques adopted in the papers about debugging refer to the  V\&V phase, with few exceptions (e.g., the work in P40 proposes a technique that can be applied directly during coding, P81 in Evolution and Maintainance, and P83 in operation). 
The authors of these papers pinpointed a \textit{confounding bias} that is unaddressed by association-based techniques. Specifically, in SFL, they identified bias between the occurrence of specific runtime events (e.g., coverage of a given statement) and the manifestation of program failures, arising from the interactions between statements in a program (P24). In the context of delta debugging, the confounding is introduced by the partial state replacement problem (P7), when performing causal state minimization (i.e., determine the minimal subset of state differences between a buggy and correct execution needed to reproduce the failure). CR demonstrated to be very effective in mitigating these challenges, usually by applying CI on a causal model built starting from available knowledge, such as a program dependency graph.

In the other cases, the fault localization task is focused on looking for the root cause of undesired behaviors through RCA. In particular, RCA is commonly used to find the faulty service in service-based systems by exploiting information about services (P49), such as service dependencies (P31, P32, P38, P50, P70, P73),  and in configurable systems (P18 and P37). More specific applications regard: the localization of root causes through the analysis of data for network events (P36) and in the cloud (P55, P57, P80); localization of faults of software developed according to flow-based programming (P19 and P43); performance downgrades in databases (P45);  failure causes localization in object detection algorithms (P46), unmanned aerial vehicles (P61), and driver assistance systems (P62).%

RCA is also used to find root causes of performance degradation in datasets for synthetic data-driven systems (P4), and failures in software for smart grids (P5).

Four papers focus on anomaly detection to spot performance (P25 and P35), reliability (P71), and security issues (P76).
The first aims to detect anomalies on log events and Key Performance Indicators of enterprise systems. 
The second exploits CI for performance diagnosis in microservice-based systems to then spot the faulty microservice responsible for the performance problem. 
P71 uses Granger causality for anomaly detection in multivariate time series. P76 performs statistical causal analysis to discover relationships among security attacks.

For the nature of RCA and anomaly detection solutions, they are mostly applied to operation and post-operation Evolution and Maintenance phases since observational data and logs are needed. However, five papers apply RCA to the V\&V phase (P4, P19, P37, P43, P62). Authors of these papers recognize the strong connection between dataflow graphs and graphical causal models. Indeed, with the exception of P4 (that use evidential networks), they all use SCM or CausalDAG to perform CI, with different methodologies such as attribution analysis (P43) or advanced methods for big data (P19 and P37).

\subsubsection{Causal Reasoning for forecasting}
In the previous paragraphs,  CR was considered for its ability to understand the causes of undesired behaviors of the software system, as well as for activities supporting fault removal such as testing. In this case, we consider CR to forecast the behaviors of software systems before they are deployed in the operational environment. 
Forecasting is mainly related to software maintenance and evolution activities. 

\paragraph{Testing and analysis}
The ability of CR to determine the causal relationships between variables makes it particularly suited to run testing and analysis activities during the V\&V phase of software systems life cycle. Exploiting causal relations can help execute tests more prone to spot defects of the system under test, with reference to both traditional software systems (P78)\footnote{Testing security of Internet of Things (IoT) systems.} and learning-enabled ones (P42, P54, P65), also Autonomous Driving Systems (P52, P53, P62, P64). This aspect has been underlined by Netflix, which included CR in their science-centric experimentation platform (P9). 

The main task CR is used for is \textit{test generation}. It has been adopted to generate mutants and metamorphic relations for mutation (P1) and metamorphic (P33) testing.  
Authors in P34 envision a framework for test generation, in which the causal model is used as a surrogate to search the input space efficiently, with the aim of maximizing a given testing objective, such as the detection of performance/reliability issues and safety violations. They tested the  generation strategy on autonomous driving systems, highlighting better performance compared to ML-based testing. 

CR is also adopted to evaluate (P29) and to improve (P44) the testing quality. P29 applies the natural experiment method to empirically investigate the factors affecting testing quality. P44 presents a testing framework incorporating a causal model into the testing process, enabling the direct application of CI techniques to software testing problems.

A single paper proposes a technique concerning testing and analysis using causal graphs for Pointer Analysis (P59).
It is worth to note that only one paper (P82) concerns testing outside the V\&V phase. The authors extract causal relations from the requirements to enable the automatic derivation of test cases and requirements dependencies. This paper opens the possibility of applying CR for quality assessment of requirements.

\paragraph{Bug prediction and KPI prediction} %

The main prediction tasks concern bugs/defects and KPI. Such predictions are important for developers to take software release decisions, as they allow corrective actions before the manifestation in operation, or to estimate potential issues after release.  

P3 applies CR to predict and explain defect proneness of code commits, allowing developers to understand how to take corrective actions. Similarly, P84 performs cross-project defect prediction and P86 presents Linespots, a fault prediction algorithm based on DAG.

Besides the conventional bug prediction, CR is used for predicting indicators 
useful before the deployment of new software (P10, P13) by using software defect prediction (P66, P69 and P75%
) 
or by exploiting Software reliability growth models (P72), useful to look at possible reconfiguration (P20), or even to define new patches for online games (P14). For instance, P10 aim to improve Android applications by refactoring and predicting the effect of such actions with causal models. P13 evaluate the effect of software changes through Bayesian models for estimating causal treatment effects from collected observational data and experimenting in the automotive domain. 

KPI prediction is applied to many tasks like \textit{i) }software reuse, by exploiting causal models to estimate the factors that have a direct effect on the success of the reuse (P22); \textit{ii)} helpfulness prediction of the users' reviews for Virtual Reality Apps aiming to recommend them to consumers (P41); \textit{iii)} prediction of indicators service-based architectures (P39, P58); \textit{iv)} prediction of performance and memory degradation (P67, P68, P79); \textit{v)} analysis of software aging observed in image classification systems (P77); \textit{vi)} impact assessment of software changes (P85).

\paragraph{Threat modeling}
CR for threat modeling in the security domain has been used in P2. It uses causal models to analyze systems security in different situations for enabling countermeasures depending on the current security threat. It does not use traditional causal models, such as PO, CausalDAG, and SCM since P2 builds a Fuzzy Cognitive Map based on a threat model and then weights the edges.

\section{RQ2: How is CR used?}
\label{sec:rq2}

\subsection{Dimensions} 
This question explores the dimensions outlined in the scheme in Table \ref{tab:dimensions}, covering the \textbf{task} (inference and/or discovery), the \textbf{causal framework} employed, the \textbf{CD methodology} (how causal relationships are defined), the \textbf{CI methodology} (how authors estimate causal effects and the metrics used), and the \textbf{tools/libraries} utilized for CD/CI.

\subsection{Results} 

\subsubsection{Causal Reasoning task}
\label{sec:rq2:1}

\begin{figure}[ht]
\centering 	\includegraphics[width=.38\textwidth]{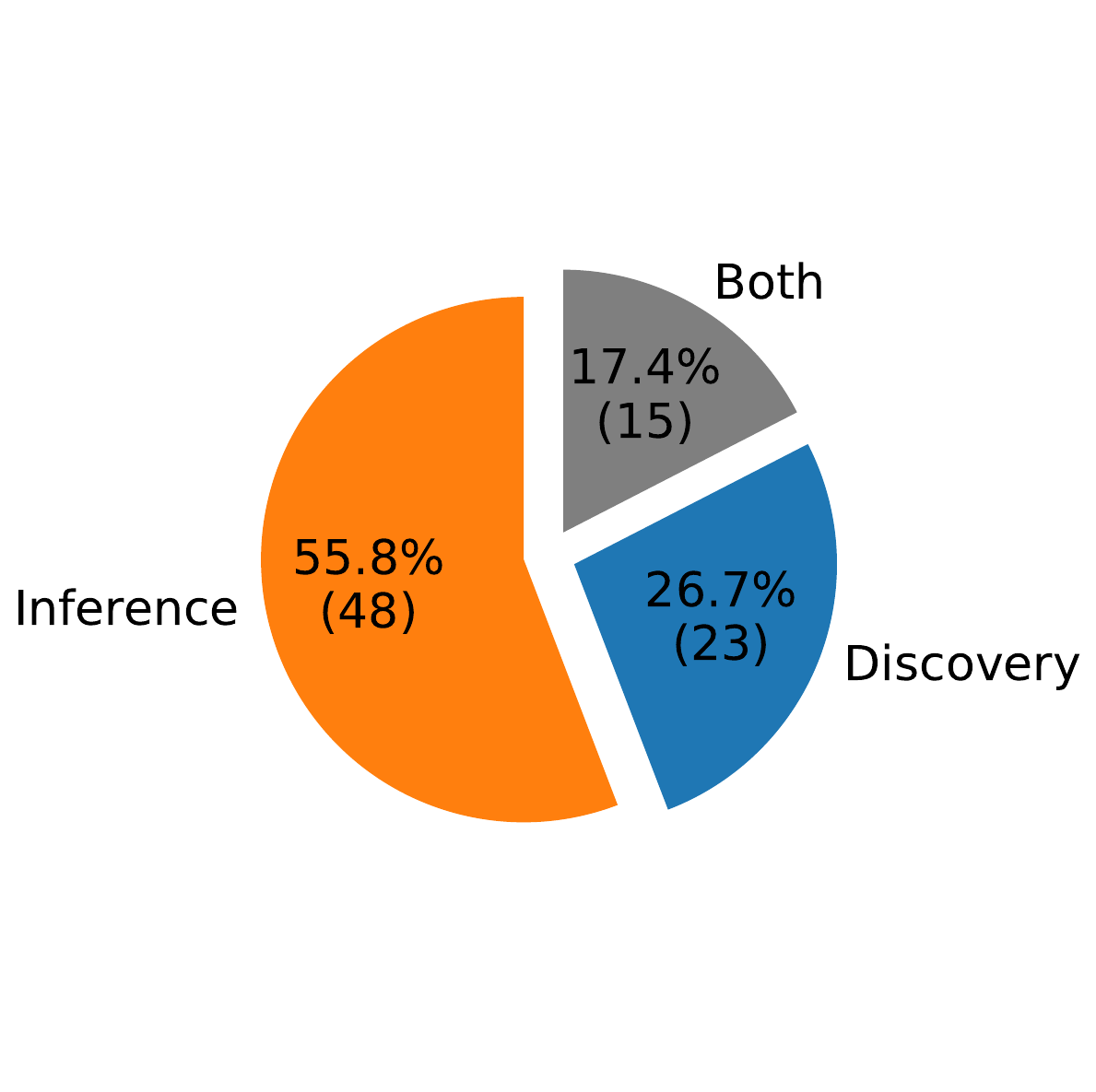}
\vspace{-46pt}
\caption{RQ2: Causal Discovery/Inference proportion.}
\label{fig:cr_tasks}
\end{figure}

Figure \ref{fig:cr_tasks} illustrates the distribution of papers performing CD, CI, or a combination of both. The majority of papers (55.8\% - 48 out of 86 papers) focus solely on inference mechanisms on a causal model, be it manually constructed or obtained from existing knowledge. Among these, 4 papers (P23, P44, P52, P53) involve manual model construction by domain experts, while the remaining derive the causal structure from a variety of sources (e.g., control flow graph, service dependency graph, message tracing, design artifacts).

A subset of papers (26.7\% - 23 out of 86 papers) employ CD algorithms to derive a causal structure but do not apply traditional CI methodologies. For root cause analysis, papers P25, P35, P38, P46, P50, and P69 utilize weighted causal graph edges, using graph algorithms (e.g., K shortest path, PageRank) to rank nodes based on their probability of being the actual cause. Other papers employ Granger causality or transfer entropy to detect causality between time series for anomaly detection (P35, P71, P76), fault prediction (P66, P69, P72, P75), KPI prediction (P74, P77), and testing (P78).
P42 introduces a causality-aware coverage criterion for Deep Neural Networks (DNNs) with the aid of CD. Coverage is computed by comparing the graph generated by different test suites to a ground truth graph. 

\begin{figure}[t]
\centering 	\includegraphics[width=.95\textwidth]{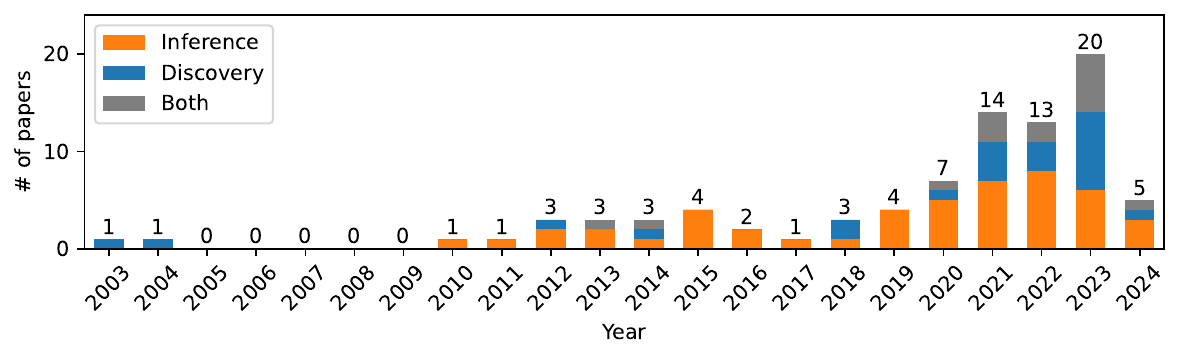}
\vspace{-9pt}
\caption{RQ2: Number of papers per task by publication year.}
\label{fig:cr_year_task}
\vspace{-9pt}
\end{figure}

A smaller percentage of papers (17.4\% - 15 out of 86) published mostly from 2022 (9 papers - Figure \ref{fig:cr_year_task}) use both CD and CI in sequence. Most of these papers (P3, P18, P20, P34, P64) utilize FCI or GFCI as CD algorithms to construct a causal model. Specifically, P3 employs GFCI to build the causal graph and the expectation-maximization algorithm for parameter estimation, aiming to construct a conditional probability table. This table is then utilized for defect prediction. In P18, root cause analysis is performed by ranking nodes based on the computation of the ACE. P20 employs counterfactuals to predict performance issues, by computing the ACE. In the case of P34 and P64, test case generation is carried out through interventions, estimating the effect with a simulation-based approach. This involves sampling the post-intervention distribution of variables using the \texttt{Do-sampler} in the \texttt{do-why} library. Finally, other papers employ granger causality as CD and regression adjustment (P67, P68, P70, P73) or neural networks (P39) as CI to perform tasks like KPI prediction and root cause analysis.

In summary, most papers use causality for inference. Indeed, combining the “Both” category with the others, we find that out of 86 papers, 63 perform CI, and 38 use CD. This indicates that the most common approach is building the causal model from available knowledge, rather than using CD algorithms on observational data. 
However, the adoption of CD is increasing, as can be seen in Figure \ref{fig:cr_year_task} that shows the number of papers by year and task.
It is worth noting that categories in this section include papers that use causal models for which CD algorithms are not directly applicable (e.g., when using the \textit{potential outcome} framework). This distinction is analyzed in the following.

\subsubsection{Causal framework}
\label{sec:rq2:2}
We identified three primary causal frameworks used in the literature (see Section \ref{sec:background}): PO, SCM, and CausalDAG. 
To classify the papers, we rely on what the authors explicitly state  in their study, or, where not stated, on our understanding of the proposal. 
While most papers refer to the three primary models, we  identified a few papers that employ alternative causal models. These include Fuzzy Cognitive Maps, Evidential Networks, Difference-in-Differences, and Causal Trees. We have grouped these models into an “Other” category.

\begin{figure}[t]
\centering 	\includegraphics[width=.83\textwidth]{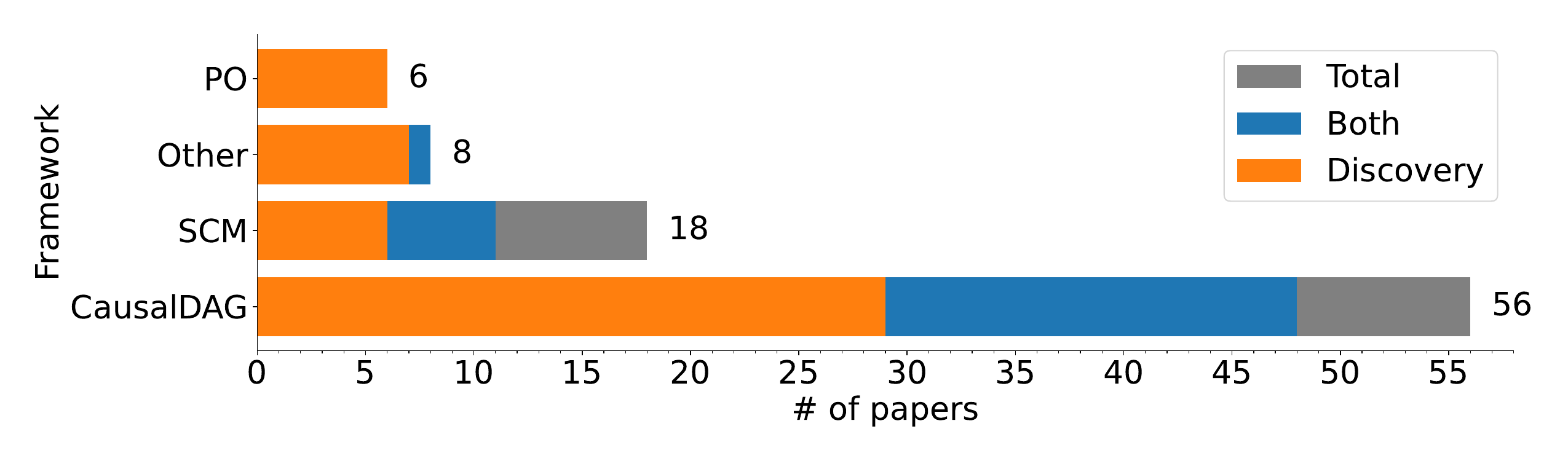}
\vspace{-12pt}
\caption{RQ2: Number of papers per causal framework.}
\label{fig:cr_model}
\vspace{-6pt}
\end{figure}

Figure \ref{fig:cr_model} shows the number of papers by models. 
It is important to note that when a paper employs multiple models, each one is counted separately, leading to a cumulative count that may exceed the total number of selected papers. The only exception is in the case of SCM and CausalDAG (since an SCM necessarily uses a CausalDAG): 
the CausalDAG category refers to the papers using only a CausalDAG but not an SCM. %

With the exception of P84 that is categorized as “Other”,\footnote{The authors of paper P84 do not explicitly refer to CausalDAG.} all the papers that use CD, as denoted by the “discovery” and “both” labels in Figure \ref{fig:cr_model}, fall under SCM and CausalDAG. This is because CD aims to construct a graphical representation of causal relations, a distinctive feature of these two models.

CausalDAG is the predominant choice, employed in estimand identification methods (e.g., Back-door criterion) and selection of the covariates to perform regression or to apply the PO framework. Most papers that are exclusively focused on CD fall into this category, as they do not require  causal effect estimation between variables to perform CI. Only two papers (P42 and P46) published in 2023 perform CD and generates an SCM; we categorized them “SCM" for causal model and as “CD" for the task, since they primarily use SCM's associated CausalDAG for their purposes.

\begin{figure}[bh]
\centering 	\includegraphics[width=.98\textwidth]{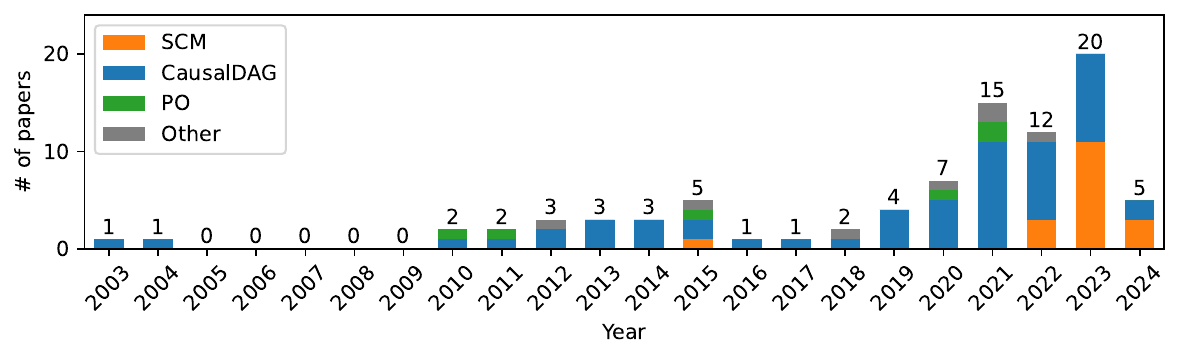}
\vspace{-8pt}
\caption{RQ2: Number of papers per causal framework by publication year.}
\label{fig:cr_year_model}
\end{figure}

Figure \ref{fig:cr_year_model} shows that all the 18 papers explicitly using SCM have been published between 2022 and 2024. The majority of them conducts both CD and CI. The only exception is P48, published in 2015, in which the authors use an SCM (though not explicitly stated) to perform debugging.
The first explicit mention to SCM was in 2010, within the approach by Baah \textit{et al.} (P28) for statistical fault localization. The authors utilized the causalDAG and the PO framework, noting that Pearl demonstrated how SCM subsumes the PO framework for counterfactual reasoning \cite{pearl2009}.

The usage of the PO framework is primarily observed in combination with other models (e.g., CausalDAG and “Other” models). Only one paper (P9), detailing the Netflix experimentation platform, explicitly employs the PO conceptual framework without referencing the use of any other model.

In summary, the conceptualization of causality by Pearl has facilitated the practical application of CI in SQA, rendering complex statistical methods more accessible in this area. Notably, we found that 54 out of 86 papers adhere to his formulation,\footnote{We consider a paper in the Pearl formulation if they are based on graphical causal models and tests (e.g., back-door criterion).} with a significant portion emerging in recent years (45 out of 54 papers from 2021 to 2023, all 2023 and 2024 papers - Figure \ref{fig:cr_year_model}). Papers categorized under “PO” or “Other” 
diverge from the strict association with Pearl's formulation; for example, P2 employs Fuzzy Cognitive Maps for threat modeling; P5 utilizes Evidential Networks for fault localization; P10 and P85 employ Difference-in-Differences for KPI prediction, while P29 for Software Testing;
P14 and P82 use causal trees and PO for KPI prediction with Heterogeneous Treatment Effect.

\subsubsection{Causal Discovery methodology}
\label{sec:rq2:3}
\begin{figure}[t]
\centering 	\includegraphics[width=.44\textwidth]{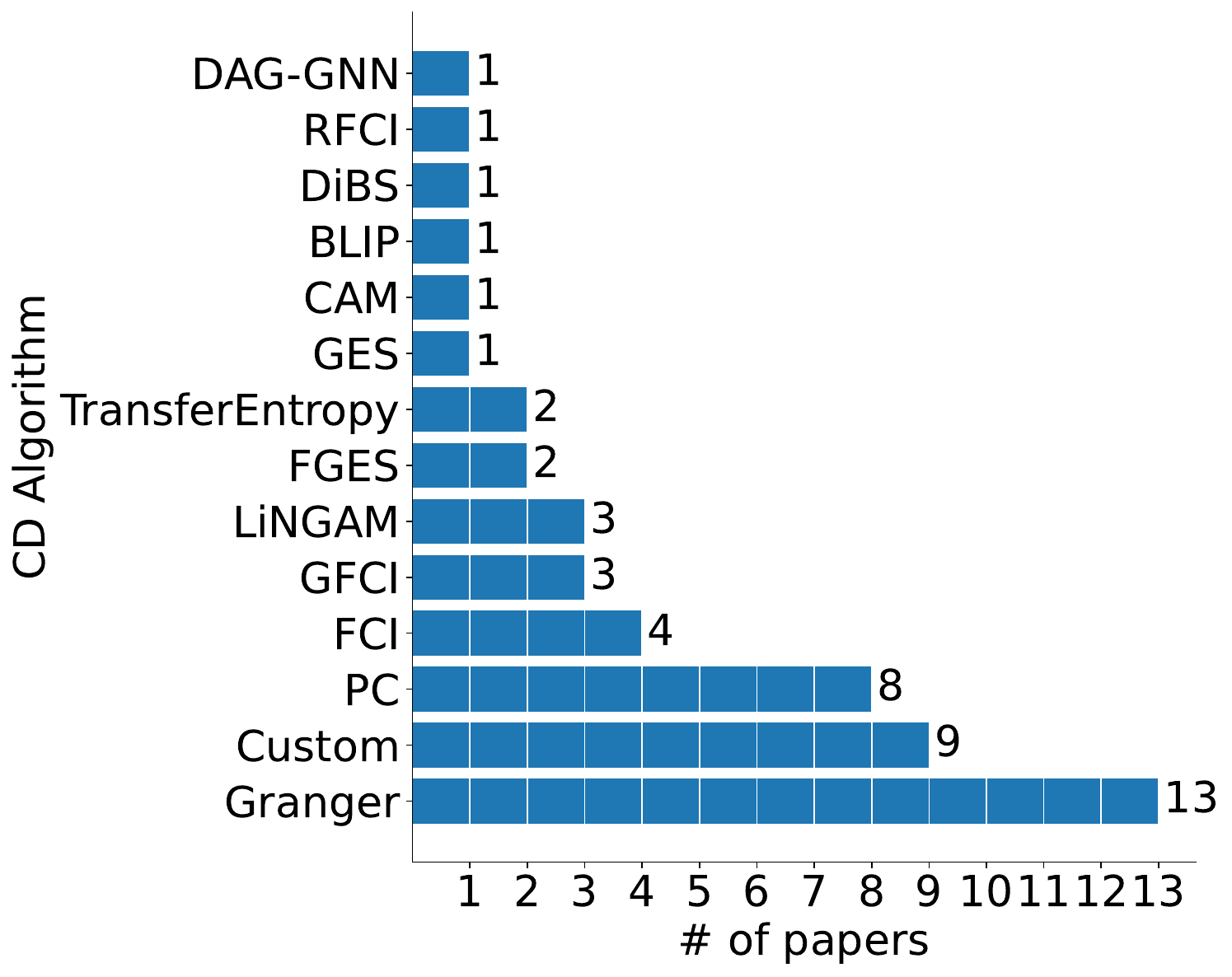}
\vspace{-6pt}
\caption{RQ2:  Number of papers per Causal Discovery algorithm.}
\label{fig:cr_cd_algo}
\vspace{-6pt}
\end{figure}

This section focuses on the CD algorithms used to automatically build a causal structure from observational data.
As shown in Figure \ref{fig:cr_tasks}, 44.1\% of the collected papers (38 papers) utilize CD algorithms.\footnote{Percentage obtained by summing the “Discovery” and “Both” categories in Figure \ref{fig:cr_tasks}.} However, by considering only papers that use graphical causal models for which CD is applicable, we find 38 out of 74 (51.3\%) papers. Figure \ref{fig:cr_cd_algo} shows the number of papers by CD algorithm. Some papers (e.g., P35, P36, and P64) use more than one algorithm: each is counted independently. Extensions of PC and LiNGAM (e.g., PC-kernel and direct/mixedLiNGAM, respectively) are considered as one.

26.5\% of papers opt for constraint-based CD algorithms. For instance, P18, P20, P34 use FCI \cite{Spirtes95}, while P25, P31, P35, P36, P50, P57, and P83 use PC\footnote{P35 employs both PC and PC-kernel.}\cite{Spirtes93}, P64 use both in addition to RFCI \cite{Colombo12}. Score-based CD algorithms such as GFCI (P3, P64, and P79), GES/FGES (P35 and P64)\cite{Chickering02}, BLIP (P45) \cite{Li22_blip}, and DiBS (P54) \cite{Lorch21} are common choices too (16.3\%). 
The authors' rationale for choosing FCI (and GFCI) lies in the ability to discover latent confounders based on conditional dependence relationships between measured variables, as well as in the compatibility with various data types.
P50 motivates the use of PC for its performance without any manual data labeling and its low complexity. 
Other papers using PC and GES either compare multiple CD algorithms or do not explicitly motivate their choice of CD algorithm.

Other papers leverage algorithms based on Functional Causal Models (8.2\%). Specifically, P35 uses LiNGAM \cite{Shimizu06} and CAM \cite{Maeda21}, while P36 and P38 use mixedLiNGAM \cite{Shimizu14} and directLiNGAM \cite{Shimizu11}, respectively.
The authors of these papers opted for LiNGAM as it consistently provides the directionality of causal relationships.

A significant number of papers (28.6\% - i.e., P35, P39, P66 to P78) employ Granger causality \cite{granger69} or transfer entropy \cite{Barnett09} to uncover causal relationships between time series. P49 is the only paper to use Continuous Optimization-based algorithms: it uses DAG-GNN \cite{Yu19} to build a CausalDAG modeling anomaly propagation paths in microservices applications and ranking metrics to localize the root cause by traversing along the graph.

\begin{figure}[t]
\centering 	\includegraphics[width=.95\textwidth]{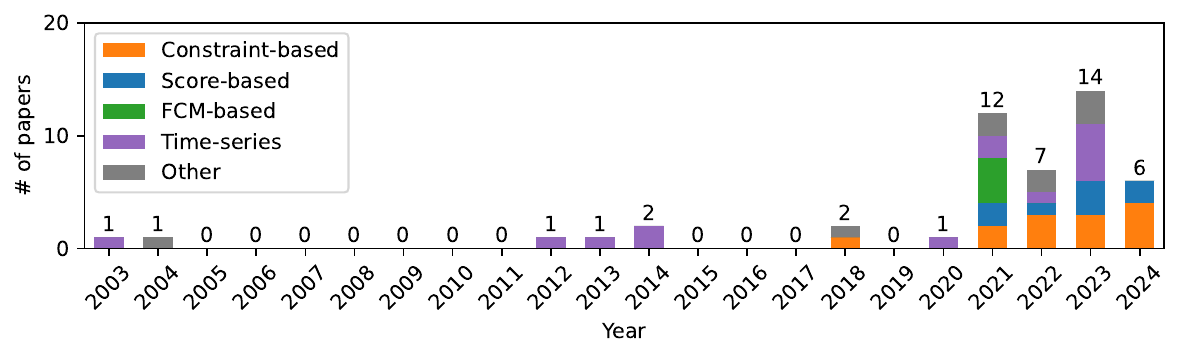}
\vspace{-9pt}
\caption{RQ2: Number of papers per Causal Discovery algorithm by publication year}
\label{fig:cr_year_cd}
\vspace{-6pt}
\end{figure}

While numerous CD algorithms are available, several papers opt for custom CD methods, often extending existing ones. For instance, P22 employs an ensemble of Message Length-based CD algorithms, P31 and P25 utilize a modified version of the PC algorithm, P41 leverages Latent Dirichlet Allocation for natural language processing and feature extraction, P47 employ CD with Ordering based Reinforcement Learning, and P82 Recursive Neural Tensor Network. Without considering algorithms for time series and P22 presented by Li and Dai in 2004 that was the first to use CD, (Figure \ref{fig:cr_year_cd}), the first paper to implement CD comes only in 2018. This is P57, in which  Wang \textit{et al.} use the PC algorithm to build a CausalDAG and random-walk to perform root cause analysis in microservices systems.

Despite the undeniable advances made in the last decade in the release of tools implementing CD algorithms \cite{wire}, by looking at the time window considered in Figure \ref{fig:cr_year_cd}, it can be seen that only a few papers used these algorithms. On the other hand, it shows an increment in their adoption, mainly starting from 2021, when the first FCM and Score-based algorithms have been used.

Finally, most papers opt to derive the causal model from experts' knowledge of the domain and software systems. While these sources of information are valuable to build the entire causal model, it is worth noting that they can be used in conjunction with many CD algorithms, by providing them with prior knowledge specifying required or forbidden relationships, as well as their direction. However, we did not find any paper that used CD algorithms in combination with prior knowledge.

\subsubsection{Causal Inference methodology}
\label{sec:rq2:4}
Out of 63 papers that perform CI, 39 explicitly give information about the inference methodology. The remaining either do not give details or perform other forms of inference (i.e., not associated with any in the classification scheme); among them, P3 uses expectation-maximization \cite{Dempster1977} to build a conditional probability table, P7 performs differential checking between buggy and correct programs, P10 and P29 compute the ATE with the Difference-in-Difference (DiD) model, P19 propose a novel algorithm that use Shapley values to compute the contribution estimate of a change, P53 use structural equation modeling and path analysis for latent hazard notification for automated driving systems, and P61 perform actual causality analysis with the HP2SAT tool proposed by Ibrahim \textit{et al.} \cite{ibrahim20}.

\begin{figure}[b]
\centering 	\includegraphics[width=.77\textwidth]{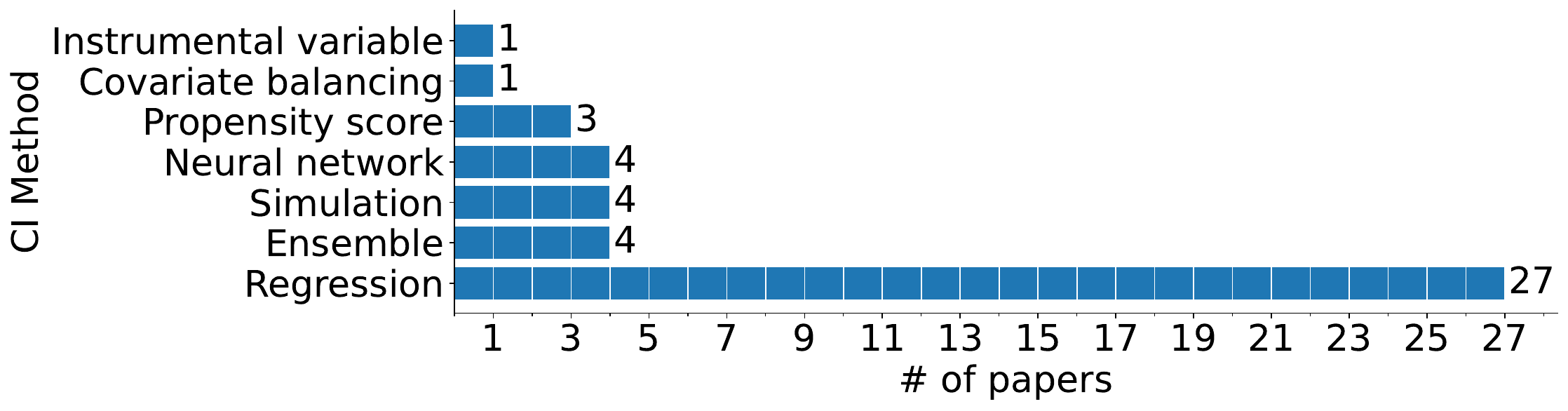}
\vspace{-6pt}
\caption{RQ2: Number of papers per Causal Inference methodology}. \textit{(Note: some papers cover multiple CI methods.)}
\label{fig:cr_cd_estimation}
\end{figure}

Figure \ref{fig:cr_cd_estimation} shows the number of papers per CI methodology.  The most used methodology is regression adjustment, with 27 out of 39 papers. Ensemble, simulation, and neural networks methods see four papers each: the first method comprehend P14 using causal trees, P21 and P26 random forest, and P44 causal forest; in the second method, P34, P51, P52, and P64 use SCM as causal model and simulate the effect of interventions/counterfactuals by sampling from post-intervention distributions - the all perform testing tasks, with the exception of P51 that targets fault localization; in the third method, P39 and P40 use Long Short-Term Memory (LSTM) neural networks, respectively for prediction of performance indicators in microservices architectures and for automatic program repair - P45 and P54 use double ML to compute the ATE.

Three papers use propensity score-based methods: P13 uses Bayesian propensity score matching to generate balanced control and treatment groups from observational data in the automotive domain, P15 uses the generalized propensity score to correctly estimate the effect numerical expression have on failures in numerical software by reducing the confounding caused by other faulty expressions. P17 computes the propensity scores to balance treatment and control group to ultimately compute \textit{suspiciousness score} and perform fault localization for debugging.

Finally, covariate balancing and instrumental variable methods have one paper each. The former is P16, an extension of  previously published P15, using covariate balancing propensity score to reduce confounding bias  in numerical software; the latter is P45 (included also in simulation), using instrumental variable to overcome latent confounders.

All papers fall in the categories “without unobserved confounders” and “advanced methods for big data”. This means that to simplify the problem authors assumed that all confounders were among the observed features. 

The most employed metrics are the ATE and CATE, with 11 papers using them (P9, P10, P13, P14, P23, P29, P41, P44, P45, P48, P54). ACE is adopted by 4 papers (P18, P20, P51, P52). The remaining papers do not explicitly state the metrics  used.
A number of papers that perform fault localization rely on other structured metrics: \textit{attribution score} (P19 and P43); \textit{suspiciousness score} (P8, P11, P15, P16, P17, P21, P24, P26, P27, P28, P30, P34, P48, P56). %
Attribution score is adopted in recent papers (both in 2023)  computing the causal attribution with the \texttt{dowhy-GCM} library; it aims at identifying and attributing changes in a distribution of a variable to changes in causal mechanisms of upstream nodes \cite{Budhathoki21}. Suspiciousness score is adopted in papers mainly ranging between 2010 and 2017, in which the score is used to guide developers to faults, by highlighting suspicious statements in the code.

\begin{figure}[t]
\centering 	\includegraphics[width=.93\textwidth]{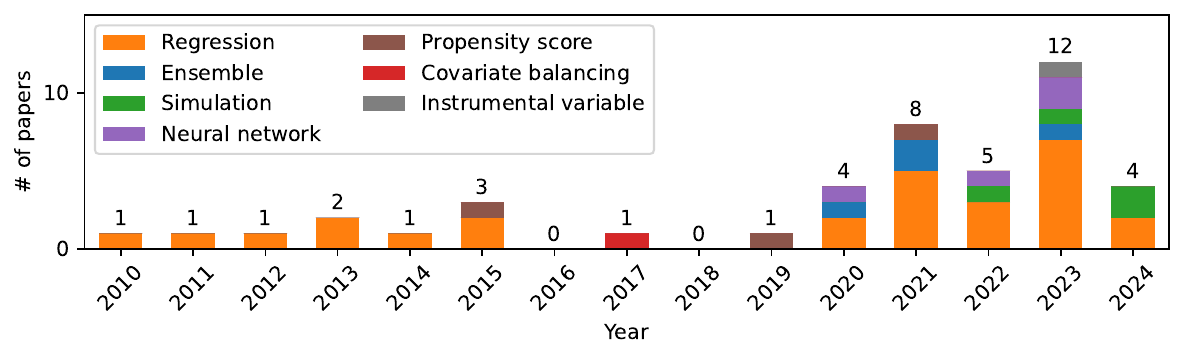}
\vspace{-9pt}
\caption{RQ2: Number of papers per Causal Inference methodology by publication year.}
\label{fig:cr_year_ci}
\vspace{-6pt}
\end{figure}

In summary, starting from 2020 we observe (Figure \ref{fig:cr_year_ci}) an increment in the use of regression adjustment, simulation, and neural network-based CI. We attribute this to the implementation of these methods in libraries and tools (e.g., \texttt{dowhy}), that ease the application of CI providing black-box capabilities to estimate causal effects from data. On the other hand, many inference methodologies, including the ones that assume unobserved confounders, are worth to be explored since they could strongly improve the inference effectiveness in many contexts. 

\subsubsection{Tools / libraries}
\label{sec:rq2:5}

\begin{figure}[b]
\centering 	\includegraphics[width=.93\textwidth]{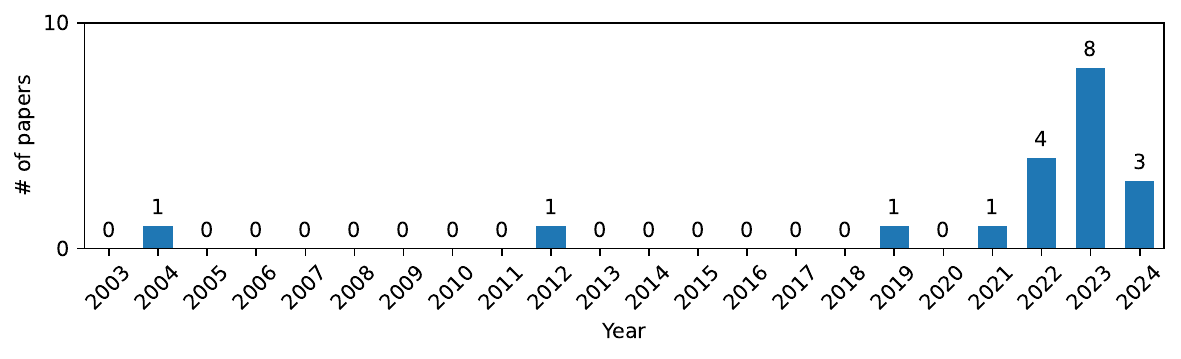}
\vspace{-9pt}
\caption{RQ2: Number of papers using third party tools by year.}
\label{fig:cr_year_tool}
\end{figure}

Figure \ref{fig:cr_year_tool} shows the number of papers explicitly mentioning the usage of third party tools. It can be seen that many papers, especially before 2021, implemented their code to perform causal tasks. Starting from 2021, an increasing number of papers use available third party tools and libraries, which we collected and reported. Table \ref{tab:cr_tools} shows the tools and libraries we found, together with their implementation language, papers using them, and their reference. Notably, the most used ones are \texttt{causal-learn} and \texttt{dowhy} for CD and CI, respectively. 

\texttt{causal-learn} is a Python wrapper of \texttt{Tetrad} (second most used CD tool) and provides a large number of constraint-, score-, and FCM-based CD algorithms. It also provides the implementation for a number of (Conditional) independence tests (e.g., Fisher, Chi-square) and score functions (e.g., BIC, BDeu, Generalized score with cross-validation and marginal likelihood) to guide these algorithms. 

\texttt{dowhy} is a Python library, initially developed by Microsoft, that covers the whole CI process: \textit{modelling} a causal problem, \textit{identifying} a target estimand, \textit{estimating} causal effect based on identified estimand, and also possibly running a series of \textit{refutation} tests on the estimate used to increase the confidence. It offers a wide number of estimation methods and supports both the analytical and the simulation-based inference (i.e., \texttt{Do-sampler}) through an extension called \texttt{DoWhy-GCM}. This extension provides an easy and automatic way to answer causal questions, such as simulating the impact of interventions, computing counterfactuals, estimating average causal effects, and attributing distributional changes. \texttt{dowhy} has been designated as the most complete tool in recent studies carried out by Nogueira \textit{et al.} \cite{wire}.

\texttt{causal-learn} and \texttt{dowhy} are now provided and maintained within \texttt{py-why}, an open source ecosystem for causal ML, encompassing also \texttt{EconML} (a tool used by paper P45, mainly employed for  decision making in economics). It provides a number of ML-based techniques to estimate individualized causal responses from observational or experimental data.

Other valuable CD tools are: \texttt{CausalDiscoveryToolbox}, which also provide implementations for constraint-, score-, and FCM-based CD algorithms;  \texttt{lingam}, which provides the implementation of LiNGAM; and \texttt{pycausal}, another Python wrapper of \texttt{Tetrad} but not maintained anymore.

Other CI tools include: \texttt{Bayes-Net} and \texttt{pyagrum}, two tools for Bayesian Networks and other Probabilistic Graphical Models respectively implemented as Matlab toolbox and as a Python/C++ library; \texttt{JFCM}, a Java library that implements Fuzzy Cognitive Maps; \texttt{causality}, a Python library implementing Pearl's CI methods on DAGs; \texttt{DeepIV}, a package for counterfactual prediction using deep instrument variable methods that builds on Keras \cite{keras}; \texttt{MatchIt} is an \Rlogo~package that provides implementation of some CI methods, such as propensity score-based and covariate balancing; and \texttt{SPSS Amos}, an IBM proprietary tool for structural equation modeling.

In summary, starting from 2020, when many tools and libraries that ease the application of CD and CI have been released and used by researchers, an increasing number of papers rely on them. Among them,  \texttt{causal-learn} and \texttt{dowhy}, given by the same provider, resulted to be the most employed.

\begin{table}[t]
\centering
\caption{Causal Discovery / Causal Inference tools.}
\label{tab:cr_tools}
\resizebox{.76\textwidth}{!}{%
\begin{tabular}{l|l|l|l|l}
\toprule
\multicolumn{1}{l|}{Tool} & Task & \multicolumn{1}{l|}{Language} & \multicolumn{1}{l|}{Papers} & Ref.\\\midrule
\texttt{causal-learn} & Discovery & Python & P18, P20, P31, P49, P83 & \cite{causallearn}\\\hline
\texttt{Tetrad} & Discovery & Java & P3, P34, P64 & \cite{tetrad}\\\hline
\texttt{CausalDiscoveryToolbox} & Discovery & Python/\Rlogo & P35 & \cite{causaldiscoverytoolbox} \\\hline
\texttt{lingam} & Discovery & Python & P35 & \cite{lingamtool} \\\hline
\texttt{pycausal} & Discovery & Python & P64 & \cite{pycausal} \\\hline
\texttt{MatchIt} & Inference & \Rlogo & P17 & \cite{matchit} \\\hline
\texttt{EconML} & Inference & Python & P45 & \cite{bayesnet} \\\hline
\texttt{DeepIV} & Inference & Python & P45 & \cite{deepiv} \\\hline
\texttt{Bayes-Net} & Inference & Matlab & P22 & \cite{bayesnet} \\\hline
\texttt{DoWhy} & Inference & Python & P19, P34, P41, P43, P64 & \cite{dowhy} \\\hline
\texttt{JFCM} & Inference & Java & P2 & \cite{jfcm} \\\hline
\texttt{pyagrum} & Inference & Python/C++ & P52 & \cite{agrum} \\\hline
\texttt{causality} & Inference & Python & P18 & \cite{causalitytool} \\\hline
\texttt{IBM SPSS Amos} & Inference & N/A & P45 & \cite{amostoolibm}\\ \bottomrule
\end{tabular}
}
\vspace{-6pt}
\end{table}

\section{RQ3: Are the solutions using CR experimentally validated?}
\label{sec:rq3}

With this research question, we aim to evaluate to what degree the solutions using CR are experimentally validated. For this purpose, we refer to “Experiment”, “Industrial application” and “Domain” dimensions in Table \ref{tab:dimensions}. 

Except for P9, all the papers provide an experimental evaluation, with P35 also providing a comparison with association-based techniques. Although P9 does not provide experiments, the authors describe Netflix's approach to integrating data science and engineering practices through an advanced experimentation platform (Netflix XP). P3, P10, P18, P19 provide an implemented tool, while P5 and P22 provide code supporting their work.

\begin{figure}[t]
\centering 	\includegraphics[width=.63\textwidth]{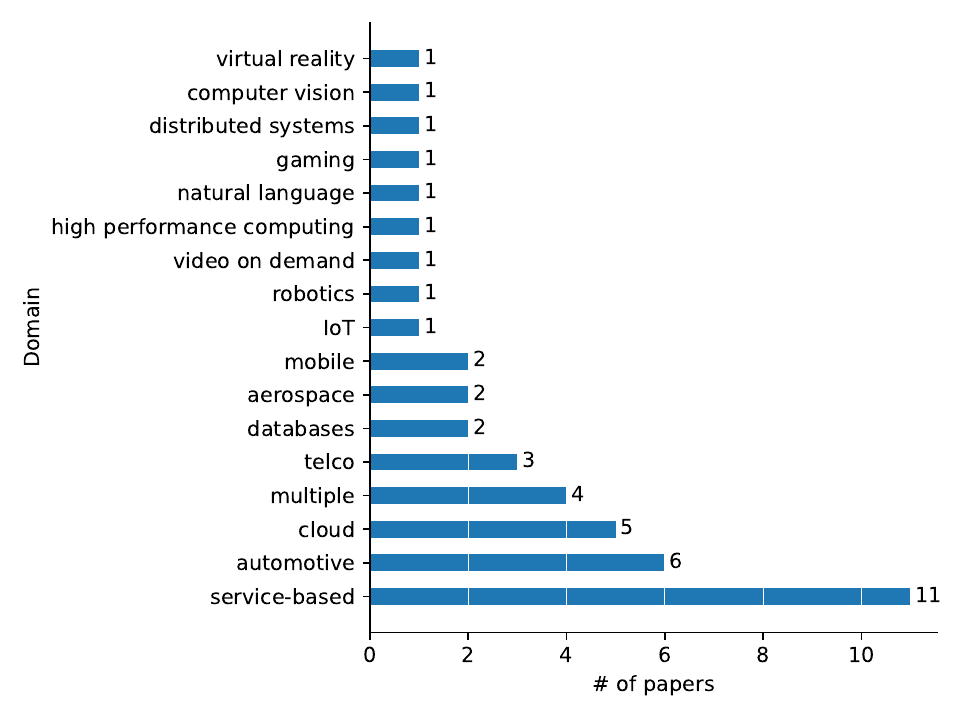}
\vspace{-9pt}
\caption{RQ3: Number of papers per application domain.}
\label{fig:domain}
\vspace{-6pt}
\end{figure}

\begin{figure}[t]
\centering 	\includegraphics[width=.77\textwidth]{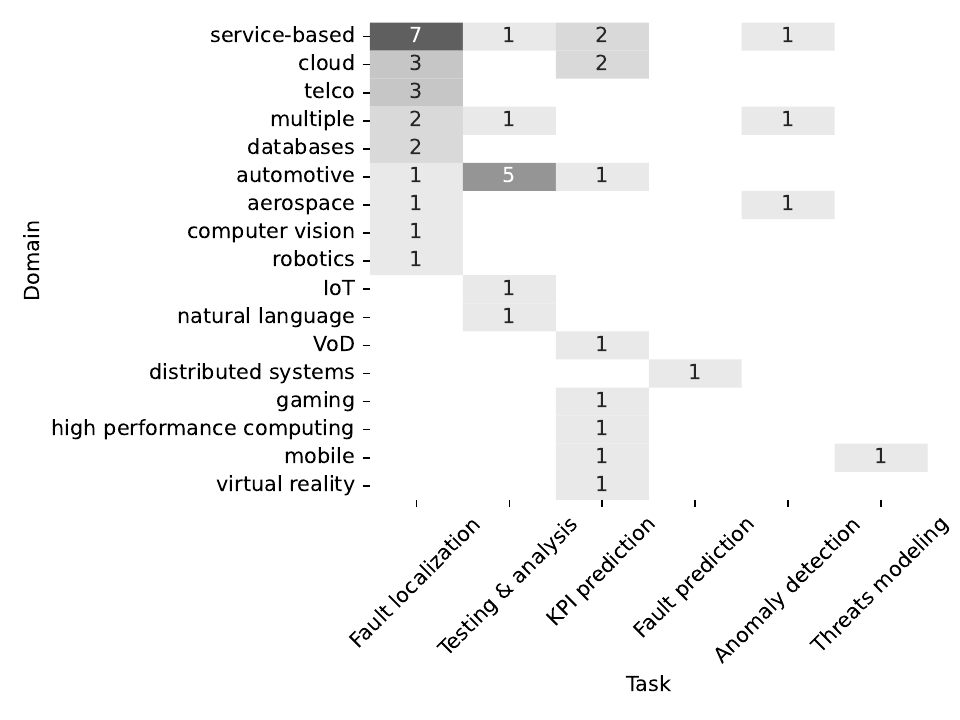}
\vspace{-6pt}
\caption{RQ3: Number of papers per SQA task and application domain.}
\label{fig:tasksdomain}
\vspace{-6pt}
\end{figure}

Figure \ref{fig:domain} shows that the most explored domains are web-services/microservices (P9, P31, P32, P35, P38, P39, P49, P50, P57, P58, P70), automotive (P13, P34, P52, 53, 62, 64), cloud (P55, P73, P77, P79, P80), and telco (P5, P6, P12). We note that many proposals do not apply to a specific domain, or they target multiple domains. Figure \ref{fig:tasksdomain} shows that papers focusing on service-based systems, on cloud and on telco usually performing fault localization tasks. Papers on automotive inspect testing techniques (with P62 focusing on both testing and fault localization).

Few papers validated their proposals by applying CI in industrial contexts, like Volvo (P13) for KPI prediction. %
P13 proposes BOAT (Bayesian propensity score matching for ObservAtional Testing) for automotive software engineering. It generates balanced control and treatment groups from an observational online evaluation and estimates causal treatment effects attributable to software changes. The authors point out how this method can be complementary to agile methodologies to enable responsiveness to change and allow development teams to make data-driven decisions.
P25 present a causal-inference technique for fault localization, employing a dependence-based causal model with the matching of test executions based on their dynamic dependencies, that accounts for the effects of dynamic data and control dependences and significantly reduces confounding bias during fault localization. The paper reports empirical results on 16 programs, including the ones in the Siemens suite (Print-tokens, Print-tokens2, Replace, Schedule, Schedule2, Tcas, and Tot-info), indicating that the new technique performs better than existing fault-localization techniques.
P57 introduces CloudRanger, a tool designed for root cause analysis in cloud-native systems. Its effectiveness is demonstrated through the analysis of real incidents that occurred in IBM Bluemix. P70 and P73 present DyCause, a root cause analysis tool tailored for microservices systems. DyCause is evaluated both in a controlled simulation environment and in a real-world cloud system, specifically IBM Cloud.

In some cases, the industrial partner is not explicitly reported for the used case study, but stated as part of the author information. In P25 the authors (from \textit{Walmart Applied AI Group}) describe LADDERS (Log-based Anomaly Detection and Diagnosis for Enterprise Systems) as an end-to-end system exploiting CI for anomaly detection in the case of system malfunction or failure. In P31 (from \textit{Adobe}) the authors propose Root Cause Discovery, a hierarchical and localized CD algorithm, to detect the root cause of a failure in complex microservice architectures.

There are cases of data provided by organizations like the University of Carolina (P18) in the robotics domain for fault localization and RCA, and data collected from commercial products like Oculus (P41) for virtual reality, Android Apps on the Google Play store (P10), and League of Legends game (P14) aiming at running CI (and also discovery in P41) for KPI prediction.
Many papers refer to open-source datasets and software (P3, P11, P15, P21, P74, P75, P81) for their experiments, such as SockShop\footnote{https://microservices-demo.github.io/.} (P31, P38) and TrainTicket by Zhou \textit{et al.} \cite{Zhou21} (P39), even both (P35), for microservices, and CARLA simulator by Dosovitskiy \textit{et al.} \cite{Dosovitskiy17} (P34) for automotive.

We finally applied the research methods classification schema provided by Petersen \cite{Petersen15} and Wieringa \textit{et al.} \cite{Wieringa06}. Specifically, we considered two categories: \textit{1)} \textit{evaluation research}, namely papers that evaluate the proposal in a real-word industrial context such as by case study, field study, field experiment, and survey; \textit{2)} \textit{validation research}, namely papers that validate a novel solution not used in practice through research methods such as laboratory experiments, simulation, prototyping, and mathematical analysis. We found 11 papers belonging to the first category (i.e., evaluation research), while 75 to the second (i.e., validation research).

\section{Discussion}
\label{sec:discussion}
This Section highlights the main findings of our analysis, and the entailed open challenges for future research.

\subsection{ Main findings}\label{sec:discussion_overview}
Table \ref{findings} lists the main findings, grouped by research question, derived from results reported in the previous Sections. 
CR found a noticeable application in those tasks pertaining to \textit{fault removal} (e.g., via fault localization), but a good share of papers target testing (which is often preparatory to fault removal) and prediction of KPIs of interest for disparate tasks such as to improve reuse (P22), “helpfulness” (P42) and application-specific indicators (P40) (as well as of bugs in one case P3). 
Fault prediction is a further area of application, mostly by using Granger causality to assess cause-effect relations between the predictor metrics' time series and the fault time series (P66, P69, P72, P75). 

While most papers target software reliability in operation, performance issues are often of interest too, e.g., in those papers that perform root cause analysis starting from observed performance issues (P32, P36, P39). It is worth to highlight the (13) papers targeting security (e.g., P2, P23 for threat modelling) and safety.

\renewcommand{\arraystretch}{1.1}
		\begin{table}[t]
			\caption{Main findings of this study.}
			\label{findings}
			\begin{tabular}{p{14.2cm}|c}\hline
					\toprule 
					{\cellcolor[gray]{.9} What is CR used for (RQ1)} & {\cellcolor[gray]{.9}Section} \\ \midrule
					
					\begin{tabular}[c]{@{}p{14.cm}@{}}
						More than half of the papers (44/86)  use CR for \textbf{fault localization} (debugging or RCA) \\ 
					\end{tabular}&
					\begin{tabular}[c]{@{}r@{}} \ref{sec:rq1}  \end{tabular}\\
					\hline
					
					\begin{tabular}[c]{@{}p{14.cm}@{}}
						CR is used mostly to improve or assess/predict \textbf{reliability} (63/86) and \textbf{performance} (12/86), rarely for \textbf{security} (7/86) and \textbf{safety} (6/86).  
					\end{tabular}&\begin{tabular}[c]{@{}r@{}} \ref{sec:rq1} \end{tabular}\\\midrule
					
					\multicolumn{2}{l}{\cellcolor[gray]{.9} How is CR used (RQ2)} \\ \midrule
					\begin{tabular}[c]{@{}p{12.8cm}@{}}
						Most papers use only \textbf{CI} (48/86) rather than only \textbf{CD} (23/74 where 74 is the subset of papers using graphical causal models, for which CD is applicable). A small share of 15 papers exploit both CI and CD. This result in a total of 63 (i.e., 48+15) over 86 papers using CI and 38 (i.e., 23+15) over 74 papers using CD.  
					\end{tabular}&\begin{tabular}[c]{@{}r@{}} \ref{sec:rq2:1} \end{tabular}\\\hline
					\begin{tabular}[c]{@{}p{14.cm}@{}}
						\textbf{Causal DAG}s are the preferred way to deal with causality (56/86), 18 use an \textbf{SCM}. These mostly exploit the Pearlian's formulation (54/86), and are the most recent ones (from 2021 to 2024).   
					\end{tabular}&\begin{tabular}[c]{@{}r@{}} \ref{sec:rq2:2} \end{tabular}\\\hline
					\begin{tabular}[c]{@{}p{14.cm}@{}}
						Although \textbf{CD} algorithm are underused 38/74 papers employing CD, the trend is increasing (30/31 papers appeared in 2020-2024). Despite the availability of new CD algorithms, researchers opt for old algorithms (e.g., constraint-based like PC and FCI) or build their custom version.
					\end{tabular}&\begin{tabular}[c]{@{}r@{}} \ref{sec:rq2:3} \end{tabular}\\\hline
					
					\begin{tabular}[c]{@{}p{14.cm}@{}}
						\textbf{CI} algorithms rely in most cases on regression-based estimation (27/39 papers reporting the inference algorithm). About 38\% of  papers using CI (24/63)  use it as a black box through libraries. All papers assume that  confounders are among the observed features.%
					\end{tabular}&\begin{tabular}[c]{@{}r@{}} \ref{sec:rq2:4} \end{tabular}\\\hline
					
					\begin{tabular}[c]{@{}p{14.cm}@{}}
						Several \textbf{tools and libraries} are being released since 2020; 19/86 papers use these tools, 16 of which are published in 2021-2023.
					\end{tabular}&
					\begin{tabular}[c]{@{}r@{}} \ref{sec:rq2:5} 
					\end{tabular}\\\midrule
					
					\multicolumn{2}{l}{\cellcolor[gray]{.9} Are the solutions using CR experimentally validated} (RQ3) \\ \midrule
					\begin{tabular}[c]{@{}p{14.cm}@{}}
						Except for P9, all the papers provide an experimental evaluation. The most explored domains are \textbf{web-services/microservices}, \textbf{automotive}, \textbf{cloud} and \textbf{telco}. The presence of just 11 industrial subjects in 86 papers shows low maturity or technology readiness of CR solutions.
					\end{tabular}&\begin{tabular}[c]{@{}r@{}} \ref{sec:rq3} \end{tabular}\\\hline
					
					\bottomrule
				\end{tabular}
			\end{table}

			From the analysis done in RQ2, CI methods (often using regression adjustment estimation techniques) through causal DAGs are the predominant solution. The spread of the Pearl's formulation is also further favouring this type of model, making the use of causality more intuitive. Causal structure discovery algorithms have a great potential in supporting the creation of causal models, although human knowledge plays always a prominent role in causal modelling, since it encodes the necessary assumptions going beyond a mere data-driven approach. Given the recent advances in causal structure discovery \cite{Vowels22} and tools readily available, the use of these algorithms is expected to increase. A sign of an increasing trend stems from the fact that 30/38 papers using CD are published in 2020-2024.
			
\begin{figure}[t]
				\centering 	\includegraphics[width=0.52\textwidth]{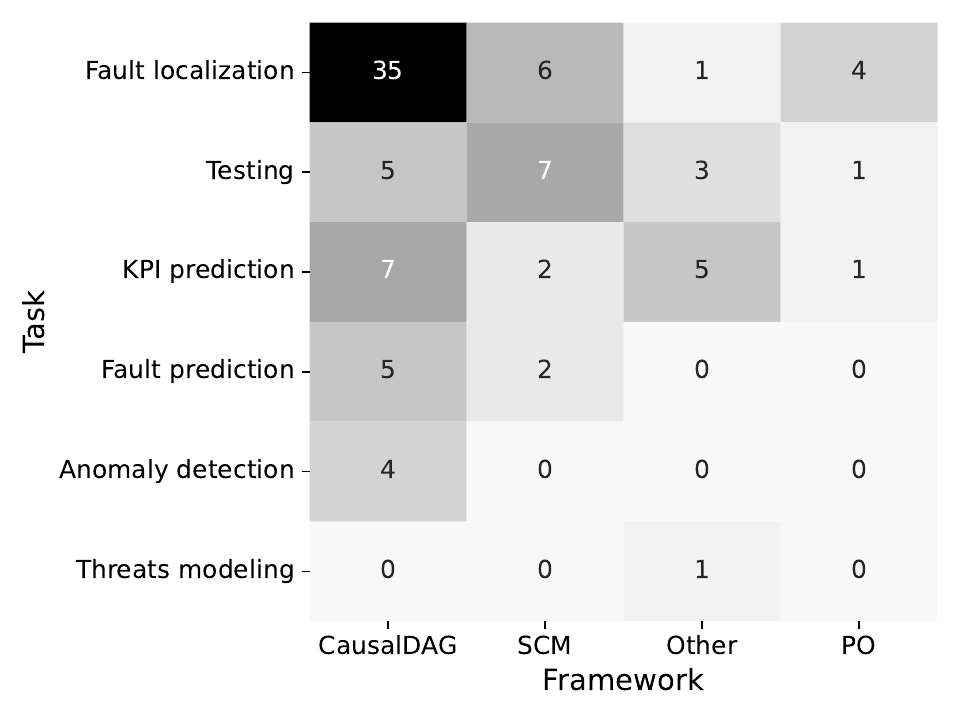}
				\vspace{-6pt}
				\caption{Number of papers per CR framework and SQA task.}
				\label{fig:task_framework}
				\vspace{-6pt}
			\end{figure}

			Figure \ref{fig:task_framework} shows the number of the paper occurrences mapped to the SQA tasks (RQ1) and CR frameworks (RQ2). Graphical causal models, like CausalDAG and SCM, have been applied to every SQA task. The only exception is the Threats modeling task, which is addressed only by one paper using fuzzy CausalDAG (P2), thus still employing graphical modeling. In general, the most used CR framework for fault localization, KPI and fault prediction, and anomaly detection is CausalDAG, while SCM is the most used for testing. The PO framework is mainly employed for fault localization, yet it finds adoption also for testing and KPI prediction. The overall results point out that graphical causal models are the most adopted CR frameworks, and are enough versatile to be used for a variety of SQA tasks.
			
			As for RQ3, almost all papers provide an experimental evaluation. Most of them rely on open-source artifacts for the experiments, and just few papers (12) experiment on industrial subjects. Except for automotive and microservices, the technology readiness of CR solutions for SQA seems still at an early stage.

			It is also interesting to look at the cross-count in Table \ref{cross-analysis} referred to the two main attributes of RQ1 and RQ2, namely how CR is used for which task.  The usage of CD algorithms, although much more limited, has been explored for almost all the tasks; CI is by far the most used method for fault localization, in which the causal model is build in most cases (48/86) without the support of a CD algorithm.  
			
			\begin{table}[hb]
				\centering
				\caption{Number of papers by CR method and SQA task.}
				\label{cross-analysis}
				\renewcommand{\arraystretch}{1.03}
				\begin{tabular}{l|c|c|c}
					\toprule
					& \multicolumn{3}{c}{\textbf{CR method}}  \\
					\textbf{SQA task} &{Inference} &  {Discovery} & {Both}\\\hline
					{Fault localization}  &30 &7 & 7\\\hline
					{Testing and analysis}&10 &3 &3\\\hline
					{KPI prediction}&6 &4 &4\\\hline
					{Fault prediction}&2 &5  &1 \\\hline
					{Anomaly detection}&- &4 &- \\\hline
					{Threats modeling}&1 &- &- \\\bottomrule
				\end{tabular}
			\end{table}

			\subsection{Open challenges and research opportunities}
			
			CR is a research area with a large applicability, as it represents one of the most accredited methodologies to go beyond a purely data-driven approach \cite{Pearl18}. Causal models can be automatically extracted from data (CD) and fine-tuned (\textit{fitting}). This aspect can ease their effective and efficient implementation and their adoption in industrial and commercial environments. Moreover, the possibility to improve such models with previous knowledge by defining causal-effect relationships that, differently from DNNs, can be integrated with a human-readable format is an aspect not completely inspected in the literature. This ability allows mixing a data-driven model construction with the domain knowledge that human experts can encode in the model in the form of required or forbidden cause-effect relations. 
			
			Readability and interpretability of causal models by humans allow researchers to envision their adoption with human-in-the-loop. 
			Proper infrastructures/algorithms can determine when to ask for human feedback during learning. 
			The aspects of incorporating prior/domain knowledge and receiving human feedback for the discovery and inference phases is mentioned only in a few papers, mainly in their “future work" sections (P20, P22, P28, P37).
			
			The possibility for users to query causal models for understanding and forecasting is very close to what users currently do with generative artificial intelligence. However, the possibility of interacting directly with a human-readable model can be crucial for ethical aspects, as “constraints” can be defined by adding/removing causal relationships.
			
			CR solutions can be used to improve the usability of software systems by supporting developers or even the final users. For instance, they can support front-end developers in designing user interfaces, such as moving or replacing a widget, or final users suggesting step-by-step how to interact with the user interface.
			
			CR can also be adopted in conjunction with ML. CR can be used to define strategies for automating and reducing the cost of fine-tuning ML models and Deep Neural Networks \cite{kaddour2022causal}. It also naturally fits the explainability and understandability tasks. Since it is adopted for RCA and Anomaly detection, CR could be very effective for explainability/understandability of ML-based systems behaviors.

			From the above analysis, further possible research directions emerge: 
			
			\begin{itemize}
				\item Due to its high level of abstraction, CR is potentially applicable to numerous quality-related tasks in software engineering. At date, fault localization is the main targeted task, testing and KPI prediction being addressed more recently. While both tasks can be efficiently supported by CR, testing could in principle gain more from the use of CR. In fact, while \textit{prediction} is well handled by Machine Learning models, testing could benefit more from CR: the conceptual steps involved in a testing process go beyond an observations-based prediction task -- tests generation, for instance, is about \textit{planning} rather than predicting, in the form of what-if scenarios (i.e., what output would we get if we submit a specific input). 
				Given the cause-effect relation between test inputs and outputs, CR can be exploited for tests generation (as in P35) as well as for tests selection and prioritization. 
				
				\item There is wide room of applicability in the early phases of the life cycle, e.g., during design. CR has been exploited for threats modelling in only one study; using it to support design, e.g., in conjunction with behavioural design diagrams, and requirements analysis is a direction worth to be explored, especially in reactive event-driven systems. For instance, given a model of the environment (possibly enriched with historical data), the use of CR allows to model and infer the outcome of what-if scenarios at low cost, hence helping to draw requirements and make design choices. This is largely unexplored in the current software engineering literature. 
				
				\item Further tasks could be better addressed with help of CR. For instance, anomaly detection could leverage the attribution analysis easily done on causal models (e.g., via libraries like \texttt{DoWhy}) which helps with the identification of the variables responsible for observed anomalies \cite{Budhathoki22} and distributional changes \cite{Budhathoki21}. 
				Much work is done on debugging, but we found just one work on code smells (P23), whereas CR could be used to relate code smells with performance or correctness issues and to assess the expected effect of refactoring in the code (again, in a what-if style). Similarly, attribution can be applied to identify the component in a microservices architecture that caused an anomaly. %
			\end{itemize}
			
			Besides the potential application of CR in software engineering tasks, an important research effort is required in a better exploitation of the advances in the CR literature. 
			For instance: 
			\begin{itemize}
				
				\item Based on the results, while research in software engineering is increasingly applying CD algorithms, the approaches often: \textit{i)} rely on outdated methodologies (e.g., PC, FCI), and \textit{ii)} do not leverage prior/domain knowledge, which is one of the key advantages of graphical causal models. It is worth stressing that integrating CD with additional knowledge sources (e.g., human expertise or architectural frameworks like service-dependency graphs in service-based systems) could significantly enhance the extraction of causal models that better align with the observed data. Additionally, this integration would allow for the validation of assumptions in certain application contexts where there is lack of knowledge about the data generating process (e.g., while two variables may be assumed causally independent, a causal discovery algorithm might reveal dependencies based on the data). In general, it is crucial to align more closely with the AI community and explore recent advancements in CD methods \cite{wire}.

				\item Combining different strategies to learn a causal graph more efficiently, such as using reinforcement learning \cite{zhu2020causal}, active interventions  \cite{scherrer2022learning}, or active learning via deep reinforcement learning \cite{AMIRINEZHAD202222}. This would foster the use of CD algorithms to capture relevant relations in the system under test depending on the addressed task (e.g., design, testing, fault localization).

				\item As for the inference, several strategies could be explored to decide about which intervention is more fruitful. For instance, 
				a common strategy is active learning, which selects interventions that minimize the uncertainty about the true graph \cite{zhang23,Steyvers03,acharya18}. Other possibilities include sampling or search-based strategies to navigate the space of possible interventions intelligently towards an objective  \cite{34}.

					\item Most classified papers fall within the categories of “without unobserved confounders” and “advanced methods for big data”. This means that, mostly to simplify the problem, the authors assumed that all confounders were among the observed features. The adoption of strategies that also deal with unobserved confounders would reduce the bias in cases where this assumption is not satisfied. %
					
					\item A promising research path is the use of \textit{causal reinforcement learning} (CRL) \cite{zeng2023survey}. This framework encompasses under the same theory reinforcement learning, where actual physical interventions are done in an environment to get a reward, with causal modelling where interventions are simulated on a causal model of the environment. These are shown to be two sides of the same coin. Its exploitation for software engineering would be helpful to deal with sequential tasks, such as sequential tests generation. Today's strategies include reinforcement learning to this aim, but empowering it with causal knowledge is expected to boost performance, as further knowledge can be given to the agent for a smarter exploration of the test sequences space. %

					\item Libraries and tools for CD and inference are being developed at a fast pace. This is paramount to boost research in this field. The competence of software engineering community could indeed push also this area of research, fostering the development/maintenance of open source libraries and tools to ease the adoption of CR by software engineering practitioners. The main directions we identify to pursue this trend are: \textit{i)} the increasing adoption of CR for software engineering would naturally trigger new requirements that CR tools will have to satisfy. It could be important to foster the communities to share their effort in implementing or improving such requirements, e.g., open sourcing all their code and contributing to already existing libraries, like \texttt{py-why} ecosystem, with \texttt{causal-learn} for discovery and \texttt{dowhy} for inference; 
					\textit{ii)} besides scientific publications, is to set up, wider cooperative efforts in developing CR solutions tailored for software engineering, for instance though funded research projects;
					\textit{iii)} at a higher level, while are witnessing to CR applied to software engineering, an opposite trend could be promoted (e.g., through workshops and similar initiatives) namely software engineering for CR. This would resemble what happened with the bidirectional trend of ML and software engineering.

					\item One of the challenges emerged from RQ3 is the lack of industrial applications. The use of CR in software engineering  is a  relatively young field, the level of technological readiness of the solutions is still low. Industry-academia collaboration is required to favour the adoption and spread of these solutions. %

				\end{itemize}
				
				We believe that the fruitful application of CR to SQA tasks by the community is a short/medium term effort, partly already in place, that would favour the spread of CR as an alternative tool (e.g., with respect to or in conjunction with purely ML-based solutions to support software engineering). The better exploitation of CR, we believe, is a longer term effort, as it requires a deeper understanding of potential pros and cons.

\section{Threats to validity}
\label{sec:threats}

In the following, the main threats to validity are reported. 

The process of selecting the primary studies was performed by three people, however, some papers might have been included or excluded incorrectly. 
We followed the search evaluation steps in accordance with Petersen’s guidelines \cite{Petersen15} and Kitchenham’s suggestions \cite{Kitchenham13}. 
The test-set approach for search validation suggested that 19/21 papers were included in the list of selected papers, showing substantial confidence in the search. However, similarly to the remaining 2 papers, a few could be missing.

The search was conducted by querying multiple data sources, so as to cover a high number of publishers. Additionally, the terms used in the search string were very general. This allowed to be very conservative in the first search (which included 770 papers after merging) in order not to miss relevant studies.  The resulting set was then manually filtered to discard irrelevant results with a set of documented inclusion and exclusion criteria, used to refine the search unambiguously (693 studies were discarded). An additional threat to the validity of the search process stems from the terms chosen for the query string. Specifically, the terms related to SQA were not systematically selected, but are based on our experience and judgment. However, we believe that the terms we used reflect areas with well-established research communities, such as dependability, security, and performance engineering.

Another threat regards the quality of selected studies. We have not performed a systematic quality assessment of the papers; instead, an informal author's judgment has been applied. To mitigate this threat: \textit{i)} only peer-reviewed studies have been considered (excluding white papers, editorials, etc.); \textit{ii)} studies were searched among those indexed by the most used digital libraries in computer engineering and computer science, which filter out several low-quality conferences and journals; \textit{iii)} inclusion/exclusion criteria applied on the 770 studies after full-text reading assured to keep only pertinent studies.

Data extraction and classification could also be biased by a subjective interpretation. The classification scheme has been derived by iteratively refining an initial scheme as more and more studies were analyzed, so as to guarantee that the data extraction process was aligned with the research questions. In general, the best practices for mapping studies by Petersen \textit{et al.} have been followed in each phase of the study as documented above \cite{Petersen15}.

 \section{Related work} 
 \label{sec:related}

Our work is inspired by and extends the recent study by Siebert \cite{Siebert23} mentioned in Section \ref{sec:introduction}, which is the closest to ours. While Siebert provided a map of 25 papers useful to outline the main research trends up to 2022, the current status of this research area demands now a broader and more detailed picture to give researchers the tools to access this field. This work considers SQA dimensions along with software engineering tasks, phases and domains, and considers various CI frameworks besides the Perlian's one, while extending the temporal range up to 2024 with a search process returning 86 papers (Siebert analyzed 25 papers). The higher number of papers and finer granularity of the analysis yields an updated and broader view of the current state in the area.

A related attempt to review the applications of CI to software engineering is in the short paper by Chadbourne and Eisty \cite{Chadbourne2023}. 
The work selects 45 papers matching CI and software engineering keywords, and broadly classify them based on software engineering field and CR method. Unlike Siebert, this paper considers also Granger causality; on the other hand, \textit{i)} it also considers Bayesian network (not necessarily \textit{causal Bayesian networks}) as CI framework; \textit{ii)} it does not distinguishing between CD and CI; \textit{iii)} it adopts a rough-grain classification of software engineering fields that mixes quality attributes, tasks and phases in 6 categories, with the idea of giving a map, without focusing on the analysis of the retrieved contributions to the area. Being published in 2023, the contribution serves as further evidence of the increasing interest in CR by software engineering researchers and for %
a quick prospect of the papers over the defined macro-categories.

While we corroborate the preliminary findings of both works, our work expands them through a finer-grain analysis on: \textit{i)} the search process (hence on number of papers), \textit{ii)} the classification with respect to software engineering topics covered, \textit{iii)} and the classification with respect to the causal frameworks and methods.

A further related study, although not a secondary study, is the work by Furia \textit{et al.} \cite{Furia2023}. In their article, the authors promote the use of structural causal models to analyze Empirical Software Engineering (ESE) data -- hence as a valuable tool in the ESE research area. 
Their experiment show that the use of causal analysis on observational data helps answer research questions more precisely and more robustly than purely predictive techniques.

Besides these papers, there have been efforts in reviewing the application of probabilistic (non-causal) approaches in software engineering. Particularly, it is worth to mention BN as the closest framework to causal analysis and as one of the main tools adopted by the software engineering community. In this regard, there have been secondary studies with systematic literature reviews on the application of BN for: requirements engineering (surveyed by Del Águila \cite{Aguila}); for software quality prediction (surveyed by Tosun \cite{Tosun} and more recently by Misirli \textit{et al.} \cite{Misirli}); for decision-making in software engineering (Nageswarao and Geethanjali \cite{Nageswarao}); 
for software project management (De Sousa \textit{et al.} \cite{DeSousa}); for software development effort prediction (Radlinski \cite{radlinski2010survey}); for supporting decisions in value-based software engineering (Mendes \cite{Mendes}). In contrast to these studies, we do not consider BN in their broader application but only through a causal interpretation. To achieve this, our study only considers papers using either CD, CI, or both; hence, we exclude papers that do not explicitly address SQA tasks through the use of CR.

\section{Conclusions}
\label{sec:conclusions}
This article presented an analysis of the existing literature on CR for SQA. We selected and analyzed 86 papers looking at which activities CR is used for within the software life cycle (RQ1), how CR is used (RQ2), and how the CR-based solutions are validated (RQ3).

Our review reveals that CR has the potential to significantly enhance key SQA tasks, particularly in fault localization and software testing. CR methods show promise in improving precision for identifying software faults and predicting critical quality attributes such as reliability, performance, and security. However, despite this potential, the use of CR in SQA remains at an early stage, with limited empirical validation, especially in industrial settings.

Further research is required to advance the maturity and readiness of CR-based solutions for industrial applications. Specific areas needing attention include the development of scalable tools and frameworks, addressing challenges related to data quality, and validating CR methods in real-world environments. As interest in CR continues to grow, the field presents exciting opportunities for both researchers and practitioners. However, significant work is still needed to bridge the gap between experimental findings and practical applications in the software industry.

We release data and scripts to replicate this study in the online repository.\footref{note1} %

\section*{Acknowledgment}
This project has received funding from the European Union Horizon 2020 research and innovation programme under the Marie Sk{\l}odowska-Curie grant agreement No 871342 “uDEVOPS”.

\bibliographystyle{elsarticle-num}

\begin{thebibliography}{100}
\expandafter\ifx\csname url\endcsname\relax
  \def\url#1{\texttt{#1}}\fi
\expandafter\ifx\csname urlprefix\endcsname\relax\def\urlprefix{URL }\fi
\expandafter\ifx\csname href\endcsname\relax
  \def\href#1#2{#2} \def\path#1{#1}\fi

\bibitem{Avizienis04}
A.~Avizienis, J.-C. Laprie, B.~Randell, C.~Landwehr, Basic concepts and
  taxonomy of dependable and secure computing, IEEE Transactions on Dependable
  and Secure Computing 1~(1) (2004) 11--33.
\newblock \href {https://doi.org/10.1109/TDSC.2004.2}
  {\path{doi:10.1109/TDSC.2004.2}}.

\bibitem{Pearl18}
J.~Pearl, D.~Mackenzie, The Book of Why: The New Science of Cause and Effect,
  1st Edition, Basic Books, Inc., USA, 2018.

\bibitem{pearl2009}
J.~Pearl, Causality: Models, Reasoning and Inference, 2nd Edition, Cambridge
  University Press, USA, 2009.

\bibitem{Siebert23}
J.~Siebert, Applications of statistical causal inference in software
  engineering, Information and Software Technology 159~(C) (2023).
\newblock \href {https://doi.org/10.1016/j.infsof.2023.107198}
  {\path{doi:10.1016/j.infsof.2023.107198}}.

\bibitem{Wright1921}
S.~Wright, Correlation and causation, Journal of Agricultural Research 20
  (1921) 557--580.

\bibitem{Hernan00}
M.~{\'A}. Hern{\'a}n, B.~Brumback, J.~M. Robins, {Marginal Structural Models to
  Estimate the Causal Effect of Zidovudine on the Survival of HIV-Positive
  Men}, Epidemiology 11~(5) (2000).
\newblock \href {https://doi.org/10.1097/00001648-200009000-00012}
  {\path{doi:10.1097/00001648-200009000-00012}}.

\bibitem{Imbens04}
G.~W. Imbens, {Nonparametric Estimation of Average Treatment Effects Under
  Exogeneity: A Review}, The Review of Economics and Statistics 86~(1) (2004)
  4--29.
\newblock \href {https://doi.org/10.1162/003465304323023651}
  {\path{doi:10.1162/003465304323023651}}.

\bibitem{Morgan2014}
S.~L. Morgan, C.~Winship, Counterfactuals and Causal Inference: Methods and
  Principles for Social Research, 2nd Edition, Analytical Methods for Social
  Research, Cambridge University Press, 2014.

\bibitem{Mani01}
S.~Mani, G.~Cooper, Causal discovery from medical textual data, Proceedings
  AMIA Symposium (2001) 542.

\bibitem{Rajeev99}
R.~H. D., S.~W., Causal effects in nonexperimental studies: Reevaluating the
  evaluation of training programs, Journal of the American Statistical
  Association 94~(448) (1999) 1053--1062.

\bibitem{LaLonde86}
R.~J. LaLonde, Evaluating the econometric evaluations of training programs with
  experimental data, The American Economic Review 76~(4) (1986) 604--620.

\bibitem{wire}
A.~R. Nogueira, A.~Pugnana, S.~Ruggieri, D.~Pedreschi, J.~Gama, Methods and
  tools for causal discovery and causal inference, WIREs Data Mining and
  Knowledge Discovery 12~(2) (2022) e1449.
\newblock \href {https://doi.org/10.1002/widm.1449}
  {\path{doi:10.1002/widm.1449}}.

\bibitem{stanford-causal-models}
C.~Hitchcock, {Causal Models}, in: E.~N. Zalta, U.~Nodelman (Eds.), The
  {Stanford} Encyclopedia of Philosophy, {S}ummer 2024 Edition, Metaphysics
  Research Lab, Stanford University, 2024.

\bibitem{sharmaBook}
E.~Kiciman, A.~Sharma, {Causal Reasoning: Fundamentals and Machine Learning
  Applications}, https://causalinference.gitlab.io/ (2019).

\bibitem{Neyman90}
J.~Splawa-Neyman, D.~M. Dabrowska, T.~P. Speed, On the application of
  probability theory to agricultural experiments. essay on principles. section
  9., Statistical Science 5~(4) (1990) 465--472.

\bibitem{Neyman1923}
J.~Neyman, Sur les applications de la théorie des probabilités aux
  experiences agricoles: Essai des principes, Roczniki Nauk Rolniczych 10
  (1923) 1–51.

\bibitem{Rubin74}
D.~B. Rubin, Estimating causal effects of treatments in randomized and
  nonrandomized studies, Journal of Educational Psychology 66~(5) (1974)
  688--701.
\newblock \href {https://doi.org/10.1037/h0037350}
  {\path{doi:10.1037/h0037350}}.

\bibitem{Guo2020}
R.~Guo, L.~Cheng, J.~Li, P.~R. Hahn, H.~Liu, {A Survey of Learning Causality
  with Data: Problems and Methods}, ACM Computing Surveys 53~(4) (2020).
\newblock \href {https://doi.org/10.1145/3397269} {\path{doi:10.1145/3397269}}.

\bibitem{Holland86}
P.~W. Holland, Statistics and causal inference, Journal of the American
  Statistical Association 81~(396) (1986) 945--960.
\newblock \href {https://doi.org/10.1080/01621459.1986.10478354}
  {\path{doi:10.1080/01621459.1986.10478354}}.

\bibitem{Cox1958}
D.~R. Cox, Planning of Experiments, Wiley, New York, 1958.

\bibitem{Rubin80}
D.~B. Rubin, Randomization analysis of experimental data: The fisher
  randomization test comment, Journal of the American Statistical Association
  75~(371) (1980) 591--593.

\bibitem{Pearl1999}
J.~Pearl, Probabilities of causation: Three counterfactual interpretations and
  their identification, Synthese 121~(1) (1999) 93--149.
\newblock \href {https://doi.org/10.1023/A:1005233831499}
  {\path{doi:10.1023/A:1005233831499}}.

\bibitem{Kingma22}
D.~P. Kingma, M.~Welling, {Auto-Encoding Variational Bayes} (2022).
\newblock \href {http://arxiv.org/abs/1312.6114} {\path{arXiv:1312.6114}}.

\bibitem{Peters17}
J.~Peters, D.~Janzing, B.~Schlkopf, Elements of Causal Inference: Foundations
  and Learning Algorithms, The MIT Press, 2017.

\bibitem{Lechner11}
M.~Lechner, The estimation of causal effects by difference-in-difference
  methods, Foundations and Trends® in Econometrics 4~(3) (2011) 165--224.
\newblock \href {https://doi.org/10.1561/0800000014}
  {\path{doi:10.1561/0800000014}}.

\bibitem{Kosko1986}
B.~Kosko, Fuzzy cognitive maps, International Journal of Man-Machine Studies
  24~(1) (1986) 65--75.
\newblock \href {https://doi.org/https://doi.org/10.1016/S0020-7373(86)80040-2}
  {\path{doi:https://doi.org/10.1016/S0020-7373(86)80040-2}}.

\bibitem{Barbrook-Johnson22}
P.~Barbrook-Johnson, A.~S. Penn, Fuzzy cognitive mapping, in: Systems Mapping:
  How to build and use causal models of systems, Springer International
  Publishing, Cham, 2022, pp. 79--95.
\newblock \href {https://doi.org/10.1007/978-3-031-01919-7_6}
  {\path{doi:10.1007/978-3-031-01919-7_6}}.

\bibitem{Athey16}
S.~Athey, G.~Imbens, Recursive partitioning for heterogeneous causal effects,
  Proceedings of the National Academy of Sciences 113~(27) (2016) 7353--7360.
\newblock \href {https://doi.org/10.1073/pnas.1510489113}
  {\path{doi:10.1073/pnas.1510489113}}.

\bibitem{Wager17}
S.~Wager, S.~Athey, Estimation and inference of heterogeneous treatment effects
  using random forests, Journal of the American Statistical Association
  113~(523) (2018) 1228–1242.
\newblock \href {https://doi.org/10.1080/01621459.2017.1319839}
  {\path{doi:10.1080/01621459.2017.1319839}}.

\bibitem{Tran19}
C.~Tran, E.~Zheleva, Learning triggers for heterogeneous treatment effects,
  Proceedings of the AAAI Conference on Artificial Intelligence 33~(01) (2019)
  5183--5190.
\newblock \href {https://doi.org/10.1609/aaai.v33i01.33015183}
  {\path{doi:10.1609/aaai.v33i01.33015183}}.

\bibitem{Yao21}
L.~Yao, Z.~Chu, S.~Li, Y.~Li, J.~Gao, A.~Zhang, A survey on causal inference,
  ACM Transactions on Knowledge Discovery from Data 15~(5) (2021).
\newblock \href {https://doi.org/10.1145/3444944} {\path{doi:10.1145/3444944}}.

\bibitem{imbens20}
G.~W. Imbens, {Potential Outcome and Directed Acyclic Graph Approaches to
  Causality: Relevance for Empirical Practice in Economics}, Journal of
  Economic Literature 58~(4) (2020) 1129--79.
\newblock \href {https://doi.org/10.1257/jel.20191597}
  {\path{doi:10.1257/jel.20191597}}.

\bibitem{Rosenbaum84}
P.~R. Rosenbaum, D.~B. Rubin, Reducing bias in observational studies using
  subclassification on the propensity score, Journal of the American
  Statistical Association 79~(387) (1984) 516--524.

\bibitem{Caliendo08}
M.~Caliendo, S.~Kopeinig, Some practical guidance for the implementation of
  propensity score matching, Journal of Economic Surveys 22~(1) (2008) 31--72.
\newblock \href {https://doi.org/10.1111/j.1467-6419.2007.00527.x}
  {\path{doi:10.1111/j.1467-6419.2007.00527.x}}.

\bibitem{vanderwal11}
W.~M. van~der Wal, R.~B. Geskus, {ipw: An R Package for Inverse Probability
  Weighting}, Journal of Statistical Software 43~(13) (2011) 1–23.
\newblock \href {https://doi.org/10.18637/jss.v043.i13}
  {\path{doi:10.18637/jss.v043.i13}}.

\bibitem{Thistlethwaite1960}
D.~L. Thistlethwaite, D.~T. Campbell, Regression-discontinuity analysis: An
  alternative to the ex post facto experiment, Journal of Educational
  Psychology 51 (1960) 309--317.
\newblock \href {https://doi.org/10.1037/h0044319}
  {\path{doi:10.1037/h0044319}}.

\bibitem{Theil61}
H.~Theil, Economic Forecasts and Policy, North-Holland Publishing Company,
  Amsterdam, 1961.

\bibitem{louizos17}
C.~Louizos, U.~Shalit, J.~Mooij, D.~Sontag, R.~Zemel, M.~Welling, Causal effect
  inference with deep latent-variable models, in: Proceedings of the 31st
  International Conference on Neural Information Processing Systems (NeurIPS),
  Curran Associates Inc., 2017, p. 6449–6459.

\bibitem{dowhyGCM}
P.~Blöbaum, P.~Götz, K.~Budhathoki, A.~A. Mastakouri, D.~Janzing,
  \href{https://arxiv.org/abs/2206.06821}{Dowhy-gcm: An extension of dowhy for
  causal inference in graphical causal models} (2022).
\newblock \href {https://doi.org/10.48550/ARXIV.2206.06821}
  {\path{doi:10.48550/ARXIV.2206.06821}}.
\newline\urlprefix\url{https://arxiv.org/abs/2206.06821}

\bibitem{Glymour19}
C.~Glymour, K.~Zhang, P.~Spirtes, Review of causal discovery methods based on
  graphical models, Frontiers in Genetics 10 (06 2019).
\newblock \href {https://doi.org/10.3389/fgene.2019.00524}
  {\path{doi:10.3389/fgene.2019.00524}}.

\bibitem{Sjoberg03}
D.~I.~K. Sj{\o}berg, B.~Anda, E.~Arisholm, T.~Dyb{\aa}, M.~J{\o}rgensen,
  A.~Karahasanovi{\'{c}}, M.~Vok{\'a}{\v{c}}, {Challenges and Recommendations
  When Increasing the Realism of Controlled Software Engineering Experiments},
  in: R.~Conradi, A.~I. Wang (Eds.), Empirical Methods and Studies in Software
  Engineering -- Experiences from ESERNET, Vol. 2765 of Lecture Notes in
  Computer Science, Springer, 2003, pp. 24--38.
\newblock \href {https://doi.org/10.1007/978-3-540-45143-3_3}
  {\path{doi:10.1007/978-3-540-45143-3_3}}.

\bibitem{Eberhardt2017}
F.~Eberhardt, Introduction to the foundations of causal discovery,
  International Journal of Data Science and Analytics 3~(2) (2017) 81--91.
\newblock \href {https://doi.org/10.1007/s41060-016-0038-6}
  {\path{doi:10.1007/s41060-016-0038-6}}.

\bibitem{wang24}
L.~Wang, S.~Huang, S.~Wang, J.~Liao, T.~Li, L.~Liu, A survey of causal
  discovery based on functional causal model, Engineering Applications of
  Artificial Intelligence 133 (2024) 108258.
\newblock \href {https://doi.org/10.1016/j.engappai.2024.108258}
  {\path{doi:10.1016/j.engappai.2024.108258}}.

\bibitem{Spirtes21}
P.~Spirtes, C.~Glymour, R.~Scheines, {Causation, Prediction, and Search}, 2nd
  Edition, Adaptive Computation and Machine Learning, MIT Press, Cambridge, MA,
  USA, 2001.

\bibitem{Colombo12}
D.~Colombo, M.~H. Maathuis, M.~Kalisch, T.~S. Richardson, {Learning
  high-dimensional directed acyclic graphs with latent and selection
  variables}, The Annals of Statistics 40~(1) (2012) 294--321.
\newblock \href {https://doi.org/10.1214/11-AOS940}
  {\path{doi:10.1214/11-AOS940}}.

\bibitem{Malinsky18}
D.~Malinsky, D.~Danks, Causal discovery algorithms: A practical guide,
  Philosophy Compass 13~(1) (2018).
\newblock \href {https://doi.org/10.1111/phc3.12470}
  {\path{doi:10.1111/phc3.12470}}.

\bibitem{Chickering02}
D.~M. Chickering, Learning equivalence classes of bayesian-network structures,
  J. Mach. Learn. Res. 2 (2002) 445–498.
\newblock \href {https://doi.org/10.1162/153244302760200696}
  {\path{doi:10.1162/153244302760200696}}.

\bibitem{Ramsey17}
J.~Ramsey, M.~Glymour, R.~Sanchez-Romero, C.~Glymour, {A million variables and
  more: the Fast Greedy Equivalence Search algorithm for learning
  high-dimensional graphical causal models, with an application to functional
  magnetic resonance images}, International Journal of Data Science and
  Analytics 3 (2017).
\newblock \href {https://doi.org/10.1007/s41060-016-0032-z}
  {\path{doi:10.1007/s41060-016-0032-z}}.

\bibitem{Ogarrio16}
J.~M. Ogarrio, P.~Spirtes, J.~Ramsey, {A Hybrid Causal Search Algorithm for
  Latent Variable Models}, in: A.~Antonucci, G.~Corani, C.~P. Campos (Eds.),
  Proceedings of the Eighth International Conference on Probabilistic Graphical
  Models, Vol.~52 of Proceedings of Machine Learning Research, PMLR, Lugano,
  Switzerland, 2016, pp. 368--379.

\bibitem{Shimizu06}
S.~Shimizu, P.~O. Hoyer, A.~Hyv\"arinen, A.~Kerminen, A linear non-gaussian
  acyclic model for causal discovery, Journal of Machine Learning Research
  7~(72) (2006) 2003--2030.

\bibitem{Spirtes2016}
P.~Spirtes, K.~Zhang, Causal discovery and inference: concepts and recent
  methodological advances, Applied Informatics 3~(1) (2016) 3.
\newblock \href {https://doi.org/10.1186/s40535-016-0018-x}
  {\path{doi:10.1186/s40535-016-0018-x}}.

\bibitem{Zheng18}
X.~Zheng, B.~Aragam, P.~K. Ravikumar, E.~P. Xing, {DAGs with NO TEARS:
  Continuous Optimization for Structure Learning}, in: S.~Bengio, H.~Wallach,
  H.~Larochelle, K.~Grauman, N.~Cesa-Bianchi, R.~Garnett (Eds.), Advances in
  Neural Information Processing Systems, Vol.~31, Curran Associates, Inc.,
  2018, pp. 1--12.

\bibitem{Yu19}
Y.~Yu, J.~Chen, T.~Gao, M.~Yu, {DAG}-{GNN}: {DAG Structure Learning with Graph
  Neural Networks}, in: K.~Chaudhuri, R.~Salakhutdinov (Eds.), Proceedings of
  the 36th International Conference on Machine Learning (ICML), Vol.~97 of
  Proceedings of Machine Learning Research, PMLR, 2019, pp. 7154--7163.

\bibitem{Mooij20}
J.~M. Mooij, S.~Magliacane, T.~Claassen, Joint causal inference from multiple
  contexts, Journal of Machine Learning Research 21~(99) (2020) 1--108.

\bibitem{Tsamardinos06}
I.~Tsamardinos, L.~E. Brown, C.~F. Aliferis, The max-min hill-climbing bayesian
  network structure learning algorithm, Machine Learning 65~(1) (2006) 31--78.
\newblock \href {https://doi.org/10.1007/s10994-006-6889-7}
  {\path{doi:10.1007/s10994-006-6889-7}}.

\bibitem{Moraffah21}
R.~Moraffah, P.~Sheth, M.~Karami, A.~Bhattacharya, Q.~Wang, A.~Tahir,
  A.~Raglin, H.~Liu, Causal inference for time series analysis: problems,
  methods and evaluation, Knowledge and Information Systems 63~(12) (2021)
  3041--3085.
\newblock \href {https://doi.org/10.1007/s10115-021-01621-0}
  {\path{doi:10.1007/s10115-021-01621-0}}.

\bibitem{granger69}
C.~W.~J. Granger, Investigating causal relations by econometric models and
  cross-spectral methods, Econometrica 37~(3) (1969) 424--438.

\bibitem{Barnett09}
L.~Barnett, A.~B. Barrett, A.~K. Seth, {Granger Causality and Transfer Entropy
  Are Equivalent for Gaussian Variables}, Phys. Rev. Lett. 103 (2009) 238701.
\newblock \href {https://doi.org/10.1103/PhysRevLett.103.238701}
  {\path{doi:10.1103/PhysRevLett.103.238701}}.

\bibitem{ISO9000}
ISO-9000, ISO 9000 : International Standards for Quality Management, 2nd
  Edition, International Organization for Standardization, Gen{\`e}ve,
  Switzerland, 1992.

\bibitem{ISO24765}
ISO-24765, {24765-2017 - ISO/IEC/IEEE International Standard - Systems and
  software engineering--Vocabulary}, IEEE, 2017.
\newblock \href {https://doi.org/10.1109/IEEESTD.2017.8016712}
  {\path{doi:10.1109/IEEESTD.2017.8016712}}.

\bibitem{IEEE730}
IEEE-730, {IEEE Standard for Software Quality Assurance Processes}, IEEE, 2014.
\newblock \href {https://doi.org/10.1109/IEEESTD.2014.6835311}
  {\path{doi:10.1109/IEEESTD.2014.6835311}}.

\bibitem{IEEE1044}
IEEE-1044, {IEEE Standard Classification for Software Anomalies}, IEEE, 2010.
\newblock \href {https://doi.org/10.1109/IEEESTD.2010.5399061}
  {\path{doi:10.1109/IEEESTD.2010.5399061}}.

\bibitem{IEEE610}
IEEE-610, {IEEE Standard Glossary of Software Engineering Terminology}, IEEE,
  1990.
\newblock \href {https://doi.org/10.1109/IEEESTD.1990.101064}
  {\path{doi:10.1109/IEEESTD.1990.101064}}.

\bibitem{iso25010}
{{ISO}/{IEC} 25010}, {ISO}/{IEC} 25010:2023, Systems and software engineering
  — Systems and software Quality Requirements and Evaluation (SQuaRE) —
  Product quality model, International Organization for Standardization, 2023.

\bibitem{Fleiss03}
J.~Fleiss, B.~Levin, M.~Paik, The Measurement of Interrater Agreement, John
  Wiley \& Sons, 2003, Ch.~18, pp. 598--626.
\newblock \href {https://doi.org/10.1002/0471445428.ch18}
  {\path{doi:10.1002/0471445428.ch18}}.

\bibitem{Petersen15}
K.~Petersen, S.~Vakkalanka, L.~Kuzniarz, Guidelines for conducting systematic
  mapping studies in software engineering: An update, Information and Software
  Technology 64 (2015) 1--18.
\newblock \href {https://doi.org/10.1016/j.infsof.2015.03.007}
  {\path{doi:10.1016/j.infsof.2015.03.007}}.

\bibitem{Kitchenham13}
B.~Kitchenham, P.~Brereton, A systematic review of systematic review process
  research in software engineering, Information and Software Technology 55~(12)
  (2013) 2049–2075.
\newblock \href {https://doi.org/10.1016/j.infsof.2013.07.010}
  {\path{doi:10.1016/j.infsof.2013.07.010}}.

\bibitem{1}
S.~Oh, S.~Lee, S.~Yoo, Effectively sampling higher order mutants using causal
  effect, in: 2021 IEEE International Conference on Software Testing,
  Verification and Validation Workshops (ICSTW), 2021, pp. 19--24.
\newblock \href {https://doi.org/10.1109/ICSTW52544.2021.00017}
  {\path{doi:10.1109/ICSTW52544.2021.00017}}.

\bibitem{2}
M.~Salehie, L.~Pasquale, I.~Omoronyia, R.~Ali, B.~Nuseibeh, Requirements-driven
  adaptive security: Protecting variable assets at runtime, in: 2012 20th IEEE
  International Requirements Engineering Conference (RE), 2012, pp. 111--120.
\newblock \href {https://doi.org/10.1109/RE.2012.6345794}
  {\path{doi:10.1109/RE.2012.6345794}}.

\bibitem{3}
Y.~Hu, W.~Luo, Z.~Hu, A practical approach to explaining defect proneness of
  code commits by causal discovery, Engineering Applications of Artificial
  Intelligence 123 (2023) 106187.
\newblock \href {https://doi.org/10.1016/j.engappai.2023.106187}
  {\path{doi:10.1016/j.engappai.2023.106187}}.

\bibitem{4}
S.~Galhotra, A.~Fariha, R.~Louren\c{c}o, J.~Freire, A.~Meliou, D.~Srivastava,
  {DataPrism: Exposing Disconnect between Data and Systems}, in: SIGMOD '22:
  Proceedings of the 2022 International Conference on Management of Data, ACM,
  2022, p. 217–231.
\newblock \href {https://doi.org/10.1145/3514221.3517864}
  {\path{doi:10.1145/3514221.3517864}}.

\bibitem{5}
E.~Piatkowska, C.~Gavriluta, P.~Smith, F.~P. Andrén, Online reasoning about
  the root causes of software rollout failures in the smart grid, in: 2020 IEEE
  International Conference on Communications, Control, and Computing
  Technologies for Smart Grids (SmartGridComm), 2020, pp. 1--7.
\newblock \href {https://doi.org/10.1109/SmartGridComm47815.2020.9303005}
  {\path{doi:10.1109/SmartGridComm47815.2020.9303005}}.

\bibitem{6}
Y.~Yu, X.~Li, K.~Bu, Y.~Chen, J.~Yang, Falcon: Differential fault localization
  for sdn control plane, Computer Networks 162 (2019) 106851.
\newblock \href {https://doi.org/10.1016/j.comnet.2019.07.007}
  {\path{doi:10.1016/j.comnet.2019.07.007}}.

\bibitem{7}
W.~N. Sumner, X.~Zhang, Comparative causality: Explaining the differences
  between executions, in: 2013 35th International Conference on Software
  Engineering (ICSE), 2013, pp. 272--281.
\newblock \href {https://doi.org/10.1109/ICSE.2013.6606573}
  {\path{doi:10.1109/ICSE.2013.6606573}}.

\bibitem{8}
R.~Gore, P.~F. Reynolds, Reducing confounding bias in predicate-level
  statistical debugging metrics, in: 2012 34th International Conference on
  Software Engineering (ICSE), 2012, pp. 463--473.
\newblock \href {https://doi.org/10.1109/ICSE.2012.6227169}
  {\path{doi:10.1109/ICSE.2012.6227169}}.

\bibitem{9}
N.~Diamantopoulos, J.~Wong, D.~I. Mattos, I.~Gerostathopoulos, M.~Wardrop,
  T.~Mao, C.~McFarland, Engineering for a science-centric experimentation
  platform, in: 2020 ACM/IEEE 42nd International Conference on Software
  Engineering: Software Engineering in Practice (ICSE-SEIP), ACM, 2020, p.
  191–200.
\newblock \href {https://doi.org/10.1145/3377813.3381349}
  {\path{doi:10.1145/3377813.3381349}}.

\bibitem{10}
O.~Hamdi, A.~Ouni, E.~A. AlOmar, M.~Ó~Cinnéide, M.~W. Mkaouer, An empirical
  study on the impact of refactoring on quality metrics in {Android}
  applications, in: 2021 IEEE/ACM 8th International Conference on Mobile
  Software Engineering and Systems (MobileSoft), 2021, pp. 28--39.
\newblock \href {https://doi.org/10.1109/MobileSoft52590.2021.00010}
  {\path{doi:10.1109/MobileSoft52590.2021.00010}}.

\bibitem{11}
G.~Shu, B.~Sun, A.~Podgurski, F.~Cao, Mfl: Method-level fault localization with
  causal inference, in: 2013 IEEE Sixth International Conference on Software
  Testing, Verification and Validation, 2013, pp. 124--133.
\newblock \href {https://doi.org/10.1109/ICST.2013.31}
  {\path{doi:10.1109/ICST.2013.31}}.

\bibitem{12}
X.~Li, Y.~Yu, K.~Bu, Y.~Chen, J.~Yang, R.~Quan, Thinking inside the box:
  Differential fault localization for sdn control plane, in: 2019 IFIP/IEEE
  Symposium on Integrated Network and Service Management (IM), 2019, pp.
  353--359.

\bibitem{13}
Y.~Liu, D.~I. Mattos, J.~Bosch, H.~H. Olsson, J.~Lantz, Bayesian propensity
  score matching in automotive embedded software engineering, in: 2021 28th
  Asia-Pacific Software Engineering Conference (APSEC), 2021, pp. 233--242.
\newblock \href {https://doi.org/10.1109/APSEC53868.2021.00031}
  {\path{doi:10.1109/APSEC53868.2021.00031}}.

\bibitem{14}
Y.~He, C.~Tran, J.~Jiang, K.~Burghardt, E.~Ferrara, E.~Zheleva, K.~Lerman,
  Heterogeneous effects of software patches in a multiplayer online battle
  arena game, in: Proceedings of the 16th International Conference on the
  Foundations of Digital Games, FDG '21, ACM, 2021, pp. 1--9.
\newblock \href {https://doi.org/10.1145/3472538.3472550}
  {\path{doi:10.1145/3472538.3472550}}.

\bibitem{15}
Z.~Bai, G.~Shu, A.~Podgurski, {NUMFL: Localizing Faults in Numerical Software
  Using a Value-Based Causal Model}, in: 2015 IEEE 8th International Conference
  on Software Testing, Verification and Validation (ICST), 2015, pp. 1--10.
\newblock \href {https://doi.org/10.1109/ICST.2015.7102597}
  {\path{doi:10.1109/ICST.2015.7102597}}.

\bibitem{16}
Z.~Bai, G.~Shu, A.~Podgurski, Causal inference based fault localization for
  numerical software with {NUMFL}, Software Testing, Verification and
  Reliability 27~(6) (2017) e1613, e1613 stvr.1613.
\newblock \href {https://doi.org/10.1002/stvr.1613}
  {\path{doi:10.1002/stvr.1613}}.

\bibitem{17}
F.~Feyzi, S.~Parsa, Inforence: effective fault localization based on
  information-theoretic analysis and statistical causal inference, Frontiers of
  Computer Science 13~(4) (2019) 735--759.
\newblock \href {https://doi.org/10.1007/s11704-017-6512-z}
  {\path{doi:10.1007/s11704-017-6512-z}}.

\bibitem{18}
M.~A. Hossen, S.~Kharade, B.~Schmerl, J.~Cámara, J.~M. O'Kane, E.~C.
  Czaplinski, K.~A. Dzurilla, D.~Garlan, P.~Jamshidi, {CaRE: Finding Root
  Causes of Configuration Issues in Highly-Configurable Robots}, IEEE Robotics
  and Automation Letters 8~(7) (2023) 4115--4122.
\newblock \href {https://doi.org/10.1109/LRA.2023.3280810}
  {\path{doi:10.1109/LRA.2023.3280810}}.

\bibitem{19}
A.~Paleyes, S.~Guo, B.~Scholkopf, N.~D. Lawrence, Dataflow graphs as complete
  causal graphs, in: 2023 IEEE/ACM 2nd International Conference on AI
  Engineering – Software Engineering for AI (CAIN), 2023, pp. 7--12.
\newblock \href {https://doi.org/10.1109/CAIN58948.2023.00010}
  {\path{doi:10.1109/CAIN58948.2023.00010}}.

\bibitem{20}
M.~S. Iqbal, R.~Krishna, M.~A. Javidian, B.~Ray, P.~Jamshidi, Unicorn:
  Reasoning about configurable system performance through the lens of
  causality, in: Proceedings of the Seventeenth European Conference on Computer
  Systems, EuroSys '22, ACM, 2022, p. 199–217.
\newblock \href {https://doi.org/10.1145/3492321.3519575}
  {\path{doi:10.1145/3492321.3519575}}.

\bibitem{21}
A.~Podgurski, Y.~Küçük, Counterfault: Value-based fault localization by
  modeling and predicting counterfactual outcomes, in: IEEE International
  Conference on Software Maintenance and Evolution (ICSME), 2020, pp. 382--393.
\newblock \href {https://doi.org/10.1109/ICSME46990.2020.00044}
  {\path{doi:10.1109/ICSME46990.2020.00044}}.

\bibitem{22}
G.~Li, H.~Dai, What will affect software reuse: A causal model analysis,
  International Journal of Software Engineering and Knowledge Engineering
  14~(03) (2004) 351--364.
\newblock \href {https://doi.org/10.1142/S021819400400166X}
  {\path{doi:10.1142/S021819400400166X}}.

\bibitem{23}
E.~Garmash, A.~Cheshkov, Exploring the effect of null usage in source code, in:
  2021 International Conference on Code Quality (ICCQ), 2021, pp. 1--14.
\newblock \href {https://doi.org/10.1109/ICCQ51190.2021.9392959}
  {\path{doi:10.1109/ICCQ51190.2021.9392959}}.

\bibitem{24}
G.~K. Baah, A.~Podgurski, M.~J. Harrold, Mitigating the confounding effects of
  program dependences for effective fault localization, in: Proceedings of the
  19th ACM SIGSOFT Symposium and the 13th European Conference on Foundations of
  Software Engineering (ESEC/FSE), ESEC/FSE, ACM, 2011, p. 146–156.
\newblock \href {https://doi.org/10.1145/2025113.2025136}
  {\path{doi:10.1145/2025113.2025136}}.

\bibitem{25}
S.~A. Mondal, P.~Rv, S.~Rao, A.~Menon, {LADDERS: Log Based Anomaly Detection
  and Diagnosis for Enterprise Systems}, Annals of Data Science (2023)
  1--19.~\href {https://doi.org/10.1007/s40745-023-00471-7}
  {\path{doi:10.1007/s40745-023-00471-7}}.

\bibitem{26}
Y.~Küçük, T.~A.~D. Henderson, A.~Podgurski, Improving fault localization by
  integrating value and predicate based causal inference techniques, in: 2021
  IEEE/ACM 43rd International Conference on Software Engineering (ICSE), 2021,
  pp. 649--660.
\newblock \href {https://doi.org/10.1109/ICSE43902.2021.00066}
  {\path{doi:10.1109/ICSE43902.2021.00066}}.

\bibitem{27}
S.-F. Sun, A.~Podgurski, Properties of effective metrics for coverage-based
  statistical fault localization, in: 2016 IEEE International Conference on
  Software Testing, Verification and Validation (ICST), 2016, pp. 124--134.
\newblock \href {https://doi.org/10.1109/ICST.2016.31}
  {\path{doi:10.1109/ICST.2016.31}}.

\bibitem{28}
G.~K. Baah, A.~Podgurski, M.~J. Harrold, {Causal Inference for Statistical
  Fault Localization}, in: Proceedings of the 19th International Symposium on
  Software Testing and Analysis (ISSTA), ACM, 2010, p. 73–84.
\newblock \href {https://doi.org/10.1145/1831708.1831717}
  {\path{doi:10.1145/1831708.1831717}}.

\bibitem{29}
J.~Li, X.~Cui, Y.~Wang, F.~Xie, An empirical study of software testing quality
  based on natural experiments, in: IEEE 22nd International Conference on
  Software Quality, Reliability, and Security Companion, 2022, pp. 499--508.
\newblock \href {https://doi.org/10.1109/QRS-C57518.2022.00080}
  {\path{doi:10.1109/QRS-C57518.2022.00080}}.

\bibitem{30}
Z.~Bai, S.-F. Sun, A.~Podgurski, The importance of being positive in causal
  statistical fault localization: Important properties of baah et al.'s csfl
  regression model, in: 2015 IEEE/ACM 1st International Workshop on Complex
  Faults and Failures in Large Software Systems (COUFLESS), 2015, pp. 7--13.
\newblock \href {https://doi.org/10.1109/COUFLESS.2015.9}
  {\path{doi:10.1109/COUFLESS.2015.9}}.

\bibitem{31}
A.~Ikram, S.~Chakraborty, S.~Mitra, S.~Saini, S.~Bagchi, M.~Kocaoglu, Root
  cause analysis of failures in microservices through causal discovery, in:
  S.~Koyejo, S.~Mohamed, A.~Agarwal, D.~Belgrave, K.~Cho, A.~Oh (Eds.),
  Advances in Neural Information Processing Systems, Vol.~35, Curran
  Associates, Inc, 2022, pp. 31158--31170.

\bibitem{32}
L.~Li, J.~Liu, Z.~Zhou, H.~Luo, W.~Liu, J.~Li, Causal inference based service
  dependency graph for statistical service fault localization, in: 2014 10th
  International Conference on Semantics, Knowledge and Grids, 2014, pp. 41--48.
\newblock \href {https://doi.org/10.1109/SKG.2014.21}
  {\path{doi:10.1109/SKG.2014.21}}.

\bibitem{33}
A.~G. Clark, M.~Foster, N.~Walkinshaw, R.~M. Hierons, Metamorphic testing with
  causal graphs, in: 2023 IEEE Conference on Software Testing, Verification and
  Validation (ICST), 2023, pp. 153--164.
\newblock \href {https://doi.org/10.1109/ICST57152.2023.00023}
  {\path{doi:10.1109/ICST57152.2023.00023}}.

\bibitem{34}
L.~Giamattei, R.~Pietrantuono, S.~Russo, Reasoning-based software testing, in:
  2023 IEEE/ACM 45th International Conference on Software Engineering: New
  Ideas and Emerging Results (ICSE-NIER), 2023, pp. 66--71.
\newblock \href {https://doi.org/10.1109/ICSE-NIER58687.2023.00018}
  {\path{doi:10.1109/ICSE-NIER58687.2023.00018}}.

\bibitem{35}
L.~Wu, J.~Tordsson, E.~Elmroth, O.~Kao, Causal inference techniques for
  microservice performance diagnosis: Evaluation and guiding recommendations,
  in: 2021 IEEE International Conference on Autonomic Computing and
  Self-Organizing Systems (ACSOS), 2021, pp. 21--30.
\newblock \href {https://doi.org/10.1109/ACSOS52086.2021.00029}
  {\path{doi:10.1109/ACSOS52086.2021.00029}}.

\bibitem{36}
R.~Jarry, S.~Kobayashi, K.~Fukuda, A quantitative causal analysis for network
  log data, in: 2021 IEEE 45th Annual Computers, Software, and Applications
  Conference (COMPSAC), 2021, pp. 1437--1442.
\newblock \href {https://doi.org/10.1109/COMPSAC51774.2021.00213}
  {\path{doi:10.1109/COMPSAC51774.2021.00213}}.

\bibitem{37}
C.~Dubslaff, K.~Weis, C.~Baier, S.~Apel, Causality in configurable software
  systems, in: 2022 IEEE/ACM 44th International Conference on Software
  Engineering (ICSE), ACM, 2022, p. 325–337.
\newblock \href {https://doi.org/10.1145/3510003.3510200}
  {\path{doi:10.1145/3510003.3510200}}.

\bibitem{38}
L.~Wu, J.~Tordsson, J.~Bogatinovski, E.~Elmroth, O.~Kao, Microdiag:
  Fine-grained performance diagnosis for microservice systems, in: IEEE/ACM
  International Workshop on Cloud Intelligence, 2021, pp. 31--36.
\newblock \href {https://doi.org/10.1109/CloudIntelligence52565.2021.00015}
  {\path{doi:10.1109/CloudIntelligence52565.2021.00015}}.

\bibitem{39}
S.~Ji, W.~Wu, Y.~Pu, {Multi-indicators prediction in microservice using Granger
  causality test and Attention LSTM}, in: 2020 IEEE World Congress on Services
  (SERVICES), 2020, pp. 77--82.
\newblock \href {https://doi.org/10.1109/SERVICES48979.2020.00030}
  {\path{doi:10.1109/SERVICES48979.2020.00030}}.

\bibitem{40}
J.~Wang, S.~Si, Z.~Zhu, X.~Qu, Z.~Hong, J.~Xiao, Leveraging causal inference
  for explainable automatic program repair, in: 2022 International Joint
  Conference on Neural Networks (IJCNN), 2022, pp. 1--6.
\newblock \href {https://doi.org/10.1109/IJCNN55064.2022.9892168}
  {\path{doi:10.1109/IJCNN55064.2022.9892168}}.

\bibitem{41}
M.~Zhang, Y.~Qian, Y.~Jiang, Y.~Wang, Y.~Liu, {Helpfulness Prediction for VR
  Application Reviews: Exploring Topic Signals for Causal Inference}, in: IEEE
  International Symposium on Mixed and Augmented Reality Adjunct
  (ISMAR-Adjunct), IEEE, 2022, pp. 17--21.
\newblock \href {https://doi.org/10.1109/ISMAR-Adjunct57072.2022.00014}
  {\path{doi:10.1109/ISMAR-Adjunct57072.2022.00014}}.

\bibitem{42}
Z.~Ji, P.~Ma, Y.~Yuan, S.~Wang, Cc: Causality-aware coverage criterion for deep
  neural networks, in: 2023 IEEE/ACM 45th International Conference on Software
  Engineering (ICSE), 2023, pp. 1788--1800.
\newblock \href {https://doi.org/10.1109/ICSE48619.2023.00153}
  {\path{doi:10.1109/ICSE48619.2023.00153}}.

\bibitem{43}
A.~Paleyes, N.~D. Lawrence, Causal fault localisation in dataflow systems, in:
  Proceedings of the 3rd Workshop on Machine Learning and Systems, EuroMLSys
  '23, ACM, 2023, p. 140–147.
\newblock \href {https://doi.org/10.1145/3578356.3592593}
  {\path{doi:10.1145/3578356.3592593}}.

\bibitem{44}
A.~G. Clark, M.~Foster, B.~Prifling, N.~Walkinshaw, R.~M. Hierons, V.~Schmidt,
  R.~D. Turner, Testing causality in scientific modelling software, ACM
  Transactions on Software Engineering and Methodology 33~(1) (2023).
\newblock \href {https://doi.org/10.1145/3607184} {\path{doi:10.1145/3607184}}.

\bibitem{45}
Z.~Ji, P.~Ma, S.~Wang, Perfce: Performance debugging on databases with chaos
  engineering-enhanced causality analysis, in: 2023 38th IEEE/ACM International
  Conference on Automated Software Engineering (ASE), IEEE, 2023, pp.
  1454--1466.
\newblock \href {https://doi.org/10.1109/ASE56229.2023.00106}
  {\path{doi:10.1109/ASE56229.2023.00106}}.

\bibitem{46}
L.~Yuanxin, L.~Rui, M.~Yuxi, X.~Yunzhi, M.~Lingzhong, Fca: A causal inference
  based method for analyzing the failure causes of object detection algorithms,
  in: 2023 IEEE 23rd International Conference on Software Quality, Reliability,
  and Security Companion (QRS-C), 2023, pp. 247--256.
\newblock \href {https://doi.org/10.1109/QRS-C60940.2023.00069}
  {\path{doi:10.1109/QRS-C60940.2023.00069}}.

\bibitem{47}
B.~Johnson, Y.~Brun, A.~Meliou, Causal testing: understanding defects' root
  causes, in: 2020 ACM/IEEE 42nd International Conference on Software
  Engineering (ICSE), ACM, 2020, p. 87–99.
\newblock \href {https://doi.org/10.1145/3377811.3380377}
  {\path{doi:10.1145/3377811.3380377}}.

\bibitem{48}
X.~Wang, S.~Jiang, X.~Ju, H.~Cao, Y.~Liu, Mitigating the dependence confounding
  effect for effective predicate-based statistical fault localization, in: IEEE
  39th Annual Computer Software and Applications Conference, IEEE, 2015, pp.
  105--114.
\newblock \href {https://doi.org/10.1109/COMPSAC.2015.37}
  {\path{doi:10.1109/COMPSAC.2015.37}}.

\bibitem{49}
R.~Xin, P.~Chen, Z.~Zhao, {CausalRCA}: Causal inference based precise
  fine-grained root cause localization for microservice applications, Journal
  of Systems and Software 203~(C) (2023).
\newblock \href {https://doi.org/10.1016/j.jss.2023.111724}
  {\path{doi:10.1016/j.jss.2023.111724}}.

\bibitem{50}
Q.~Zhang, T.~Jia, Z.~Wu, Q.~Wu, L.~Jia, D.~Li, Y.~Tao, Y.~Xiao, Fault
  localization for microservice applications with system logs and monitoring
  metrics, in: 2022 7th International Conference on Cloud Computing and Big
  Data Analytics (ICCCBDA), 2022, pp. 149--154.
\newblock \href {https://doi.org/10.1109/ICCCBDA55098.2022.9778893}
  {\path{doi:10.1109/ICCCBDA55098.2022.9778893}}.

\bibitem{51}
B.~Sun, J.~Sun, L.~H. Pham, J.~Shi, Causality-based neural network repair, in:
  2022 IEEE/ACM 44th International Conference on Software Engineering (ICSE),
  ACM, 2022, p. 338–349.
\newblock \href {https://doi.org/10.1145/3510003.3510080}
  {\path{doi:10.1145/3510003.3510080}}.

\bibitem{52}
R.~Maier, L.~Grabinger, D.~Urlhart, J.~Mottok, Causal models to support
  scenario-based testing of adas, IEEE Transactions on Intelligent
  Transportation Systems 25~(2) (2024) 1815--1831.
\newblock \href {https://doi.org/10.1109/TITS.2023.3317475}
  {\path{doi:10.1109/TITS.2023.3317475}}.

\bibitem{53}
Q.~Li, Y.~Su, W.~Wang, Z.~Wang, J.~He, G.~Li, C.~Zeng, B.~Cheng, Latent hazard
  notification for highly automated driving: Expected safety benefits and
  driver behavioral adaptation, IEEE Transactions on Intelligent Transportation
  Systems 24~(10) (2023) 11278--11292.
\newblock \href {https://doi.org/10.1109/TITS.2023.3280955}
  {\path{doi:10.1109/TITS.2023.3280955}}.

\bibitem{54}
Z.~Ji, P.~Ma, S.~Wang, Y.~Li, Causality-aided trade-off analysis for machine
  learning fairness, in: 2023 38th IEEE/ACM International Conference on
  Automated Software Engineering (ASE), IEEE, 2023, pp. 371--383.
\newblock \href {https://doi.org/10.1109/ASE56229.2023.00105}
  {\path{doi:10.1109/ASE56229.2023.00105}}.

\bibitem{55}
P.~Chen, Y.~Qi, D.~Hou, Causeinfer: Automated end-to-end performance diagnosis
  with hierarchical causality graph in cloud environment, IEEE Transactions on
  Services Computing 12~(2) (2019) 214--230.
\newblock \href {https://doi.org/10.1109/TSC.2016.2607739}
  {\path{doi:10.1109/TSC.2016.2607739}}.

\bibitem{56}
V.~Musco, M.~Monperrus, P.~Preux, Mutation-based graph inference for fault
  localization, in: 2016 IEEE 16th International Working Conference on Source
  Code Analysis and Manipulation (SCAM), 2016, pp. 97--106.
\newblock \href {https://doi.org/10.1109/SCAM.2016.24}
  {\path{doi:10.1109/SCAM.2016.24}}.

\bibitem{57}
P.~Wang, J.~Xu, M.~Ma, W.~Lin, D.~Pan, Y.~Wang, P.~Chen, Cloudranger: Root
  cause identification for cloud native systems, in: 2018 18th IEEE/ACM
  International Symposium on Cluster, Cloud and Grid Computing (CCGRID), 2018,
  pp. 492--502.
\newblock \href {https://doi.org/10.1109/CCGRID.2018.00076}
  {\path{doi:10.1109/CCGRID.2018.00076}}.

\bibitem{58}
S.~Zhang, Y.~Liu, D.~Pei, Y.~Chen, X.~Qu, S.~Tao, Z.~Zang, X.~Jing, M.~Feng,
  Funnel: Assessing software changes in web-based services, IEEE Transactions
  on Services Computing 11~(1) (2018) 34--48.
\newblock \href {https://doi.org/10.1109/TSC.2016.2539945}
  {\path{doi:10.1109/TSC.2016.2539945}}.

\bibitem{59}
P.~Liu, Y.~Li, B.~Swain, J.~Huang, {PUS: A Fast and Highly Efficient Solver for
  Inclusion-based Pointer Analysis}, in: 2022 IEEE/ACM 44th International
  Conference on Software Engineering (ICSE), 2022, pp. 1781--1792.
\newblock \href {https://doi.org/10.1145/3510003.3510075}
  {\path{doi:10.1145/3510003.3510075}}.

\bibitem{60}
S.~Gao, C.~Gao, C.~Wang, J.~Sun, D.~Lo, Y.~Yu, {Two Sides of the Same Coin:
  Exploiting the Impact of Identifiers in Neural Code Comprehension}, in: 2023
  IEEE/ACM 45th International Conference on Software Engineering (ICSE), IEEE,
  2023, p. 1933–1945.
\newblock \href {https://doi.org/10.1109/ICSE48619.2023.00164}
  {\path{doi:10.1109/ICSE48619.2023.00164}}.

\bibitem{61}
E.~Zibaei, R.~Borth, Building causal models for finding actual causes of
  unmanned aerial vehicle failures, Frontiers in Robotics and AI 11 (02 2024).
\newblock \href {https://doi.org/10.3389/frobt.2024.1123762}
  {\path{doi:10.3389/frobt.2024.1123762}}.

\bibitem{62}
Z.~Zhong, Z.~Hu, S.~Guo, X.~Zhang, Z.~Zhong, B.~Ray, Detecting multi-sensor
  fusion errors in advanced driver-assistance systems, in: Proceedings of the
  31st ACM SIGSOFT International Symposium on Software Testing and Analysis
  (ISSTA), ACM, 2022, p. 493–505.
\newblock \href {https://doi.org/10.1145/3533767.3534223}
  {\path{doi:10.1145/3533767.3534223}}.

\bibitem{63}
A.~Fariha, S.~Nath, A.~Meliou, Causality-guided adaptive interventional
  debugging, in: Proceedings of the 2020 ACM SIGMOD International Conference on
  Management of Data, ACM, 2020, p. 431–446.
\newblock \href {https://doi.org/10.1145/3318464.3389694}
  {\path{doi:10.1145/3318464.3389694}}.

\bibitem{64}
L.~Giamattei, A.~Guerriero, R.~Pietrantuono, S.~Russo, Causality-driven testing
  of autonomous driving systems, ACM Transactions on Software Engineering and
  Methodology 33~(3) (2024) 1--35.
\newblock \href {https://doi.org/10.1145/3635709} {\path{doi:10.1145/3635709}}.

\bibitem{65}
M.~M. Rahman, I.~Ceka, C.~Mao, S.~Chakraborty, B.~Ray, W.~Le, Towards causal
  deep learning for vulnerability detection, in: 2024 IEEE/ACM 46th
  International Conference on Software Engineering (ICSE), ACM, 2024, pp.
  1--11.
\newblock \href {https://doi.org/10.1145/3597503.3639170}
  {\path{doi:10.1145/3597503.3639170}}.

\bibitem{66}
C.~Couto, C.~Silva, M.~T. Valente, R.~Bigonha, N.~Anquetil, Uncovering causal
  relationships between software metrics and bugs, in: 2012 16th European
  Conference on Software Maintenance and Reengineering, 2012, pp. 223--232.
\newblock \href {https://doi.org/10.1109/CSMR.2012.31}
  {\path{doi:10.1109/CSMR.2012.31}}.

\bibitem{67}
P.~Zheng, Y.~Zhou, M.~R. Lyu, Y.~Qi, Granger causality-aware prediction and
  diagnosis of software degradation, in: 2014 IEEE International Conference on
  Services Computing, 2014, pp. 528--535.
\newblock \href {https://doi.org/10.1109/SCC.2014.76}
  {\path{doi:10.1109/SCC.2014.76}}.

\bibitem{68}
D.~Cotroneo, F.~Frattini, R.~Natella, R.~Pietrantuono, Performance degradation
  analysis of a supercomputer, in: 2013 IEEE International Symposium on
  Software Reliability Engineering Workshops (ISSREW), 2013, pp. 263--268.
\newblock \href {https://doi.org/10.1109/ISSREW.2013.6688904}
  {\path{doi:10.1109/ISSREW.2013.6688904}}.

\bibitem{69}
G.~Denaro, R.~Heydarov, A.~Mohebbi, M.~Pezzè, Prevent: An unsupervised
  approach to predict software failures in production, IEEE Transactions on
  Software Engineering 49~(12) (2023) 5139--5153.
\newblock \href {https://doi.org/10.1109/TSE.2023.3327583}
  {\path{doi:10.1109/TSE.2023.3327583}}.

\bibitem{70}
Y.~Pan, M.~Ma, X.~Jiang, P.~Wang, Dycause: Crowdsourcing to diagnose
  microservice kernel failure, IEEE Transactions on Dependable and Secure
  Computing 20~(6) (2023) 4763--4777.
\newblock \href {https://doi.org/10.1109/TDSC.2022.3233915}
  {\path{doi:10.1109/TDSC.2022.3233915}}.

\bibitem{71}
S.~Xing, J.~Niu, T.~Ren, {GCFormer: Granger Causality based Attention Mechanism
  for Multivariate Time Series Anomaly Detection}, in: 2023 IEEE International
  Conference on Data Mining (ICDM), IEEE, 2023, pp. 1433--1438.

\bibitem{72}
V.~Nagaraju, S.~Pritchard, L.~Fiondella, Adaptive incremental learning for
  software reliability growth models, in: S.~Yamamoto, H.~Mori (Eds.), Human
  Interface and the Management of Information: Applications in Complex
  Technological Environments, Springer, 2022, pp. 352--366.

\bibitem{73}
Y.~Pan, M.~Ma, X.~Jiang, P.~Wang, Faster, deeper, easier: crowdsourcing
  diagnosis of microservice kernel failure from user space, in: Proceedings of
  the 30th ACM SIGSOFT International Symposium on Software Testing and Analysis
  (ISSTA), ACM, 2021, p. 646–657.
\newblock \href {https://doi.org/10.1145/3460319.3464805}
  {\path{doi:10.1145/3460319.3464805}}.

\bibitem{74}
T.~Sharma, P.~Singh, D.~Spinellis, An empirical investigation on the
  relationship between design and architecture smells, Empirical Software
  Engineering 25~(5) (2020) 4020--4068.
\newblock \href {https://doi.org/10.1007/s10664-020-09847-2}
  {\path{doi:10.1007/s10664-020-09847-2}}.

\bibitem{75}
C.~Couto, P.~Pires, M.~T. Valente, R.~S. Bigonha, N.~Anquetil, Predicting
  software defects with causality tests, Journal of Systems and Software 93
  (2014) 24--41.
\newblock \href {https://doi.org/10.1016/j.jss.2014.01.033}
  {\path{doi:10.1016/j.jss.2014.01.033}}.

\bibitem{76}
W.~Lee, X.~Qin, Statistical causality analysis of infosec alert data, in:
  V.~Kumar, J.~Srivastava, A.~Lazarevic (Eds.), Managing Cyber Threats: Issues,
  Approaches, and Challenges, Springer, 2005, pp. 101--127.
\newblock \href {https://doi.org/10.1007/0-387-24230-9_4}
  {\path{doi:10.1007/0-387-24230-9_4}}.

\bibitem{77}
E.~Andrade, R.~Pietrantuono, F.~Machida, D.~Cotroneo, A comparative analysis of
  software aging in image classifiers on cloud and edge, IEEE Transactions on
  Dependable and Secure Computing 20~(1) (2023) 563--573.
\newblock \href {https://doi.org/10.1109/TDSC.2021.3139201}
  {\path{doi:10.1109/TDSC.2021.3139201}}.

\bibitem{78}
R.~Pietrantuono, M.~Ficco, F.~Palmieri, {Testing the Resilience of MEC-Based
  IoT Applications Against Resource Exhaustion Attacks}, IEEE Transactions on
  Dependable and Secure Computing 21~(2) (2024) 804--818.
\newblock \href {https://doi.org/10.1109/TDSC.2023.3263137}
  {\path{doi:10.1109/TDSC.2023.3263137}}.

\bibitem{79}
E.~Andrade, F.~Machida, R.~Pietrantuono, D.~Cotroneo, Memory degradation
  analysis in private and public cloud environments, in: 2021 IEEE
  International Symposium on Software Reliability Engineering Workshops
  (ISSREW), 2021, pp. 33--39.
\newblock \href {https://doi.org/10.1109/ISSREW53611.2021.00041}
  {\path{doi:10.1109/ISSREW53611.2021.00041}}.

\bibitem{80}
P.~Kayongo, J.~Hoffswell, S.~Saini, S.~Garg, E.~Koh, H.~Wang, T.~Jacobs, Visre:
  A unified visual analysis dashboard for proactive cloud outage management,
  in: 2022 Working Conference on Software Visualization (VISSOFT), IEEE, 2022,
  pp. 5--16.
\newblock \href {https://doi.org/10.1109/VISSOFT55257.2022.00010}
  {\path{doi:10.1109/VISSOFT55257.2022.00010}}.

\bibitem{81}
H.~Wang, Y.~Lin, Z.~Yang, J.~Sun, Y.~Liu, J.~Dong, Q.~Zheng, T.~Liu, Explaining
  regressions via alignment slicing and mending, IEEE Transactions on Software
  Engineering 47~(11) (2021) 2421--2437.
\newblock \href {https://doi.org/10.1109/TSE.2019.2949568}
  {\path{doi:10.1109/TSE.2019.2949568}}.

\bibitem{82}
J.~Fischbach, T.~Springer, J.~Frattini, H.~Femmer, A.~Vogelsang, D.~Mendez,
  Fine-grained causality extraction from natural language requirements using
  recursive neural tensor networks, in: 2021 IEEE 29th International
  Requirements Engineering Conference Workshops (REW), 2021, pp. 60--69.
\newblock \href {https://doi.org/10.1109/REW53955.2021.00016}
  {\path{doi:10.1109/REW53955.2021.00016}}.

\bibitem{83}
P.~Ma, Z.~Ji, P.~Yao, S.~Wang, K.~Ren, Enabling runtime verification of causal
  discovery algorithms with automated conditional independence reasoning, in:
  2024 IEEE/ACM 46th International Conference on Software Engineering (ICSE),
  ACM, 2024, pp. 1--13.
\newblock \href {https://doi.org/10.1145/3597503.3623348}
  {\path{doi:10.1145/3597503.3623348}}.

\bibitem{84}
W.~Yaning, H.~Song, J.~Haijin, An information flow-based feature selection
  method for cross-project defect prediction, International Journal of
  Performability Engineering 14~(6) (2018) 1263.
\newblock \href {https://doi.org/10.23940/ijpe.18.06.p17.12631274}
  {\path{doi:10.23940/ijpe.18.06.p17.12631274}}.

\bibitem{85}
S.~Zhang, Y.~Liu, D.~Pei, Y.~Chen, X.~Qu, S.~Tao, Z.~Zang, Rapid and robust
  impact assessment of software changes in large internet-based services, in:
  Proceedings of the 11th ACM Conference on Emerging Networking Experiments and
  Technologies (CoNEXT), ACM, 2015, pp. 1--13.
\newblock \href {https://doi.org/10.1145/2716281.2836087}
  {\path{doi:10.1145/2716281.2836087}}.

\bibitem{86}
M.~Scholz, R.~Torkar, An empirical study of linespots: A novel past-fault
  algorithm, Software Testing, Verification and Reliability 31~(8) (2021)
  e1787.
\newblock \href {https://doi.org/10.1002/stvr.1787}
  {\path{doi:10.1002/stvr.1787}}.

\bibitem{pezze}
M.~Pezz{\`e}, M.~Young, Software Testing and Analysis: Process, Principles, and
  Techniques, Wiley India Pvt. Limited, 2008.

\bibitem{Spirtes95}
P.~Spirtes, C.~Meek, T.~Richardson, Causal inference in the presence of latent
  variables and selection bias, in: 11th Conference on Uncertainty in
  Artificial Intelligence, UAI'95, Morgan Kaufmann Publishers Inc., San
  Francisco, CA, USA, 1995, p. 499–506.

\bibitem{Spirtes93}
P.~Spirtes, C.~Glymour, R.~Scheines, Causation, Prediction, and Search,
  Vol.~81, MIT press, 1993.
\newblock \href {https://doi.org/10.1007/978-1-4612-2748-9}
  {\path{doi:10.1007/978-1-4612-2748-9}}.

\bibitem{Li22_blip}
J.~Li, D.~Li, C.~Xiong, S.~C.~H. Hoi, {BLIP:} bootstrapping language-image
  pre-training for unified vision-language understanding and generation, in:
  K.~Chaudhuri, S.~Jegelka, L.~Song, C.~Szepesv{\'{a}}ri, G.~Niu, S.~Sabato
  (Eds.), International Conference on Machine Learning, Vol. 162 of Proceedings
  of Machine Learning Research, {PMLR}, 2022, pp. 12888--12900.

\bibitem{Lorch21}
L.~Lorch, J.~Rothfuss, B.~Sch{\"o}lkopf, A.~Krause, {DiBS: Differentiable
  Bayesian Structure Learning}, Advances in Neural Information Processing
  Systems 34 (2021).

\bibitem{Maeda21}
T.~N. Maeda, S.~Shimizu, Causal additive models with unobserved variables, in:
  C.~de~Campos, M.~H. Maathuis (Eds.), 37th Conference on Uncertainty in
  Artificial Intelligence, Vol. 161 of Proceedings of Machine Learning
  Research, PMLR, 2021, pp. 97--106.

\bibitem{Shimizu14}
S.~Shimizu, K.~Bollen, Bayesian estimation of causal direction in acyclic
  structural equation models with individual-specific confounder variables and
  non-gaussian distributions, Journal of Machine Learning Research 15~(76)
  (2014) 2629--2652.

\bibitem{Shimizu11}
S.~Shimizu, T.~Inazumi, Y.~Sogawa, A.~Hyv\"{a}rinen, Y.~Kawahara, T.~Washio,
  P.~O. Hoyer, K.~Bollen, Directlingam: A direct method for learning a linear
  non-gaussian structural equation model, Journal Machine Learning Research
  12~(null) (2011) 1225–1248.

\bibitem{Dempster1977}
A.~P. Dempster, N.~M. Laird, D.~B. Rubin, Maximum likelihood from incomplete
  data via the em algorithm, Journal of the Royal Statistical Society. Series B
  (Methodological) 39~(1) (1977) 1--38.

\bibitem{ibrahim20}
A.~Ibrahim, A.~Pretschner, From checking to inference: Actual causality
  computations as optimization problems, in: D.~V. Hung, O.~Sokolsky (Eds.),
  Automated Technology for Verification and Analysis, Springer International
  Publishing, Cham, 2020, pp. 343--359.

\bibitem{Budhathoki21}
K.~Budhathoki, D.~Janzing, P.~Bloebaum, H.~Ng, Why did the distribution
  change?, in: A.~Banerjee, K.~Fukumizu (Eds.), 24th International Conference
  on Artificial Intelligence and Statistics, Vol. 130 of Proceedings of Machine
  Learning Research, PMLR, 2021, pp. 1666--1674.

\bibitem{keras}
F.~Chollet, et~al., Keras, \url{https://keras.io} (2015).

\bibitem{causallearn}
Y.~Zheng, B.~Huang, W.~Chen, J.~Ramsey, M.~Gong, R.~Cai, S.~Shimizu,
  P.~Spirtes, K.~Zhang, Causal-learn: Causal discovery in python, Journal of
  Machine Learning Research 25~(60) (2024) 1--8.

\bibitem{tetrad}
J.~Ramsey, K.~Zhan, M.~Glymour, R.~Sanchez~Romero, B.~Huang, I.~Ebert-Uphoff,
  S.~M. Samarasinghe, E.~A. Barnes, C.~Glymour, {TETRAD - A Toolbox for Causal
  Discovery}, in: 8th International Workshop on Climate Informatics, 2018, pp.
  1--4.

\bibitem{causaldiscoverytoolbox}
D.~Kalainathan, O.~Goudet, Causal discovery toolbox: Uncover causal
  relationships in python (2019).
\newblock \href {http://arxiv.org/abs/1903.02278} {\path{arXiv:1903.02278}}.

\bibitem{lingamtool}
T.~Ikeuchi, M.~Ide, Y.~Zeng, T.~N. Maeda, S.~Shimizu, Python package for causal
  discovery based on lingam, Journal of Machine Learning Research 24~(14)
  (2023) 1--8.

\bibitem{pycausal}
C.~K. Wongchokprasitti, H.~Hochheiser, J.~Espino, E.~Maguire, B.~Andrews,
  M.~Davis, C.~Inskip, bd2kccd/py-causal v1.2.1 (2019).
\newblock \href {https://doi.org/10.5281/zenodo.3592985}
  {\path{doi:10.5281/zenodo.3592985}}.

\bibitem{matchit}
D.~E. Ho, K.~Imai, G.~King, E.~A. Stuart, {MatchIt}: Nonparametric
  preprocessing for parametric causal inference, Journal of Statistical
  Software 42~(8) (2011) 1--28.
\newblock \href {https://doi.org/10.18637/jss.v042.i08}
  {\path{doi:10.18637/jss.v042.i08}}.

\bibitem{bayesnet}
K.~Murphy, {Bayes Net Toolbox for Matlab}, https://github.com/bayesnet/bnt
  (2010).

\bibitem{deepiv}
J.~Hartford, G.~Lewis, K.~Leyton-Brown, M.~Taddy, {Deep IV: A Flexible Approach
  for Counterfactual Prediction}, in: Proceedings of the 34th International
  Conference on Machine Learning {(ICML)}, Vol.~70 of Proceedings of Machine
  Learning Research, PMLR, 2017, pp. 1414--1423.

\bibitem{dowhy}
A.~Sharma, E.~Kiciman, et~al., Do{W}hy: {A Python package for causal
  inference}, https://github.com/microsoft/dowhy (2019).

\bibitem{jfcm}
D.~D. Franciscis, {JFCM Java Fuzzy Cognitive Maps},
  https://github.com/megadix/jfcm (2013).

\bibitem{agrum}
G.~Ducamp, C.~Gonzales, P.-H. Wuillemin, {aGrUM/pyAgrum : a Toolbox to Build
  Models and Algorithms for Probabilistic Graphical Models in Python}, in:
  {10th International Conference on Probabilistic Graphical Models}, Vol. 138
  of Proceedings of Machine Learning Research, Sk{{\o}}rping, Denmark, 2020,
  pp. 609--612.

\bibitem{causalitytool}
A.~Kelleher, S.~Fuhrmann, J.~Attenberg, Causality,
  https://github.com/akelleh/causality (2015).

\bibitem{amostoolibm}
IBM, {SPSS Amos},
  https://www.ibm.com/it-en/products/structural-equation-modeling-sem,
  accessed: 2024-09-25.

\bibitem{Zhou21}
X.~{Zhou}, X.~{Peng}, T.~{Xie}, J.~{Sun}, C.~{Ji}, W.~{Li}, D.~{Ding}, Fault
  analysis and debugging of microservice systems: Industrial survey, benchmark
  system, and empirical study, IEEE Transactions on Software Engineering 47~(2)
  (2021) 243--260.
\newblock \href {https://doi.org/10.1109/TSE.2018.2887384}
  {\path{doi:10.1109/TSE.2018.2887384}}.

\bibitem{Dosovitskiy17}
A.~Dosovitskiy, G.~Ros, F.~Codevilla, A.~Lopez, V.~Koltun, {CARLA}: {An} open
  urban driving simulator, in: 1st Annual Conference on Robot Learning,
  Vol.~78, Proceedings of Machine Learning Research (PMLR), 2017, pp. 1--16.

\bibitem{Wieringa06}
R.~Wieringa, N.~Maiden, N.~Mead, C.~Rolland, Requirements engineering paper
  classification and evaluation criteria: a proposal and a discussion,
  Requirements Engineering 11~(1) (2006) 102--107.
\newblock \href {https://doi.org/10.1007/s00766-005-0021-6}
  {\path{doi:10.1007/s00766-005-0021-6}}.

\bibitem{Vowels22}
M.~J. Vowels, N.~C. Camgoz, R.~Bowden, D’ya like dags? a survey on structure
  learning and causal discovery, ACM Computing Surveys 55~(4) (2022).
\newblock \href {https://doi.org/10.1145/3527154} {\path{doi:10.1145/3527154}}.

\bibitem{kaddour2022causal}
J.~Kaddour, A.~Lynch, Q.~Liu, M.~J. Kusner, R.~Silva, Causal machine learning:
  A survey and open problems (2022).
\newblock \href {http://arxiv.org/abs/2206.15475} {\path{arXiv:2206.15475}}.

\bibitem{Budhathoki22}
K.~Budhathoki, L.~Minorics, P.~Bloebaum, D.~Janzing, Causal structure-based
  root cause analysis of outliers, in: K.~Chaudhuri, S.~Jegelka, L.~Song,
  C.~Szepesvari, G.~Niu, S.~Sabato (Eds.), 39th International Conference on
  Machine Learning, Vol. 162 of Proceedings of Machine Learning Research, PMLR,
  2022, pp. 2357--2369.

\bibitem{zhu2020causal}
S.~Zhu, I.~Ng, Z.~Chen, Causal discovery with reinforcement learning (2020).
\newblock \href {http://arxiv.org/abs/1906.04477} {\path{arXiv:1906.04477}}.

\bibitem{scherrer2022learning}
N.~Scherrer, O.~Bilaniuk, Y.~Annadani, A.~Goyal, P.~Schwab, B.~Schölkopf,
  M.~C. Mozer, Y.~Bengio, S.~Bauer, N.~R. Ke, Learning neural causal models
  with active interventions (2022).
\newblock \href {http://arxiv.org/abs/2109.02429} {\path{arXiv:2109.02429}}.

\bibitem{AMIRINEZHAD202222}
A.~Amirinezhad, S.~Salehkaleybar, M.~Hashemi, Active learning of causal
  structures with deep reinforcement learning, Neural Networks 154 (2022)
  22--30.
\newblock \href {https://doi.org/10.1016/j.neunet.2022.06.028}
  {\path{doi:10.1016/j.neunet.2022.06.028}}.

\bibitem{zhang23}
J.~Zhang, L.~Cammarata, C.~Squires, T.~P. Sapsis, C.~Uhler, Active learning for
  optimal intervention design in causal models, Nature Machine Intelligence
  5~(10) (2023) 1066--1075.
\newblock \href {https://doi.org/10.1038/s42256-023-00719-0}
  {\path{doi:10.1038/s42256-023-00719-0}}.

\bibitem{Steyvers03}
M.~Steyvers, J.~B. Tenenbaum, E.~Wagenmakers, B.~Blum, Inferring causal
  networks from observations and interventions, Cognitive Science 27~(3) (2003)
  453--489.
\newblock \href {https://doi.org/10.1207/s15516709cog2703\_6}
  {\path{doi:10.1207/s15516709cog2703\_6}}.

\bibitem{acharya18}
J.~Acharya, A.~Bhattacharyya, C.~Daskalakis, S.~Kandasamy, Learning and testing
  causal models with interventions, in: S.~Bengio, H.~Wallach, H.~Larochelle,
  K.~Grauman, N.~Cesa-Bianchi, R.~Garnett (Eds.), Proceedings of the 32nd
  International Conference on Neural Information Processing Systems (NIPS),
  Vol.~31, Curran Associates Inc., 2018, p. 9469–9481.

\bibitem{zeng2023survey}
Y.~Zeng, R.~Cai, F.~Sun, L.~Huang, Z.~Hao, A survey on causal reinforcement
  learning (2023).
\newblock \href {http://arxiv.org/abs/2302.05209} {\path{arXiv:2302.05209}}.

\bibitem{Chadbourne2023}
P.~Chadbourne, N.~U. Eisty, Applications of causality and causal inference in
  software engineering, in: 2023 IEEE/ACIS 21st International Conference on
  Software Engineering Research, Management and Applications (SERA), 2023, pp.
  47--52.
\newblock \href {https://doi.org/10.1109/SERA57763.2023.10197835}
  {\path{doi:10.1109/SERA57763.2023.10197835}}.

\bibitem{Furia2023}
C.~A. Furia, R.~Torkar, R.~Feldt, Towards causal analysis of empirical software
  engineering data: The impact of programming languages on coding competitions,
  ACM Transactions on Software Engineering and Methodology 33~(1) (2024) 1--35.
\newblock \href {https://doi.org/10.1145/3611667} {\path{doi:10.1145/3611667}}.

\bibitem{Aguila}
I.~M. del {\'A}guila, J.~del Sagrado, Bayesian networks for enhancement of
  requirements engineering: a literature review, Requirements Engineering
  21~(4) (2016) 461--480.
\newblock \href {https://doi.org/10.1007/s00766-015-0225-3}
  {\path{doi:10.1007/s00766-015-0225-3}}.

\bibitem{Tosun}
A.~Tosun, A.~B. Bener, S.~Akbarinasaji, A systematic literature review on the
  applications of bayesian networks to predict software quality, Software
  Quality Journal 25~(1) (2017) 273--305.
\newblock \href {https://doi.org/10.1007/s11219-015-9297-z}
  {\path{doi:10.1007/s11219-015-9297-z}}.

\bibitem{Misirli}
A.~T. Misirli, A.~B. Bener, A mapping study on bayesian networks for software
  quality prediction, in: Proceedings of the 3rd International Workshop on
  Realizing Artificial Intelligence Synergies in Software Engineering, RAISE
  2014, ACM, 2014, p. 7–11.
\newblock \href {https://doi.org/10.1145/2593801.2593803}
  {\path{doi:10.1145/2593801.2593803}}.

\bibitem{Nageswarao}
N.~G. Nageswarao~M., A survey of bayesian network models for decision making
  system in software engineering, International Journal of Computer
  Applications 134~(8) (2016) 1--5.
\newblock \href {https://doi.org/10.5120/ijca2016906330}
  {\path{doi:10.5120/ijca2016906330}}.

\bibitem{DeSousa}
A.~L.~R. de~Sousa, C.~R.~B. de~Souza, R.~Q. Reis, A 20-year mapping of bayesian
  belief networks in software project management, IET Software 16~(1) (2022)
  14--28.
\newblock \href {https://doi.org/https://doi.org/10.1049/sfw2.12043}
  {\path{doi:https://doi.org/10.1049/sfw2.12043}}.

\bibitem{radlinski2010survey}
L.~Radlinski, A survey of bayesian net models for software development effort
  prediction, International Journal of Software Engineering and Computing 2~(2)
  (2010) 95--109.

\bibitem{Mendes}
E.~Mendes, V.~Freitas, M.~Perkusich, J.~a. Nunes, F.~Ramos, A.~Costa,
  R.~Saraiva, A.~Freire, Using bayesian network to estimate the value of
  decisions within the context of value-based software engineering: A multiple
  case study, International Journal of Software Engineering and Knowledge
  Engineering 29~(11n12) (2019) 1629--1671.
\newblock \href {https://doi.org/10.1142/S0218194019400151}
  {\path{doi:10.1142/S0218194019400151}}.

\end{thebibliography}

\end{document}